\documentclass{aa}  

\def\apjl{ApJ}
\def\apjs{ApJS}
\def\ao{Appl.~Opt.}

\def\aap{A\&A}

\usepackage{graphicx}
\usepackage{txfonts}
\usepackage{graphicx}   
\usepackage{amsmath}    
\usepackage{rotating}
\usepackage{pdflscape}
\usepackage{threeparttable}
\usepackage{color}
\usepackage{lscape}

\newcommand{\msun}{\,M$_{\sun}$}

\begin{document}

   \title{Stellar populations and the origin of thick disks in AURIGA simulations} 

   \author{Francesca Pinna
          \inst{1,}\inst{2,}\inst{3}\thanks{\email{francesca.pinna@iac.es} } 
          \and
          Daniel Walo-Martín \inst{2,}\inst{3}
          \and
          Robert J. J. Grand \inst{4,}\inst{2,}\inst{3}
          \and 
          Marie Martig \inst{4}
          \and
          Francesca Fragkoudi \inst{5}
          \and \\
          Facundo A. Gómez \inst{6,}\inst{7}
          \and
          Federico Marinacci \inst{8}
          \and
          R\"udiger Pakmor \inst{9}
          }
   \institute{Max Planck Institute for Astronomy, Koenigstuhl 17, D-69117 Heidelberg, Germany    
         \and
             Instituto de Astrofísica de Canarias, Calle Vía Láctea s/n, E-38205 La Laguna, Tenerife, Spain
        \and
Departamento de Astrofísica, Universidad de La Laguna, Av. del Astrofísico Francisco Sánchez s/n, E-38206, La Laguna, Tenerife, Spain
\and
Astrophysics Research Institute, Liverpool John Moores University, 146 Brownlow Hill, Liverpool, L3 5RF, UK
\and 
Institute for Computational Cosmology, Department of Physics, Durham University, South Road, Durham DH1 3LE, UK
\and
Instituto de Investigaci\'on Multidisciplinar en Ciencia y Tecnolog\'ia, Universidad de La Serena, Ra\'ul Bitr\'an 1305, La Serena, Chile
\and
Departamento de Astronom\'ia, Universidad de La Serena, Av. Juan Cisternas 1200 Norte, La Serena, Chile
\and
Dipartimento di Fisica e Astronomia "Augusto Righi'', Università di Bologna
via Gobetti 93/2, I-40129 Bologna, Italy
\and
Max-Planck-Institut f\"{u}r Astrophysik, Karl-Schwarzschild-Str. 1, D-85748, Garching, Germany
             }


 
  \abstract
   {Recent integral-field spectroscopy observations of edge-on galaxies have led to significant progress in our knowledge of the ages and chemical compositions of thick disks. 
   However, the origin of thick disks and their evolutionary connection with thin disks is still a matter of debate.}
   {We provide new insights into this topic by connecting the stellar populations of thick disks at redshift $z=0$ with their past formation and growth in 24 Milky Way-mass galaxies from the AURIGA zoom-in cosmological simulations. 
   We assess the role played by mergers of satellite galaxies in the mass assembly of {geometrically defined} thick disks.
   }
   {We projected each galaxy edge on and decomposed it morphologically into two disk components in order to geometrically define the thin and thick disks, as is usually done in observations {of external galaxies}. We produced age, metallicity, and [Mg/Fe] edge-on maps of the 24 galaxies. 
   We quantified the impact of satellite mergers by mapping the distribution of ex situ stars. 
   }
   {Thick disks are on average $\sim 3$~Gyr older, $\sim 0.25$~dex more metal poor, and $\sim 0.06$~dex more [Mg/Fe]-enhanced than thin disks. Their average ages range from $\sim 6$ to $\sim 9$~Gyr, metallicities from $\sim -0.15$ to $\sim 0.1$~dex, and [Mg/Fe] from $\sim 0.12$ to $\sim 0.16$~dex. These properties are the result of an early initial in situ formation, followed by a later growth driven by the combination of direct accretion of stars, some in situ star formation fueled by mergers, and dynamical heating of stars. The balance between these processes varies from galaxy to galaxy and impacts thick-disk ages and metallicities. The oldest thick disks (older than 8~Gyr) are hosted by galaxies with a low mass fraction of accreted stars (below 8\%), while the youngest thick disks (younger than 7 Gyr) are found in galaxies with higher accreted fractions (larger than 25\%). Mergers play a key role in {the mass assembly of thick disks, contributing an average accreted mass fraction of $\sim 22$\% in the analyzed thick-disk-dominated regions. In two galaxies, about half of the geometric} thick-disk mass was directly accreted. The mass fraction of accreted stars is lower than 10\% only in four thick disks. While primordial thick disks form at high redshifts in all galaxies, young metal-rich thin disks, with much lower [Mg/Fe] abundances, start to form later but at different times (at higher or lower redshifts) depending on the galaxy.
   }
   {We conclude that thick disks, although mostly formed in situ, grow thanks to the significant contribution of satellite mergers, especially through the direct accretion of stars. They result from the interplay of external processes with the internal evolution of the galaxy. 
   }


   \maketitle
%

\section{Introduction}\label{sec:intro}

Galactic internal structures are important tracers of global galaxy formation and evolution. In particular, properties of massive stellar disks naturally allow us to reconstruct, with great detail, the mass assembly of disk-dominated galaxies. 
Only bright thin disks were initially identified in galaxies, with fainter thick disks detected much later. The latter were first discovered in edge-on galaxies, defined as a second (thicker) component that was necessary to fit luminosity vertical profiles of galactic stellar disks \citep{Burstein1979}. A thick disk was initially defined, therefore, as a morphological component with a larger scale height than the thin disk. While the midplane region was dominated by the thin disk, the thick-disk dominated the light beyond a certain distance from the midplane of the galaxy. 
Only later on, when this second and thicker component was also identified in the solar neighborhood of the Milky Way, were stars in the thick disk found to have different properties from those in the thin disk. Thick-disk stars were associated with old ages, much lower metallicities, and higher velocity dispersions than those in the thin disk \citep{Gilmore1983}. 
About one decade later, it was shown that the geometrically defined thick and thin disks, in the solar neighborhood, formed two different sequences in the [Mg/Fe]-metallicity plane, and that thick disks were enhanced in $\alpha$ elements \citep[][]{Fuhrmann1998}.
Old ages, low metallicities, and high [Mg/Fe] abundances suggested that the Milky Way thick disk was formed at early times from gas that was still poorly enriched. Assuming that thick disks in external galaxies also showed similar properties, as was later shown \citep[e.g.,][]{Yoachim2008a}, understanding their origin means shedding light on the early stages of galaxy formation and evolution. 
On the other hand, young, metal-rich, and kinematically cool thin disks trace later and relatively recent phases of the evolution of a galaxy. 
Thus, tracing the full history of a disk galaxy requires understanding the processes leading to the formation of both thick and thin disks, as well as the evolutionary connection between them. 

Different scenarios have been proposed to explain the early formation of the thick disk and the later transition to star formation in the thin disk. 
One possible explanation resides in turbulent gas at high redshifts, during the epoch of frequent gas-rich mergers \citep{Brook2004}. Thick disks were formed already thick from that dynamically hot gas. This gas settled down with time, allowing for the formation of stars in much thinner layers at later times and giving birth to the thin disks. 
A different scenario suggests that stars were instead formed with low velocity dispersions, in a relatively thin disk at high redshifts. These stars were later dynamically heated through processes such as mergers or encounters with clumps, either violently or slowly with time \citep{Quinn1993, Villalobos2008, Bournaud2009, diMatteo2011}. 
Stars that are now older, such as those in the thick disk, had more time to be heated and were more exposed to violent processes (e.g., related to mergers and intense star formation) than young stars (in the thin disk). 
According to a third scenario, thick disks are mostly made up (more than 60\% of their mass) by the debris of accreted satellites of ex situ origin \citep{Abadi2003}. Most of the thin disk would have formed during a later quiescent phase, during which no mergers occurred. 

Since these scenarios, based on numerical studies, were initially proposed, 
observational works have been alternatively supporting one or another. 
Early thick-disk spectroscopic studies in edge-on galaxies used two long slits, placed on and off the galaxy midplane, respectively in the thin- and thick-disk-dominated regions. \citet{Yoachim2005,Yoachim2008a} extracted the kinematics of the thick and thin disks of a sample of nine edge-on late-type galaxies. They found a variety of kinematic properties, with high-mass galaxies (with a circular velocity above 120 km s$^{-1}$) showing similar rotation curves in the thick and thin disks, and low-mass galaxies showing strong differences. The substantial lag and counter-rotation identified in lower-mass galaxies strongly supported an accretion origin for thick disks. \citet{Yoachim2008b} observed (also in a sample of nine late-type galaxies) thick disks that were much older than the young thin disks, both with similar low metallicities. 
On the other hand, the long-slit study by \citet{Kasparova2016} showed three thick disks with diverse properties and proposed that different galaxies can form their thick disks via different scenarios. 

The advent of integral-field spectroscopy (IFS) has brought important advances in this field, providing the two-dimensional kinematic and stellar-population structure of thin and thick disks and covering the transition region between them. 
IFS observations of thick disks are challenging due to their low surface brightness, especially in late-type galaxies, and require long integration times. Furthermore, galaxies need to be seen edge on in order to dissect their vertical structure into different components. However, there are not many edge-on galaxies with deep enough archival observations. For all these reasons, there are currently few IFS thick-disk studies in the literature, and most of them include only one or two galaxies. 
The first attempt to map thick-disk properties with IFS was done with the VIsible Multi-Object Spectrograph (VIMOS) at the Very Large Telescope (VLT) at poor spatial resolution (only one Voronoi bin in the thick disk; \citealt{Comeron2015}).  The sharp age and metallicity differences between the thick and thin disk supported a fast thick-disk formation at high redshifts. 
Later studies with the Multi Unit Spectroscopic Explorer (MUSE), at a much higher resolution, revealed subtler differences between the thin and thick disks. \citet{Comeron2016} present an edge-on lenticular galaxy  with clear metallicity differences between the thick and thin disks, although very similar old ages. This suggested a fast internal formation of both thin and thick disks at high redshifts. 

\citet{Pinna2019b,Pinna2019a} added [Mg/Fe] maps, as well as star-formation histories coupled with chemical properties, to thick-disk studies. They analyzed MUSE observations of three lenticular galaxies in the Fornax cluster, located in regions of the cluster with different densities. 
In all cases, thick disks had similar properties, which suggested that they formed and grew via similar mechanisms. 
 These two studies were the first to show a younger, more metal-poor component, interpreted as having been accreted, in addition to an old and metal-poor dominant population,  which may have formed in situ.
 The authors proposed a complex thick-disk formation scenario with a dominant in situ formation combined with a significant contribution from accreted satellites. Stellar-population properties of thin disks varied from galaxy to galaxy (possibly related to different impacts of the cluster environment). Flaring in the outer disk was interpreted as a potential sign of dynamical heating through mergers.
A similar complex thick-disk formation scenario was supported by the MUSE analysis from \citet{Martig2021} of one massive disk-dominated (Sb) galaxy. Here, an important merger left a significant contribution in the thick disk, and probably drove an enhancement of star formation in the thin disk.
On the other hand, the lack of evidence for retrograde material in the MUSE kinematics of eight additional late-type galaxies (only five of them with a clear thick-disk component) suggested a predominant internal origin for thick-disk stars, still allowing for some fraction of accreted material \citep{Comeron2019}. 
Finally, a recent MUSE study of a young Sc galaxy showed a massive, thicker, metal-poor disk and an inner, small, very thin and metal-rich disk, which might be a future thin disk in its first stages of formation \citep{Sattler2023}. These results support a first buildup of a thicker disk (which occurred late in this case, since the galaxy is overall young), before the later inside-out formation of a thin disk. 

Despite the recent progress on this matter, the origin of thick disks and their evolutionary connection with thin disks are still partially unclear. 
The abovementioned recent IFS observations revealed that mergers played a key role in the mass assembly of the thick and thin disks. 
However, the relative importance of ex situ and in situ components, and therefore the balance between direct accretion of ex situ stars and in situ star formation, which can also occur from accreted gas, is still uncharted waters. 
In this context, numerical simulations are paramount since they can offer large samples of galaxies, with a wide variety of different properties and an extended coverage of thin and thick disks, resolving some of the issues related to the observational work. Furthermore, the full evolution of thick and thin disks and their properties is tracked at different snapshots of the simulation. 
Several studies have used cosmological simulations to investigate the chemical bimodality in the [$\alpha$/Fe]-metallicity plane of (thick- and thin-) disk stars. \citet{Grand2018a}, based on the six highest-resolution simulated galaxies of the AURIGA suite \citep{Grand2017}, proposed an early and fast formation of the high-$\alpha$ sequence, during an intense star-formation phase induced by gas-rich mergers. The high-$\alpha$ component is usually associated with the thick disk in both the Milky Way \citep[e.g.,][]{Hayden2015} and external galaxies \citep{Scott2021}. One of these gas-rich mergers would have provided the metal-poor gas necessary to start the formation of the low-$\alpha$ sequence, associated with the thin disk. This would have been followed by a more quiescent gas-enrichment phase driving the growth of the thin disk. \citet{Buck2020b}, using four galaxies from the Ultra High-Definition (UHD) simulation suite \citep{Buck2020a} of the NIHAO (Numerical Investigation of a Hundred Astrophysical Objects) project \citep{Wang2015}, reached a similar conclusion. 
This idea was also supported by some observational studies \citep[e.g.,][]{Scott2021}. 
VINTERGATAN, a cosmological zoom simulation of a Milky Way-mass disk galaxy, supported a similar picture as well \citep{Agertz2021}. 

\citet{Grand2020}, also based on AURIGA simulations, proposed a ``dual origin'' for the Milky Way thick disk, 
which would comprise two components. One in situ component (with similar properties to the ``starburst sequence'' shown by \citealt{An2023}) would have formed during a starburst triggered by the gas-rich merger related to the Gaia-Enceladus Sausage (see also \citealt{Ciuca2024}). 
A second component would have formed from stars from a preexisting thinner disk that was dynamically heated by the same merger. This second component explains the positive velocity-metallicity relation that was found in the Milky Way thick disk \citep{Belokurov2020}. 
The analysis based on Milky Way-mass galaxies in the EAGLE suite of cosmological simulations shows that the chemical bimodality found in the Milky Way might not be very common in disk-dominated galaxies \citep{Mackereth2018}.
This bimodality tends to appear only when accretion episodes of different intensities happen at different times. 
Early fast gas accretion and subsequent bursty star formation would lead to the formation of the high-$\alpha$ sequence. 
The geometrically defined thick disk would result from the combination between this inner high-$\alpha$ sequence and the flaring outer disk region of the low-$\alpha$ sequence \citep{Mackereth2019}. 
An alternative scenario, not involving galaxy mergers, hypothesizes that intense star formation in clumps naturally forms the high-$\alpha$ thick disk, while the low-$\alpha$ thin disk is the result of a more evenly distributed star formation \citep{Clarke2019}.

Additional studies of thick disks in galaxies, without a special focus on the Milky Way, have drawn similar conclusions as the abovementioned works. 
\citet{Martig2014a, Martig2014b} analyzed the mono-age populations in seven galaxies from \citet{Martig2012} and find that galaxies with a quiescent merger history tend to show a continuous distribution of scale heights correlated with ages.  However, mergers are capable of causing an age bimodality between a thick and a thin component. While most of the very old stars were born already dynamically hot, later generations were heated with time, partially due to mergers.
\citet{GarciadelaCruz2021}, expanding the sample to 27 simulated galaxies from \citet{Martig2012}, fitted the vertical distribution of stars in each of them with two disk components. Part of the sample, with lower galaxy and thick-disk masses and a more quiescent merger history, shows a continuous and smooth age vertical gradient (mono-age populations whose scale height increases with age). The rest of the sample, with higher galaxy masses, showed a thicker morphology, higher thick-disk masses and mass fractions, and flaring thick disks (see also \citealt{Minchev2015}). 
\citet{Park2021} also morphologically decomposed the thick from the thin disks in 19 simulated galaxies 
and find that geometrically defined thick disks are predominantly formed in situ, although they host a large fraction of accreted stars. 
The ex situ contributions in AURIGA stellar disks were quantified by \citet{Gomez2017a}, who showed that accreted stars are distributed in a thicker region, are older, and are more metal poor than the in situ disk component. 
 \citet{Yu2021} used a kinematic definition for thick disks (orbital circularities lower than 0.8) and investigated their origin in FIRE-2 (Feedback In Realistic Environments-2) simulations. While the thick disk is mostly formed in an early bursty phase of star formation, later mergers also contributed to its growth with an additional component of kinematically heated younger stars. 
 
In summary, thick-disk studies from cosmological simulations broadly support a predominant in situ formation during a bursty phase of star formation, with mergers playing a key role, potentially combined with dynamical heating and/or accretion of stars. 
However, numerical studies focused on a comparison with (IFS) observations of edge-on external galaxies, which would allow us to test the pictures drawn by observations and better interpret observed stellar-population properties, are currently lacking. 
In this work, based on a geometrical definition of the thick and thin disks, we use zoom-in magnetohydrodynamic cosmological simulations from the AURIGA project \citep{Grand2017}\footnote{Some AURIGA simulation data and the face-on maps of stellar  kinematic and populations for the sample analyzed here, are available on the project webpage: \url{https://www.mpa.mpa-garching.mpg.de/auriga/}.}.
Projecting simulations to an edge-on view allows us to perform our analysis in a way similar to that of present and future observations of edge-on galaxies. 

This paper is structured as follows. Section~\ref{sec:sim_sample} describes the simulations, the data projection and spatial binning, and the selected sample. Section~\ref{sec:analysis} explains the methods used for the analysis, including the morphological decomposition into a thick and a thin disk. We present our results in Sect.~\ref{sec:results}, discuss them in Sect.~\ref{sec:disc}, and summarize our conclusions in Sect.~\ref{sec:concl}.

\section{Our sample of 24 simulated galaxies}\label{sec:sim_sample}

We selected our galaxy sample from the original suite of AURIGA magneto-hydrodynamical zoom-in cosmological simulations of galaxy formation and evolution. We describe in this section the simulations and our selected sample.

\subsection{AURIGA zoom-in cosmological simulations} \label{sub:sim}
The AURIGA project\footnote{\url{https://www.mpa.mpa-garching.mpg.de/auriga/}} \citep{Grand2017} includes zoom-in magneto-hydrodynamical cosmological simulations of 30 Milky Way-mass late-type galaxies, with stellar masses between $10^{10}$\msun\,and slightly more than $10^{11}$\msun. These were obtained by re-simulating 30 halos from the parent largest-volume Dark Matter Only 
EAGLE simulation \citep{Schaye2015}, at intermediate resolution (L100N1504). 
These halos were selected to have a virial mass $1<M_{200}/10^{12} \rm{M}_{\sun}<2$ at $z=0$, defined as the mass contained inside the virial radius defined as $R_{200}$, the radius enclosing a mean mass volume density of 200 times the critical density for closure. 
The 30 halos were also selected to be relatively isolated at redshift $z=0$. In particular, they  were randomly chosen from those in the lowest $\tau_{\rm iso, max}$ quartile, where $\tau_{\rm iso, max}$ is the maximum value of the tidal isolation parameter $\tau_{\rm iso,i} = (M_{200,i})/(M_{200}) \times (R_{200}/R_i)^3$; here, $M_{200,i}$ and $R_i$ are respectively the virial mass of and distance to the $i$th halo in the simulation.

These simulations were obtained using an improved version of the moving-mesh code AREPO \citep{Springel2010, Rudiger2016,Weinberger2020}, which includes magneto-hydrodynamics and collisionless dynamics in a cosmological context.
The initial conditions to be used in the re-simulations, and in particular the initial distribution of dark-matter particles, were taken from the halos in the parent simulation. Then, each dark-matter particle was substituted with a pair of a dark-matter particle and a gas cell. A fixed mass was assigned to each gas cell based on the value of the adopted cosmic baryon fraction (see \citealt{Grand2017}). 
AURIGA simulations offer different resolution levels, {and in this work we use the fiducial resolution at "level 4," the only suite including the full sample of 30 galaxies. }
This level corresponds to typical masses of high-resolution dark-matter and baryonic-mass particles of $\sim 3 \times 10^5$\msun\,and $\sim 5 \times 10^4$\msun, respectively \citep{Grand2017}. 
The comoving gravitational softening length for stellar and high-resolution dark-matter particles was set to 500~$h^{-1}$cpc, with a physical gravitational softening length of 369~pc at redshift $z<1$. The softening length for gas cells was scaled by the cell size, from a minimum value of physical softening length of 369~pc to a maximum value of 1.85~kpc. This ensures that gas cells with lower densities (and thus larger volumes) have larger softening lengths than those with higher densities High-density gas cells can be smaller than their softening length since each cell has a given target mass, so high-density regions are resolved with more cells than low-density regions. 
{Cell sizes range from 80 to 300~pc in the region within the galaxies that we analyze in this work (Sect.~\ref{sub:region}). 
}

\begin{table*}[!h]
\caption{Main properties of the galaxies in our sample.}
\centering
\begin{tabular}{cccccc}
\hline\hline
Galaxy name & Hubble type $^{(1)}$ & $M_{\ast}$ ($10^{10}$\msun) $^{(2)}$& $R_{0,disk}$ (kpc) $^{(3)}$& $R_{opt}$ $^{(2)}$ (kpc) & $h_{scale}$ $^{(4)}$ (kpc) \\ 
\hline
Au1     & SBb  & 2.75   & 4.07 & 20.0 & 2.8 \\
Au2     & SBc  & 7.05   & 8.99 & 37.0 & 3.5  \\
Au3     & Sb   & 7.75   & 4    & 31.0 & 2.7  \\
Au5     & SBb  & 6.72   & 4.58 & 21.0 & 3.6 \\
Au6     & SBbc & 4.75   & 5.43 & 26.0 & 3.3 \\
Au7     & SBb  & 4.88   & 5.43 & 25.0 & 5.1 \\
Au8     & Sc   & 2.99   & 3    & 25.0 & 3.0 \\
Au9     & SBb  & 6.10   & 6.45 & 19.0 & 2.8 \\
Au10    & SBa  & 5.94   & 6.45 & 16.0 & 3.3 \\
Au12    & SBab & 6.01   & 3.73 & 19.0 & 13.1 \\
Au14    & SBb  & 10.39  & 5.09 & 26.0 & 6.2 \\
Au15    & Sbc  & 3.93   & 3    & 23.0 & 4.4 \\
Au16    & Sc   & 5.41   & 4    & 36.0 & 4.0 \\
Au17    & SBa  & 7.61   & 5.43 & 16.0 & 2.4 \\
Au18    & SBb  & 8.04   & 6.45 & 21.0 & 2.8 \\
Au19    & Sbc  & 5.32   & 3    & 24.0 & 3.8 \\
Au21    & SBb  & 7.72   & 3.39 & 24.0 & 4.3 \\
Au22    & SBa  & 6.02   & 6.28 & 13.5 & 2.7 \\
Au23    & SBbc & 9.02   & 9.50 & 25.0 & 2.9 \\
Au24    & SBc  & 6.55   & 5.09 & 30.0 & 4.2 \\
Au25    & SBb  & 3.14   & 4.41 & 21.0 & 2.1 \\
Au26    & SBa  & 10.97  & 5.43 & 18.0 & 3.2 \\
Au27    & SBbc & 9.61   & 5.77 & 26.0 & 3.7 \\
Au28    & SBa  & 10.45  & 7.46 & 17.5 & 3.8 \\
\hline
\end{tabular}
\label{tab:sample}
\begin{tablenotes}
\item {\footnotesize Notes. 
{(1) Hubble types from \citet{Walo2021}; (2) Stellar mass $M_{\ast}$ and optical radius $R_{opt}$ (defined as the radius where the $B$-band surface brightness drops below 25~mag arcsec$^{-2}$) are from \citet{Grand2017}; (3) $R_{0,disk}$, the minimum radius at which the thin disk dominates over the central component. When the galaxy is barred, this value is the length of the bar taken from \citet{Blazquez2020}, following \citet{Walo2021}. 
When the galaxy is unbarred, $R_{0,disk}$ is the approximate radius where the disk starts to dominate over the bulge, according to Fig.~4 in \citet{Grand2017}; (4) The $h_{scale}$ parameter was calculated here as the standard deviation of the vertical position of particles at $0.5 R_{opt}$, following \citet{GarciadelaCruz2021}. 
}
\normalsize
}
\end{tablenotes}
\end{table*}

 The galaxy formation model, including star formation, stellar and black-hole (BH) feedback, and magnetic fields, led to realistic galaxies matching a large variety of properties found in observed galaxies (e.g., morphologies and properties of structures such as spiral arms or bars, galaxy sizes and masses, and kinematic and chemical properties; see, e.g., \citealt{Grand2016,Grand2017}). 
The interstellar medium is characterized by two phases with cold and dense gas clouds embedded in a hot medium. These two phases are not modeled explicitly, but only as part of a sub-grid model \citep{Springel2003}. Gas becomes thermally unstable and forms stars above a density threshold of 0.13~cm$^{-3}$.
According to a \citet{Chabrier2003} initial mass function, each star-forming gas cell is either converted to a star particle or selected to be a site for type II supernova feedback.  
In the first case, the star particle represents a single stellar population (SSP) of specific masses, ages, and chemical abundances. 
In the second case, a single wind particle is launched in a random direction, and travels until it releases its energy and metals (40\% of the total metals in the original gas cell) in the final gas cell. 
The rest of the metals remain local.  
Stellar feedback from type Ia supernova and asymptotic giant branch stars are also modeled, in terms of metals and mass loss. 
BHs with a seed mass of $10^5$\msun $h^{-1}$ were introduced in the position of the densest gas cell in most-massive halos ($> 5 \times 10^{10} {\rm M_{\sun}} h^{-1}$), and acquire mass from gas cells or merging with other BHs. BH feedback injects thermal energy to surrounding gas cells following two modes: in local isotropic quasar mode, or in a randomly isotropic, nonlocal radio mode. 
A homogeneous magnetic field was introduced at the beginning of the zoom-in re-simulations and let evolve. 
The simulations were saved in 128 different snapshots corresponding to different redshifts. 
For further details on the implemented physical models or general properties of these simulations, we refer the reader to \citet{Grand2017}, who extensively described them.


\begin{landscape}
\begin{figure}
\centering
\resizebox{1.35\textwidth}{!}
{\includegraphics[scale=1.]{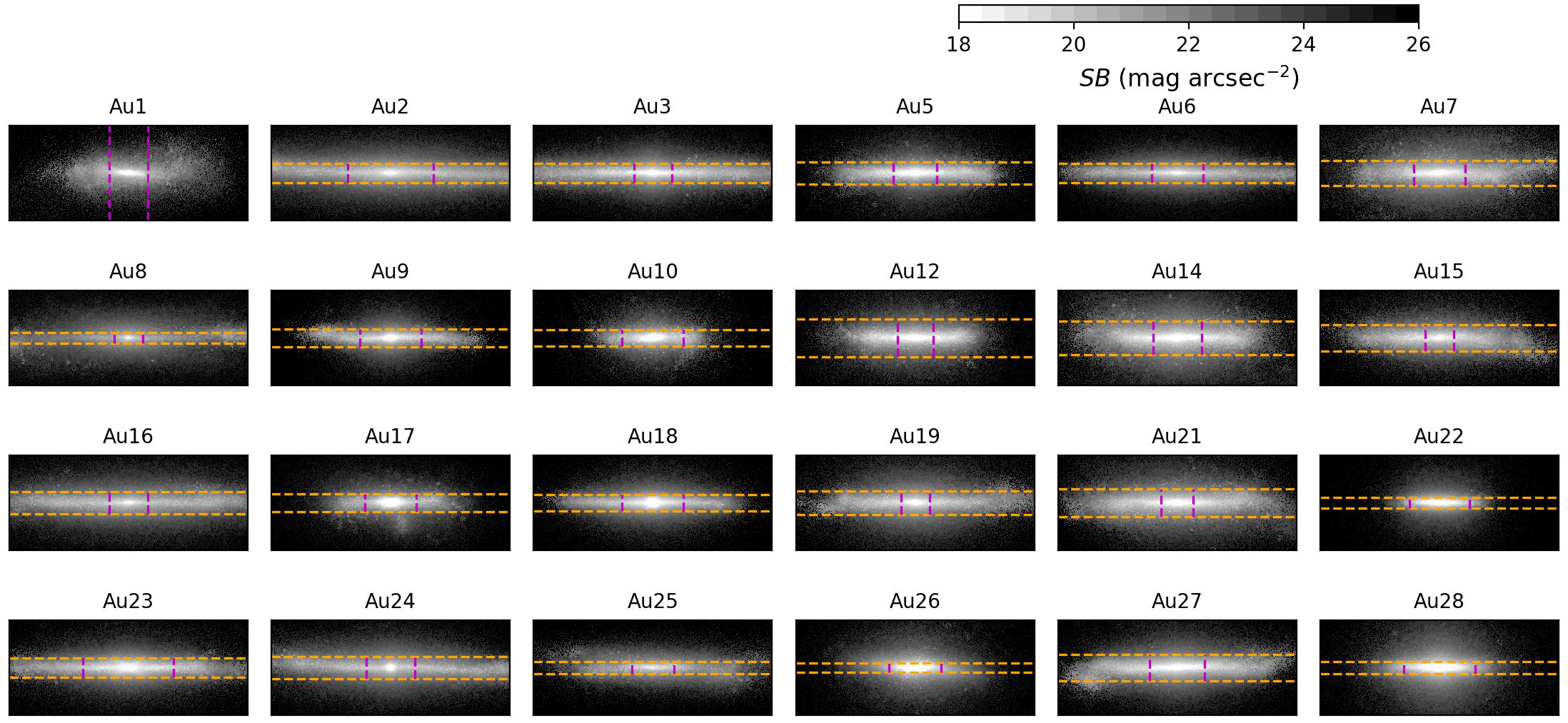}}
\caption{Mock edge-on images, in terms of surface brightness in the $V$ band (\textit{SB}), of our selected sample of 24 Milky Way-like galaxies from AURIGA simulations. For display purposes, we show for all galaxies an area of the same size, of radius 25~kpc and height 10~kpc. Surface brightness was projected into pixels of 0.1~kpc$\times$0.1~kpc size and was Voronoi binned to obtain at least 100 star particles per bin. 
The region between the two horizontal lines is where the light of the thin-disk component dominates over that of the thick disk (see Sect.~\ref{sub:morph_decomp} for details). The region between the two vertical dashed magenta lines is dominated by a central component (the bar in barred galaxies and a classical bulge in non-barred galaxies). Names of the AURIGA halos are indicated on top of each panel.
}
\label{fig:sample_images}
\end{figure}
\end{landscape}
\subsection{Projection, spatial binning, and sample selection}\label{sub:proj}

We projected to an edge-on view the 30 AURIGA galaxies, to allow for an approach as similar as possible to what was done in previous observational works on edge-on disk galaxies \citep[e.g.,][]{Pinna2019a,Pinna2019b,Martig2021}. 
To perform the projection, we followed the approach explained in \citet{Walo2021}. 
We first aligned galaxies with the axes of the reference system, by aligning with the vertical axis $Z$ the average vector of the angular momentum of young stars (younger than 3~Gyr) located within a sphere of a radius of 60~kpc around the center of potential. 
Afterward, we projected the properties of all stellar particles {to the $YZ$ plane along the line of sight (LOS),} which is parallel to the $X$ axis and lying on the $XY$ plane. The center of the coordinate system was placed on the center of the galaxy, calculated as the potential minimum of the main subhalo.

Since the number of particles decreases toward the outskirts of galaxies, we applied a Voronoi binning \citep{Cappellari2003} following \citet{Walo2021}, to ensure a minimum number of particles and a roughly constant number of particles per bin. 
We used three different spatial-resolution setups, designed for different purposes. One used a projection into pixels of size 0.1~kpc$\times$0.1~kpc, and was designed to produce mock images and allow for smooth morphological fits of the thin and the thick disks (Sect.~\ref{sub:morph_decomp}).  
We included all pixels with at least one star particle, and we performed a Voronoi binning to a target number of particles of 100 per bin, following requirements in previous work \citep{Schulze2018,Walo2020}. 
The second setup used a pixel size of 0.5~kpc$\times$0.5~kpc, including all pixels with more than two particles, and used a Voronoi binning to at least 900 particles per bin. 
This second setup was designed to allow the storage of the full distribution of the stellar-population parameters in each Voronoi bin, plus the information about the origin of star particles (ex situ or in situ, defined in  Sect.~\ref{sub:method_maps}), still in files of a reasonable size. 
With a pixel size well beyond the softening length (Sect.~\ref{sub:sim}), and a number of particles per bin well beyond the usual minimum requirement of 100 particles, this setup is optimal for the analysis of the stellar properties of the thin and thick disks. It was used for stellar-population maps and the spatial distribution of accreted stars (Sect.~\ref{sub:SPmaps} and \ref{sub:accr}). 
Finally, when mapping different simulation snapshots, we aimed at preserving spatial resolution at higher redshifts when galaxies are smaller and contain a lower number of particles. For this, we used a pixel size of 0.2~kpc$\times$0.2~kpc. We binned to 16 particles per bin and included all pixels with at least one star particle. 
However, the very low density of star particles at the highest redshift in our maps ($z=3.5$) required a pixel size of 0.5~kpc$\times$0.5~kpc coupled with a Voronoi binning to only 9 particles per bin (Sect.~\ref{sub:snap}). 

The projection and binning process provided the average properties of particles in each Voronoi bin in the given projection, including surface brightness in different bands and stellar-population parameters. 
The photometric properties of the star particles were obtained by \citet{Grand2017} using stellar-population synthesis models from \citet{Bruzual2003}, considering each star particle as a SSP. 
From the original sample of 30 Milky Way-size galaxies in the AURIGA suite of simulations, which were identified with the prefix "Au" plus numbers from 1 to 30 \citep{Grand2017}, we excluded from our analysis six of them (Au4, Au11, Au13, Au20, Au29, Au30). These galaxies showed, in their edge-on projection, strong distortions, lack of a disky shape or in one case an ongoing major merger. 
The main properties of the selected sample are indicated in Table~\ref{tab:sample}. Most galaxies in the sample are barred, while only five of them are unbarred. Stellar masses vary between $\sim 3 \times 10^{10}$\msun\,and $\sim 10^{11}$\msun.

\section{Analysis methods} \label{sec:analysis}

\subsection{Analyzed region} \label{sub:region}
Galaxies in our sample, when projected and binned as explained in Sect.~\ref{sub:proj}, are much more extended than regions that are usually observed in edge-on galaxies, due to the low surface brightness in the outskirts of edge-on galaxies (examples in \citealt{Comeron2016, Comeron2018, Comeron2019, Pinna2019a, Pinna2019b, Martig2021}). 
We selected for the analysis in this work a region that was comparable to the region typically observed in the photometric study of edge-on galaxies \citep{Comeron2018} and to the typical coverage reached in published studies using two deep MUSE pointings (at a distance between 20 and 30~Mpc, \citealt{Pinna2019a, Pinna2019b, Martig2021,  Sattler2023}). 
We selected a region with a radius of $0.8 R_{opt}$ and a vertical extension of $2 h_{scale}$ (Table~\ref{tab:sample}) above and below the midplane. 
$R_{opt}$ is the optical radius, {defined as the radius where the $B$-band surface brightness drops below 25~mag arcsec$^{-2}$}, while the parameter $h_{scale}$ was calculated as the standard deviation of the vertical positions of stellar particles at $0.5 R_{opt}$, as defined in \citet{GarciadelaCruz2021}. Thus, the analyzed region has a total size of $1.6 R_{opt} \times 4h_{scale}$ in the edge-on projection. 
This region was used for the morphological decomposition in Sect.~\ref{sub:morph_decomp} and all calculations throughout the paper, while for display purposes we use a central region of 50~kpc$\times$20~kpc, equal for all galaxies and still similar to the $1.6 R_{opt} \times 4h_{scale}$ region. 
We show in Fig.~\ref{fig:sample_images} mock images (surface brightness in the $V$ band) of this 50~kpc$\times$20~kpc region for the 24 galaxies in our sample. 
{These mock images were obtained by projecting to the edge-on view the luminosity density of the star particles in the $V$ band (see Sect.~\ref{sub:proj}), and converting the maps to the surface brightness. }
The setup with 0.1~kpc$\times$0.1~kpc pixel size and a Voronoi binning with a target number of 100 star particles per bin (see Sect.~\ref{sub:proj}) was used for these images.

\subsection{Morphological decomposition into a thick and a thin disk}\label{sub:morph_decomp}
In this paper, we aim at analyzing thick and thin disks defined in a purely geometrical way, similarly to what is usually done in observations of edge-on galaxies \citep[e.g.,][]{Comeron2015, Comeron2016, Pinna2019a, Pinna2019b, Martig2021}. This definition consists of defining a region dominated by the thin disk, close to the midplane, and a region dominated by the thick disk, at larger heights, based on a morphological decomposition into two disk components (see, e.g., \citealt{Comeron2018}). 

We decomposed morphologically our edge-on projected galaxies into two disk components. For that purpose, we fitted vertical surface brightness profiles in a similar way to what is usually done in observations \citep[e.g.,][]{Comeron2018}. We included in the fit the region described in Sect.~\ref{sub:region}, of radial extension $0.8 R_{opt}$ following \citet{Comeron2018}. We excluded the region within a radius of $R_{0,disk}$ (Table~\ref{tab:sample}), dominated by the bar in barred galaxies and by a classical bulge in non-barred galaxies. 
We divided the projected $1.6 R_{opt} \times 4h_{scale}$ area of each galaxy into four quadrants with a radial range between $R_{0,disk}$ and $0.8 R_{opt}$, two on each side with respect to the galactic center, one above and one below the midplane. 
We additionally divided each quadrant into two equal radial bins, obtaining eight bins in total. 
For each one of the eight radial bins, we extracted a median vertical profile of the luminosity density in the $V$ band. 
We modeled the vertical luminosity density with a double hyperbolic secant square (following the study regarding one Galactica simulation by \citealt{Park2021}): 
\begin{equation} \label{eq:double_disk}
        \Sigma (z) = \Sigma_{0,\rm{thin}} {\rm{sech}}^2 \left(\frac{|z|}{2 h_{z,\rm{thin}}}\right) + \Sigma_{0,\rm{thick}} {\rm{sech}}^2 \left(\frac{|z|}{2 h_{z,\rm{thick}}}\right) 
,\end{equation}
where $\rm\Sigma_{0,thin}$ and $\rm\Sigma_{0,thick}$ are the luminosity densities in the midplane of the thin and the thick disks, and $ h_{z,{\rm thin}}$ and $ h_{z,{\rm thick}}$ are their respective scale heights. 

\begin{figure}
\centering
\resizebox{0.53\textwidth}{!}
{\includegraphics[scale=1, trim={0.8cm 0cm 0cm 0}]
{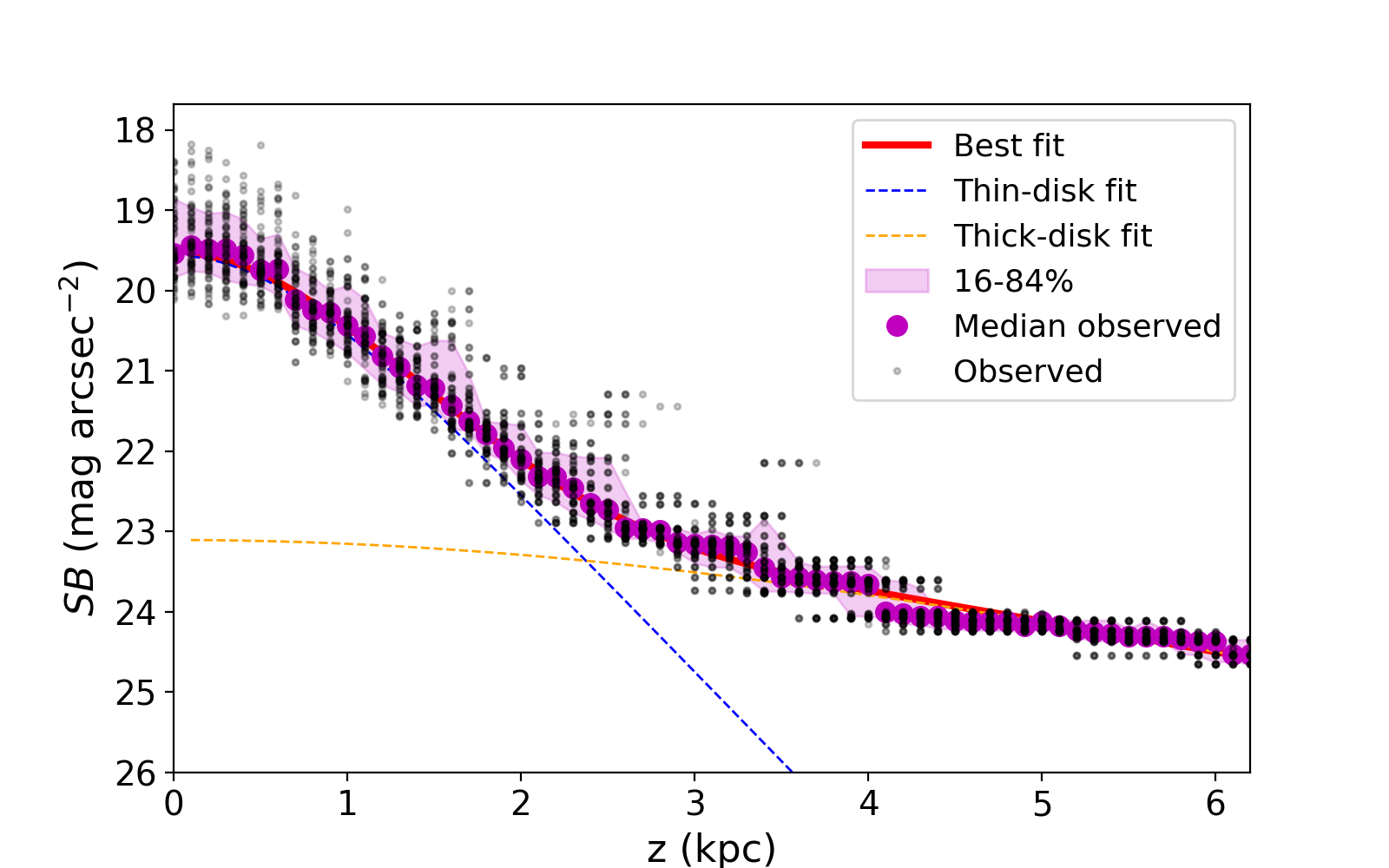}}
\caption{Fit of the vertical surface-brightness (SB) profile of one radial bin of the galaxy Au5. Each small gray circle is the measurement for one Voronoi bin, while larger magenta circles correspond to the median profile of the gray points. The magenta shading indicates the $1\sigma$ dispersion of the gray points. Dashed blue and orange lines correspond respectively to the thin- and thick-disk fits, and the solid red line is a combination of the two. 
}
\label{fig:fit}
\end{figure}

We used a Bayesian approach to fit these four parameters and estimate their uncertainties. We used the {\it emcee} Python implementation \citep{Foreman-Mackey2013} of the affine-invariant Markov chain Monte Carlo (MCMC) ensemble sampler from \citet{Goodman2010}. We used 100 walkers and 20\,000 chains. 
We maximized a likelihood $p$ of the form $\ln (p) = -0.5 \sum_n ({\rm SB}_{obs,n} - {\rm SB}_{mod,n})^2 $, where ${\rm SB}_{obs,n}$ is the observed surface brightness on the point $n$, and ${\rm SB}_{mod,n}$ is the modeled surface brightness from Eq.~\ref{eq:double_disk} for the point $n$. This is a Gaussian likelihood assuming equal variance for all the points. 
We excluded from the fits the points fainter than 26~mag arcsec$^{-2}$ and the points obtained from averaging fewer than eight Voronoi bins. 
We used uniform priors to constrain the parameters to the values possible in the analyzed region,  
the thin disk to be brighter than the thick disk in the midplane, and to have a shorter scale height than the thick disk.  
We show in Fig.~\ref{fig:fit} one fit example of the vertical surface-brightness profile of one radial bin. 
For each profile, we calculated the height $z_{tT,i}$ where the thick disk starts to dominate over the thin disk, as the height where the two hyperbolic secant squares cross each other (Fig.~\ref{fig:fit}). We averaged these values for all eight vertical profiles, weighting by $\delta z_{tT,i}^{-2}$, where $\delta z_{tT,i}^2$ is the variance of the distribution of the $z_{tT,i}$ value from the MCMC approach. With this approach, we obtained the average height $z_{tT}$ at which the thick disk starts to dominate in each galaxy. $z_{tT}$ values are indicated in Fig.~\ref{fig:sample_images} as orange, dashed and horizontal lines, and in Table~\ref{tab:thick} together with the scale heights of the thin and the thick disks. 

For one galaxy, Au1, we did not obtain good fits (no good convergence of the free parameters for most radial bins), when modeling the vertical luminosity density profiles with Eq.~\ref{eq:double_disk}. We obtained good fits when using only one hyperbolic secant square: $\Sigma (z) = \Sigma_{0} {\rm{sech}}^2 \left({|z|}/{2 h_{z}}\right)$. 
This suggests that we do not have two clear morphologically  distinct disk components within the analyzed region in this galaxy.

\subsection{Maps of galaxy properties when seen edge on} \label{sub:method_maps}
We analyze in this work the stellar-population properties that are typically analyzed in recent observational work \citep{Pinna2019a,Pinna2019b,Martig2021,Scott2021, Sattler2023}. 
 Maps of the mass-weighted projected star-particle age, [M/H] and [Mg/Fe] were obtained using the low-resolution setup (0.5~kpc$\times$0.5~kpc pixel size and Voronoi binning with a target number of 900 star particles per bin; see Sect.~\ref{sub:proj}). 
 In this work, we corrected the original values of [Mg/Fe] by applying a factor of 2.5 to magnesium yields, following \citet{vandeVoort2020}. They established this correction due to the underproduction of magnesium in the yields adopted in the simulations, if compared to observations. Since magnesium is not a cooling agent in the simulations, the need for this correction is not expected to affect the physics of the formation and evolution of the galaxy. 
Since the aim of this paper is to assess internal and external processes, we classified star particles in our simulations as in situ or ex situ, using the distinction made by \citet{Gomez2017a}. In situ stars were formed in the main subhalo, either in the same region where they are observed now or in a different region, while ex situ stars were formed in a satellite galaxy before its final disruption. 
This distinction allows us to calculate the fraction of accreted stars in each Voronoi bin and to map their spatial distribution.

\section{Results} \label{sec:results}

\subsection{Stellar-population maps} \label{sub:SPmaps}

\begin{figure}
\centering
\resizebox{0.55\textwidth}{!}
{\includegraphics[scale=1., trim={0.35cm 0.4cm 0.cm 0.6}]{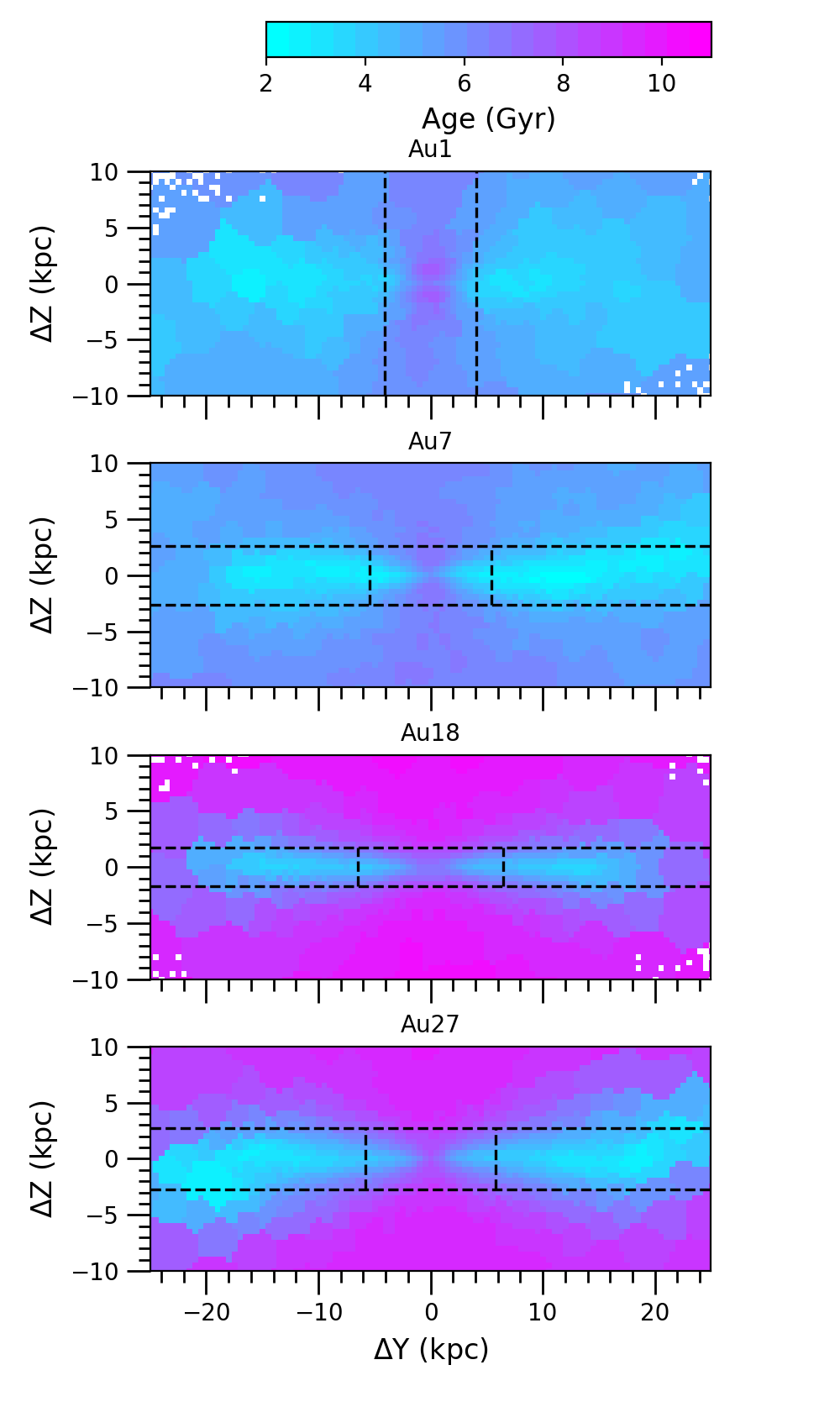}}
\caption{Four representative examples of mass-weighted stellar age maps. For display purposes, we show for all galaxies a central area of the same 50~kpc$\times$20~kpc size. Horizontal dashed black lines indicate the regions where the thin disk (within the two lines) and the thick disk dominate (above and below the region between the two lines). Vertical dashed black lines enclose the central region dominated by a bar or a classical bulge. Names of the AURIGA halos are indicated on top of each panel.
Ages were integrated and projected into pixels of 0.5~kpc$\times$0.5~kpc size and Voronoi binned to obtain at least 900 star particles per bin. 
Maps of the full sample are shown in Appendix~\ref{sub:SPmaps_all}.
}
\label{fig:sample_age_4pan}
\end{figure}

\begin{figure}
\centering
\resizebox{0.55\textwidth}{!}
{\includegraphics[scale=1., trim={0.35cm 0.4cm 0.cm 0.6}]{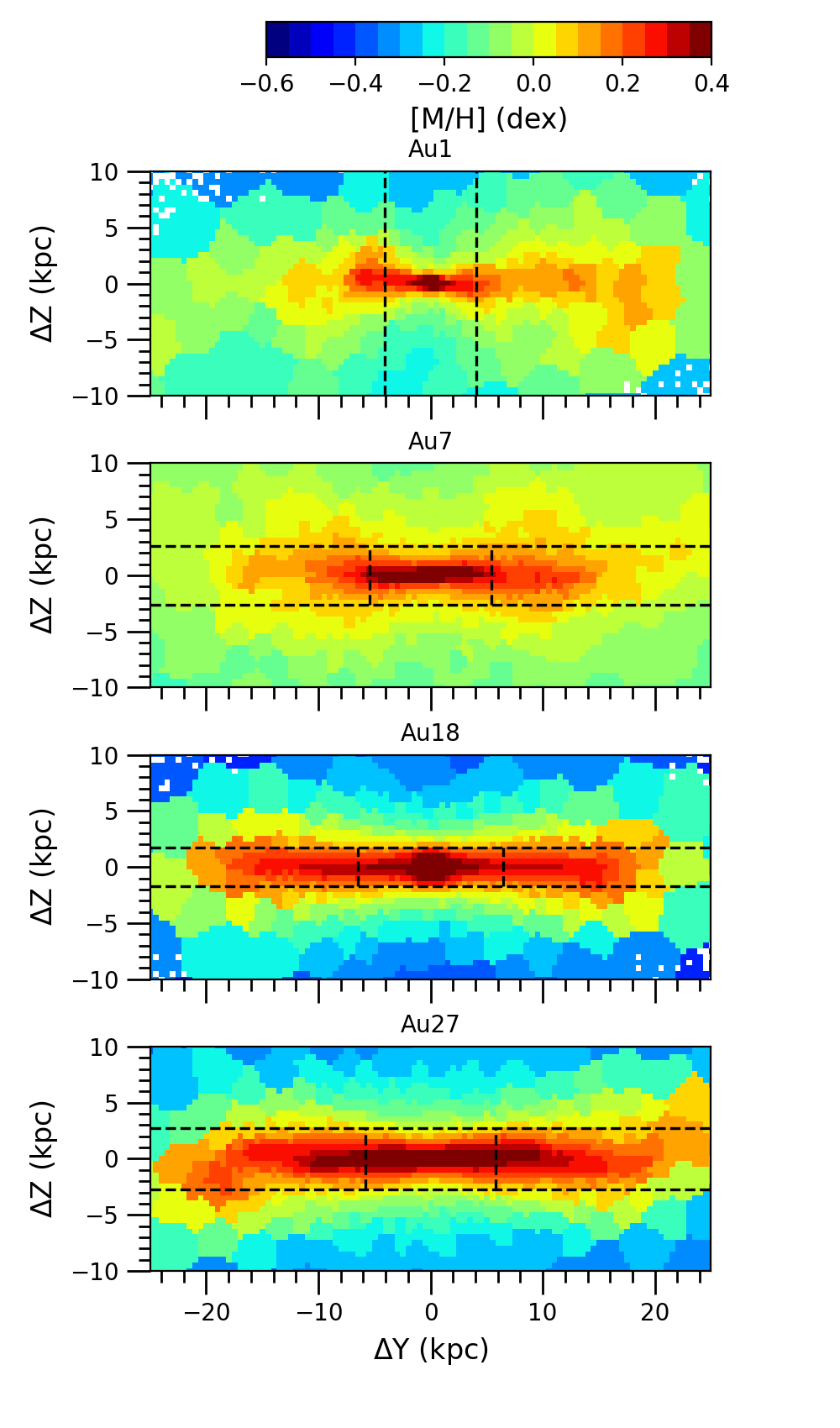}}
\caption{Four representative examples of mass-weighted total stellar metallicity [M/H] maps. For display purposes, we show for all galaxies a central area of the same 50~kpc$\times$20~kpc size. Horizontal black dashed lines indicate the regions where the thin disk (within the two lines) and the thick disk dominate (above and below the region between the two lines). Vertical dashed black lines enclose the central region dominated by a bar or a classical bulge. Names of the AURIGA halos are indicated on top of each panel. 
Metallicities were integrated and projected into pixels of 0.5~kpc$\times$0.5~kpc size and Voronoi binned to obtain at least 900 star particles per bin. 
Maps of the full sample are shown in Appendix~\ref{sub:SPmaps_all}.
}
\label{fig:sample_met_4pan}
\end{figure}

\begin{figure}
\centering
\resizebox{0.55\textwidth}{!}
{\includegraphics[scale=1., trim={0.35cm 0.4cm 0.cm 0.6}]{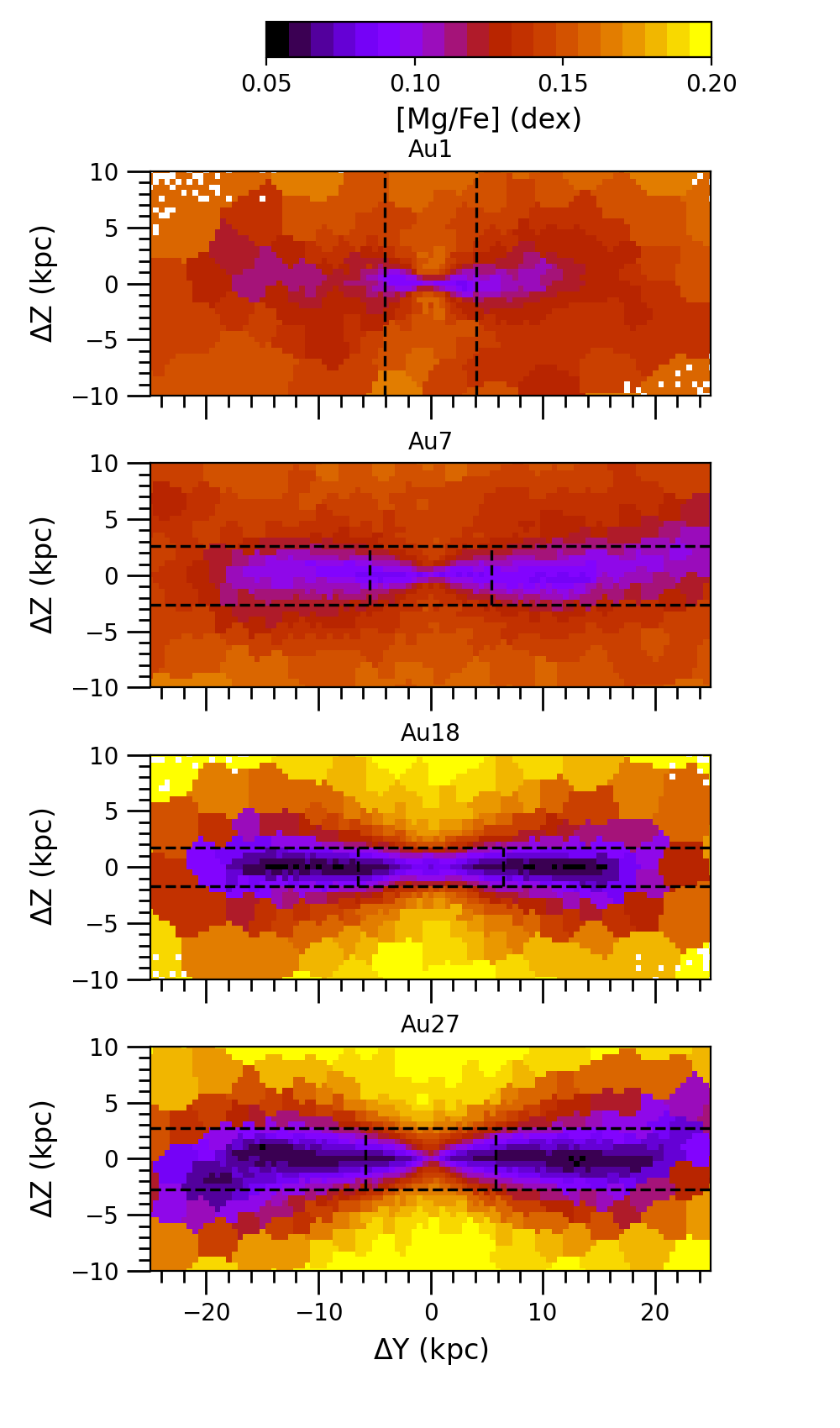}}
\caption{Four representative examples of mass-weighted stellar [Mg/Fe]-abundance maps. For display purposes, we show for all galaxies a central area of the same 50~kpc$\times$20~kpc size. Horizontal dashed black lines indicate the regions where the thin disk (within the two lines) and the thick disk dominate (above and below the region between the two lines). Vertical dashed black lines enclose the central region dominated by a bar or a classical bulge. Names of the AURIGA halos are indicated on top of each panel. 
[Mg/Fe] abundances were integrated and projected into pixels of 0.5~kpc$\times$0.5~kpc size and Voronoi binned to obtain at least 900 star particles per bin. 
Maps of the full sample are shown in Appendix~\ref{sub:SPmaps_all}.
}
\label{fig:sample_mg_4pan}
\end{figure}

Maps of the kinematic parameters are included in Appendix~\ref{app:kin_edgeon}. We describe here the mass-weighted stellar-population maps, presented for our full sample of 24 galaxies in Appendix~\ref{sub:SPmaps_all}, in Fig.~\ref{fig:sample_age} to \ref{fig:sample_mg}. We show in this section, in Fig.~\ref{fig:sample_age_4pan} to \ref{fig:sample_mg_4pan}, only four representative galaxies as examples. 
For display purposes, we show for all galaxies a region extended 50~kpc$\times$20~kpc, centered on the center of the galaxy (Sect.~\ref{sub:proj}).  In general, clear differences in the stellar-population parameters are visible between the thin-disk-dominated region (between the two horizontal lines located at $z_{tT}$, excluding the region between the two vertical lines indicating $R_{0,disk}$) and the thick-disk region (above and below the region between the horizontal lines). 
Age maps in Fig.~\ref{fig:sample_age_4pan} and \ref{fig:sample_age} show in all galaxies a much younger thin disk than the thick disk. In some cases, a strong positive radial gradient is observed in the displayed region of the thin disk, with the youngest stars concentrated in an inner region (e.g., Au10, Au12, Au17, Au22). These are mostly the smallest galaxies, since we used a fixed-size box for all galaxies. In some other galaxies, with larger $R_{opt}$, young stars cover a region out to the outer regions of our maps (e.g., Au2, Au3, and Au24). 
In general, the flared appearance in the age maps reflects that younger stars flare, because they are formed in a flaring star-forming gas disk (\citealt{Grand2016}; see also \citealt{Kawata2017} and \citealt{Benitez2018} for other simulations). 
This flaring extends to the outer regions of the geometrically defined thick disk, which presents a negative age radial gradient. While thin disks show ages as young as $1-2$~Gyr in some Voronoi bins, thick-disk regions show Voronoi bins with ages as old as 11~Gyr in the inner regions in some galaxies (e.g., Au9; see Sect.~\ref{sub:stats} for average ages). 


In Fig.~\ref{fig:sample_met_4pan} and \ref{fig:sample_met}, thin-disk regions are clearly metal rich, with negative radial gradients and values as high as $\sim 0.4$~dex in the inner region. Thick disks show more metal-poor values, mostly subsolar (often as low as -0.4~dex) or close to solar values (see also Sect.~\ref{sub:stats}). 
Thick disks in AURIGA simulations are more enhanced in [Mg/Fe] than thin disks (Fig.~\ref{fig:sample_mg_4pan} and \ref{fig:sample_mg}). While the latter show values closer to solar, thick disks show values as high as $\sim$0.2~dex or beyond in some cases (e.g., Au9). 
Our stellar-population maps (in particular metallicity maps) reveal signs of mergers in numerous galaxies: warps like in Au21 and Au27, streams as in Au14, and other distortions.

\subsection{Spatial distribution of accreted stars}\label{sub:accr}

\begin{figure}
\centering
\resizebox{0.55\textwidth}{!}
{\includegraphics[scale=1., trim={0.35cm 0.4cm 0.cm 0.6}]{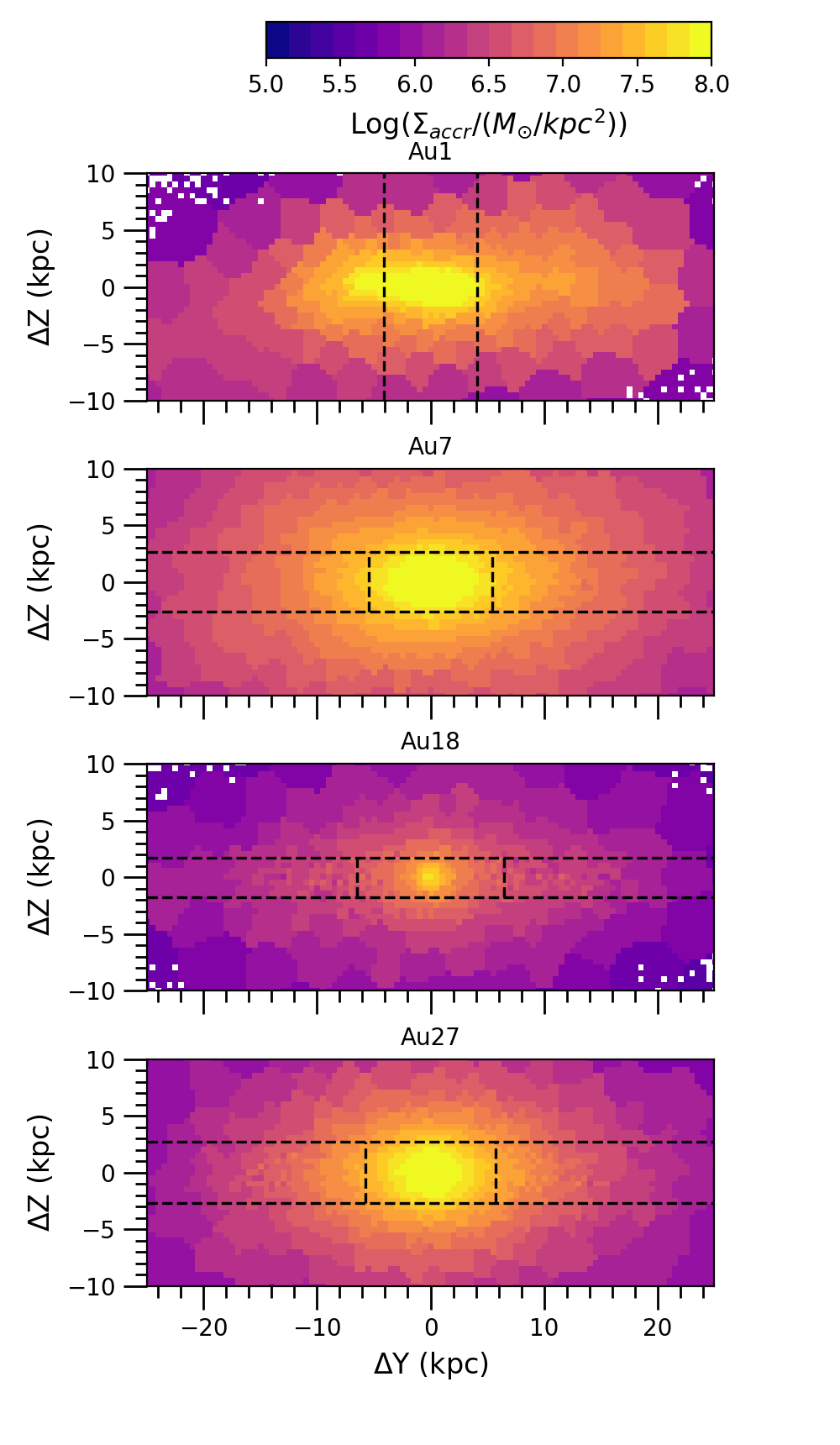}}
\caption{Four representative examples of maps of the mass surface density of accreted stars, in logarithmic scale. For display purposes, we show for all galaxies a central area of the same 50~kpc$\times$20~kpc size.  Horizontal dashed black lines indicate the regions where the thin disk (within the two lines) and the thick disk dominate (above and below the region between the two lines). Vertical dashed black lines enclose the central region dominated by a bar or a classical bulge. Names of the AURIGA halos are indicated on top of each panel. 
Pixels of 0.5~kpc$\times$0.5~kpc size, Voronoi binned to a target number of particles of 900 star particles per bin, were used to plot these maps. 
Maps of the full sample are shown in Appendix~\ref{sub:accr_maps_all}.
}
\label{fig:sample_accrmassdens_4pan}
\end{figure}

\begin{figure}
\centering
\resizebox{0.55\textwidth}{!}
{\includegraphics[scale=1., trim={0.35cm 0.4cm 0.cm 0.6}]{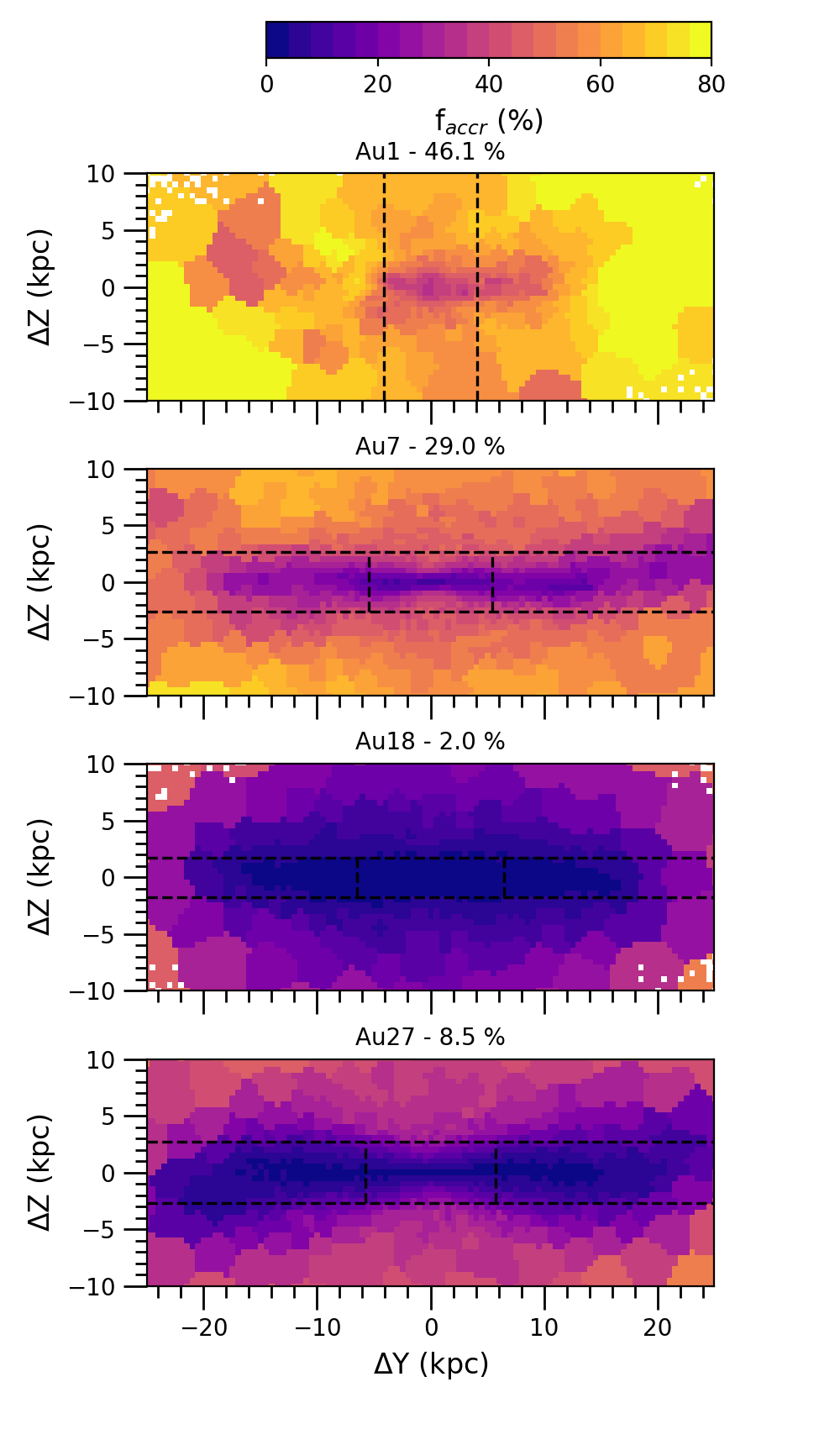}}
\caption{Four representative examples of maps of the mass fraction of accreted stars. For display purposes, we show for all galaxies a central area of the same 50~kpc$\times$20~kpc size. Horizontal dashed black lines indicate the regions where the thin disk (within the two lines) and the thick disk dominate (above and below the region between the two lines). Vertical dashed black lines enclose the central region dominated by a bar or a classical bulge. Names of the AURIGA halos and the total mass fraction of accreted stars (calculated in the analyzed region, of size $1.6 R_{opt} \times 4h_{scale}$, Sect.~\ref{sub:region}) are indicated on top of each panel. 
Pixels of 0.5~kpc$\times$0.5~kpc size, Voronoi binned to a target number of particles of 900 star particles per bin, were used to plot these maps. 
Maps of the full sample are shown in Appendix~\ref{sub:accr_maps_all}.
}
\label{fig:sample_accrmf_4pan}
\end{figure}

We show in Fig.~\ref{fig:sample_accrmassdens_4pan} maps of the mass density of the accreted stars, in the central 50~kpc$\times$20~kpc region of four galaxies in our sample. 
Maps of the full sample are shown in Fig.~\ref{fig:sample_accrmassdens}. 
These maps show that accreted stars are concentrated mostly in the inner region. In some galaxies, there are additional specific regions with a high density of accreted stars, tracing the remnants of recent minor mergers (e.g., Au17; see also below). 
Interestingly, while accreted stars are denser in the inner midplane region, their mass fraction (in each Voronoi bin) is usually higher in the thick-disk region, as shown in Fig.~\ref{fig:sample_accrmf_4pan} and \ref{fig:sample_accrmf}. In some cases, such as Au12 or Au14, we reach up to 80\% of accreted stars in some Voronoi bins in the thick-disk region. In general, ex situ stars constitute a high fraction of {mass in the thick-disk-dominated region}, while accreted fractions in the thin-disk region are low in comparison, usually below 10\% in most Voronoi bins. 
In Au17 we can identify a region (below the thin-disk-dominated region) dominated by accreted stars (high density and high mass fraction of accreted stars; see Figs.~\ref{fig:sample_accrmassdens} and \ref{fig:sample_accrmf}). 
This is the remnant of an accreted satellite, still in the process of completely merging (brighter than its surroundings in Fig.~\ref{fig:sample_images}). It can be identified in Figs.~\ref{fig:sample_age} and \ref{fig:sample_met} as a younger and more metal-rich structure than its surroundings. It corresponds to slightly lower values of [Mg/Fe] in Fig.~\ref{fig:sample_mg}, although differences with the surroundings are not obvious.  

We provide in Table~\ref{tab:accr} the total mass and mass fraction of accreted stars, computed in the full analyzed region of the galaxy (of size $1.6 R_{opt} \times 4h_{scale}$, Sect.~\ref{sub:region}). These mass fractions were also indicated on top of each panel of Fig.~\ref{fig:sample_accrmf_4pan} and \ref{fig:sample_accrmf}, together with the name of the AURIGA halo. 
Fractions of accreted stars are below 10\% for half of the sample, and relatively low in almost all galaxies (below 20\%). Exceptions are Au7 with almost 30\% of accreted stars and Au1 with $\sim$46\%. 
However, these fractions are much larger in {thick-disk regions}, $\sim$22\% on average, with $\sim$50\% of accreted stars in Au7 and $\sim$55\% in Au12. 
These fractions correspond to total accreted masses of about $7\times 10^8$ to $2\times 10^{10}$\msun\,in the galaxy ($4\times 10^8$ to $10^{10}$\msun\,in the thick disks). 
Mass densities and fractions of accreted stars are larger if a more extended region (than the one used in this work; see Sect.~\ref{sub:region}) is taken into account, since the ex situ fraction per Voronoi bin increases the farther we go from the galactic midplane. 
\citet{Gomez2017a} showed that, in AURIGA disks defined based on circularity cuts (circularities larger than values from 0.7 to 0.9), the fraction of accreted stars is lower in dynamically colder disks. They also showed that a positive radial gradient of the accreted fraction is present in most galaxies. Although we do not apply any filter based on kinematic properties to the star particles, we find similar results {(see Sect.~\ref{sub:disc:origin} for an extended discussion)}. 
\newpage

\newpage
\begin{table*}
\caption{Accreted stars in our galaxy sample.}
\centering
\begin{tabular}{lcccccr}
\hline\hline
Galaxy name &  $M_{accr,gal}$ ($10^8$\msun) & $f_{accr,gal}$ (\%) &  $M_{accr,thin}$ ($10^8$\msun) & $f_{accr,thin}$ (\%) & $M_{accr,thick}$ ($10^8$\msun) & $f_{accr,thick}$ (\%) \\
\hline
Au1$^{(1)}$   & 127.5 & 46.1 & - & - & - & - \\
Au2   & 90.7  & 10.7 & 12.0 & 5.4 & 25.7 & 13.9 \\
Au3   & 94.6  & 11.5 & 12.2 & 4.0 & 33.2 & 21.6  \\
Au5   & 40.2  & 6.0  & 7.5  & 5.0 & 19.3 & 29.6  \\
Au6   & 39.5  & 7.8  & 6.4  & 4.1 & 13.0 & 17.6  \\
Au7   & 149.0 & 29.0 & 24.3 & 27.0 & 58.9 & 49.9  \\
Au8   & 51.3  & 15.5 & 10.8 & 13.2 & 26.1 & 16.3  \\
Au9   & 27.8  & 4.6  & 2.0  & 2.9 & 11.3 & 15.8 \\
Au10  & 7.3   & 1.2  & 0.8  & 3.6 & 4.0  & 7.2  \\
Au12  & 88.0  & 14.0 & 22.9 & 12.2 & 42.0 & 55.4 \\
Au14  & 105.3 & 9.9  & 23.7 & 11.2 & 50.9 & 32.4  \\
Au15  & 42.8  & 10.7 & 14.1 & 8.7 & 16.9 & 25.9 \\
Au16  & 41.1  & 6.4  & 13.2 & 4.9 & 22.7 & 15.4  \\
Au17  & 11.4  & 1.5  & 0.9  & 1.6 & 7.6  & 10.2  \\
Au18  & 15.6  & 2.0  & 2.5  & 2.5 & 7.5  & 7.2  \\
Au19  & 98.2  & 18.0 & 22.5 & 12.2 & 31.7 & 26.8  \\
Au21  & 99.1  & 12.7 & 26.3 & 8.3 & 32.5 & 26.6  \\
Au22  & 8.2   & 1.4  & 0.4  & 2.9 & 4.2  & 5.1  \\
Au23  & 56.1  & 6.2  & 4.2  & 4.7 & 27.7 & 19.7  \\
Au24  & 61.3  & 8.5  & 16.6 & 9.1 & 29.5 & 21.6 \\
Au25  & 9.6   & 3.0  & 1.1  & 1.3 & 4.3  & 4.2   \\
Au26  & 117.7 & 10.9 & 3.8  & 5.5 & 48.3 & 24.9 \\
Au27  & 81.4  & 8.5  & 13.7 & 5.8 & 25.9 & 26.4 \\
Au28  & 203.5 & 20.0 & 3.5  & 10.6 & 89.2 & 37.6 \\
\hline
\end{tabular}
\label{tab:accr}
\begin{tablenotes}
\item {\footnotesize Notes. 
{Columns from left to right: galaxy name, ex situ mass (mass of accreted stars) in the analyzed region of the galaxy ($M_{accr,gal}$), mass fraction of accreted stars in the analyzed region of the galaxy ($f_{accr,gal}$), ex situ mass in the thin-disk region ($M_{accr,thin}$), ex situ mass fraction in the thin-disk region ($f_{accr,thin}$), ex situ mass in the thick-disk region ($M_{accr,thick}$), ex situ mass fraction in the thick-disk region ($f_{accr,thick}$). (1) Au1 has no clear double disk structure. 
}
\normalsize
}
\end{tablenotes}
\end{table*}

\subsection{Comparison of thick and thin disks}\label{sub:stats}
\begin{figure*}
\centering
\resizebox{1.\textwidth}{!}
{\includegraphics[scale=1.5]{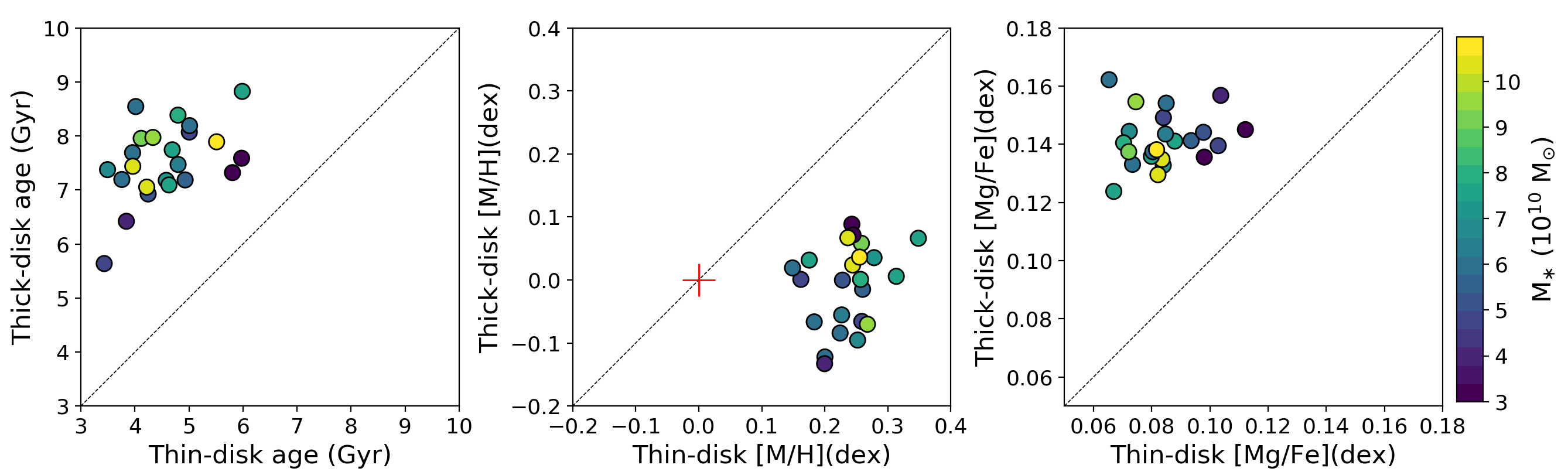}}
\caption{Thick-disk versus thin-disk stellar-population properties. From left to right: age,  [M/H], and [Mg/Fe] abundance. Points are color coded according to the total mass of the galaxy (Table~\ref{tab:sample}). The one-to-one line is indicated in black and dashed, in each panel. A red ``+'' symbol indicates solar metallicity in the middle panel. 
}
\label{fig:tTdisks_mass}
\end{figure*}

\begin{figure*}
\centering
\resizebox{1.\textwidth}{!}
{\includegraphics[scale=1.5]{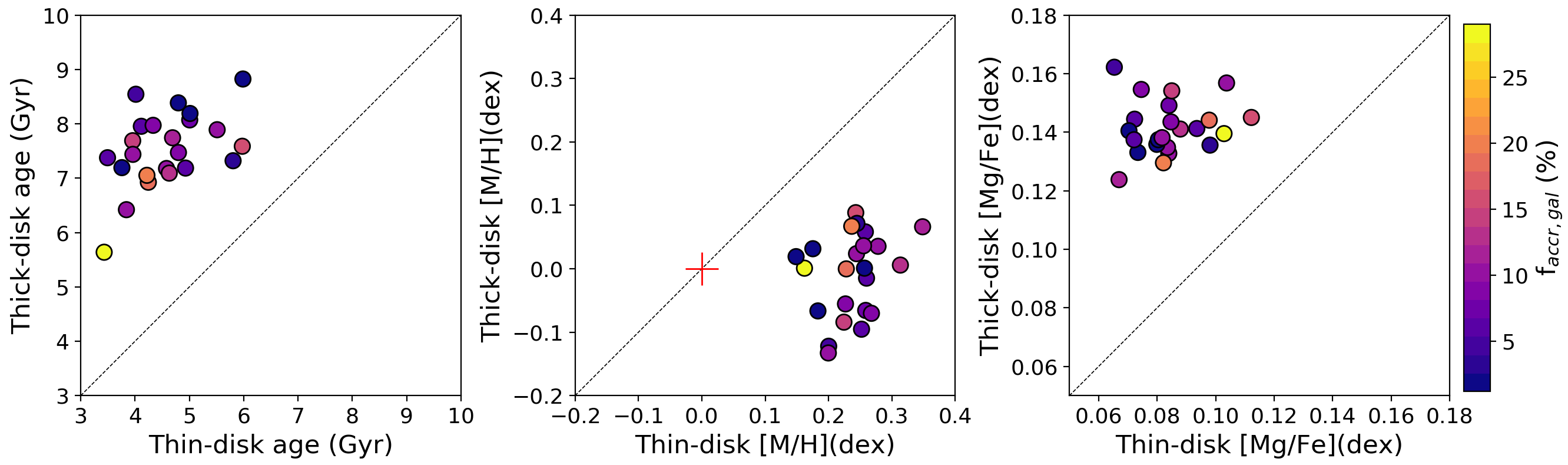}}
\caption{Thick-disk versus thin-disk stellar-population properties. From left to right: age, [M/H], and [Mg/Fe] abundance. Points are color coded according to the total mass fraction of accreted stars in the analyzed region of the galaxy (Sect.~\ref{sub:region}). The one-to-one line is indicated in black and dashed in each panel. In the middle panel, the red ``+'' symbol indicates solar metallicity. 
}
\label{fig:tTdisks_faccr}
\end{figure*}

We show in Figs.~\ref{fig:tTdisks_mass} and \ref{fig:tTdisks_faccr} a comparison of mass-weighted average thick-disk age, [M/H] and [Mg/Fe] with the respective properties in the thin disks. Each point corresponds to one of the 24 galaxies in our sample. 
Thick disks are older, more metal poor and [Mg/Fe] enhanced than thin disks in all 24 galaxies in our sample, as suggested by maps in Sect.~\ref{sub:SPmaps}. They are from $\sim 1.5$ to $\sim 4.5$~Gyr older (on average $\sim 3$~Gyr older), from $\sim 0.13$ to $\sim 0.35$~dex (on average $\sim 0.25$~dex) more metal poor, and from $\sim 0.04$ to $\sim 0.09$~dex (on average $\sim 0.06$) more [Mg/Fe]-enhanced than thin disks.
However, the ages of thick disks are distributed over a wide range and some thick disks are younger than other thin disks in the sample (left panels in Fig.~\ref{fig:tTdisks_mass} and \ref{fig:tTdisks_faccr}). 
The average [M/H] of thick disks range from subsolar to slightly super-solar values, while thin-disk average [M/H] is always super-solar, higher than the most metal-rich thick disk in the sample. 

In Fig.~\ref{fig:tTdisks_mass}, we have color coded each point according to the total stellar mass of the galaxy (Table~\ref{tab:sample}). There is no clear trend of the thick-disk properties with galaxy stellar mass (with a $R^2$ correlation coefficient lower than 0.11). However, galaxies with the lowest masses show more similar properties in the thin and the thick disks (points closer to the one-to-one line). 
In Fig.~\ref{fig:tTdisks_faccr}, we have color coded points by the total mass fraction of accreted stars ($f_{accr,gal}$ in Table~\ref{tab:accr}). 
While the total accretion fraction does not show a clear trend with thick-disk [M/H] and [Mg/Fe], it shows an anticorrelation with thick-disk age ($R^2 \sim 0.5$), with the oldest thick disks located in galaxies with the lowest accreted fractions. We find a similar age trend, but with a shallower slope, also with the accretion fractions in the thick disk. This means that the youngest thick disks tend to be hosted by galaxies with the largest fractions of accreted stars. 
This is usually the outcome of more recent mergers, involving satellite galaxies that have had more time to grow in mass {(e.g., in Au7, see also Fig.~\ref{fig:snap_au7}), and heating stars originally formed in preexisting thin disks (e.g., in Au19, see Fig.~\ref{fig:snap_au19}). 
Star-formation histories of these thick disks (which will be published in a companion paper) support this scenario, and also show that mergers provide additional dynamically hot gas, which leads to some later star formation at heights corresponding to the geometric thick disks (see also the discussion in Sect.~\ref{sub:disc:origin}). 
}
On the other hand, thin-disk [Mg/Fe] abundances (mildly) correlate with the accretion fraction in the right panel of Fig.~\ref{fig:tTdisks_faccr} ($R^2 \sim 0.3$). Galaxies with less significant accretion reach lower [Mg/Fe] values in their thin disks. 
We also checked for trends of thin- and thick-disk stellar-population properties with the presence of bars or with galaxy morphology in general, but no trends were found. 
\subsection{Time evolution}\label{sub:snap}
\begin{figure*}
\centering
\resizebox{1.\textwidth}{!}
{\includegraphics[scale=1.5]{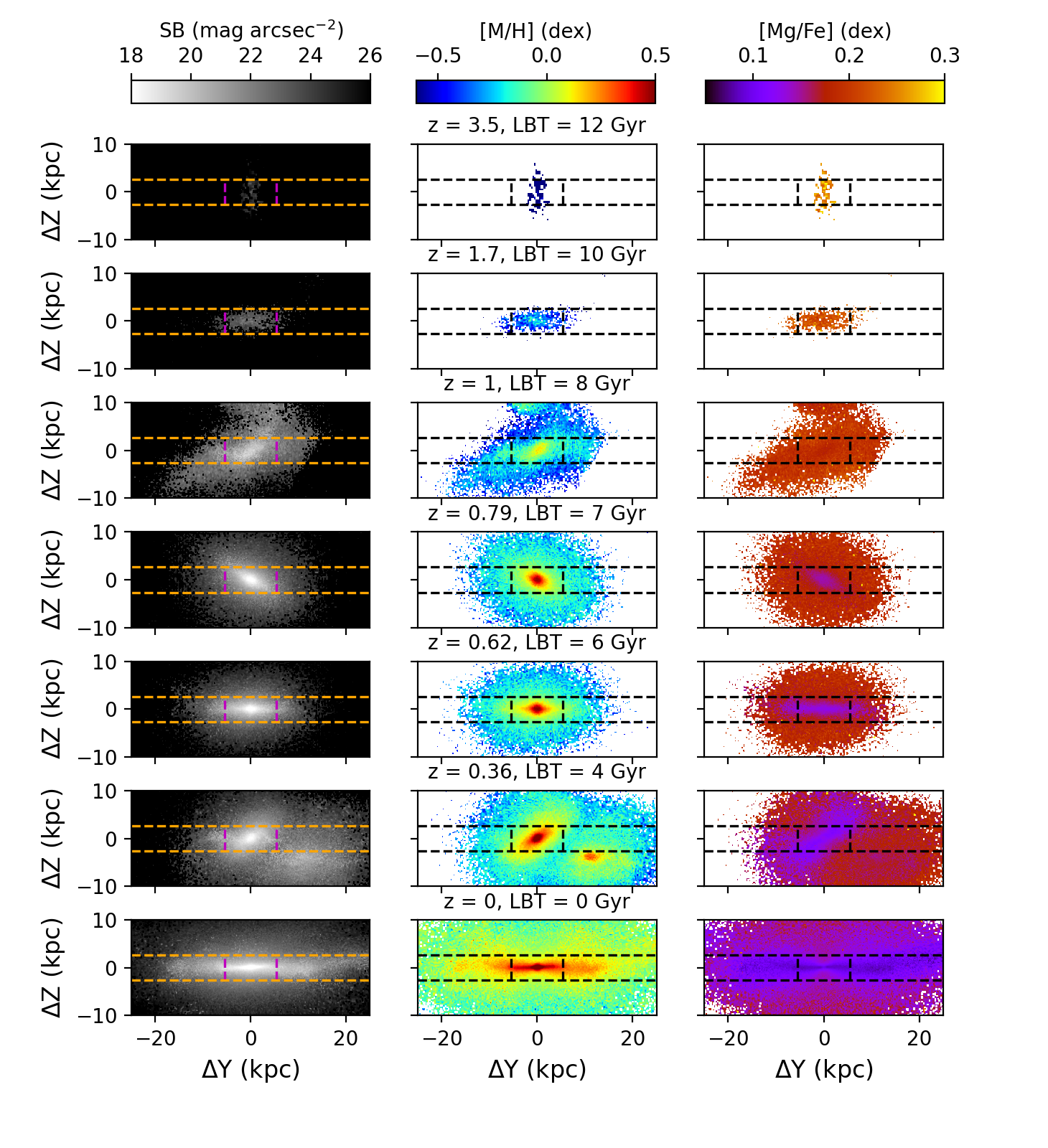}}
\caption{Time evolution of the galaxy Au7 seen edge on. Surface brightness is shown in the left column, total metallicity [M/H] in the middle column, and [Mg/Fe] abundance in the right column. 
From top to bottom, the stellar component is shown for seven different snapshots. On top of the middle panels, the redshift and the look-back time (LBT) corresponding to the snapshot shown in that row are indicated. For comparison, horizontal dashed lines separate the regions where the thin or the thick disks dominate at $z=0$, while the vertical lines indicate the region where the central component (bar or classical bulge) dominates at $z=0$. 
}
\label{fig:snap_au7}
\end{figure*}

We analyze in this section the evolution of thick and thin disks in different snapshots of the simulations across time. 
We mapped all galaxies in our sample at seven different snapshots from redshift $z=3.5$ to $z=0$. Galaxies were aligned and projected edge on in each snapshot, according to the criteria explained in Sect.~\ref{sub:proj}. Since the alignment of the galaxies is based on the angular momentum of young stars, which at high redshift is not as clear as at $z=0$, it did not give an optimal result when the galaxy was not yet clearly rotating or in the presence of a merging satellite with (differently) rotating young stars (e.g., $z=3.5$ in Fig.~\ref{fig:snap_au7}; see also Appendix~\ref{app:snap}, e.g., Au26 at $z=1.7$). 
For display purposes, we show for all galaxies a central area of the same 50~kpc$\times$20~kpc size in each simulation snapshot. 
Since at high redshift the number of stellar particles is much lower, it was needed to adapt the pixel size (to 0.2 and 0.5~kpc) and the Voronoi binning for different redshifts (to 9 and 16 particles per bin, see Sect.~\ref{sub:proj}). 

\begin{figure*}
\centering
\resizebox{1.\textwidth}{!}
{\includegraphics[scale=1.5]{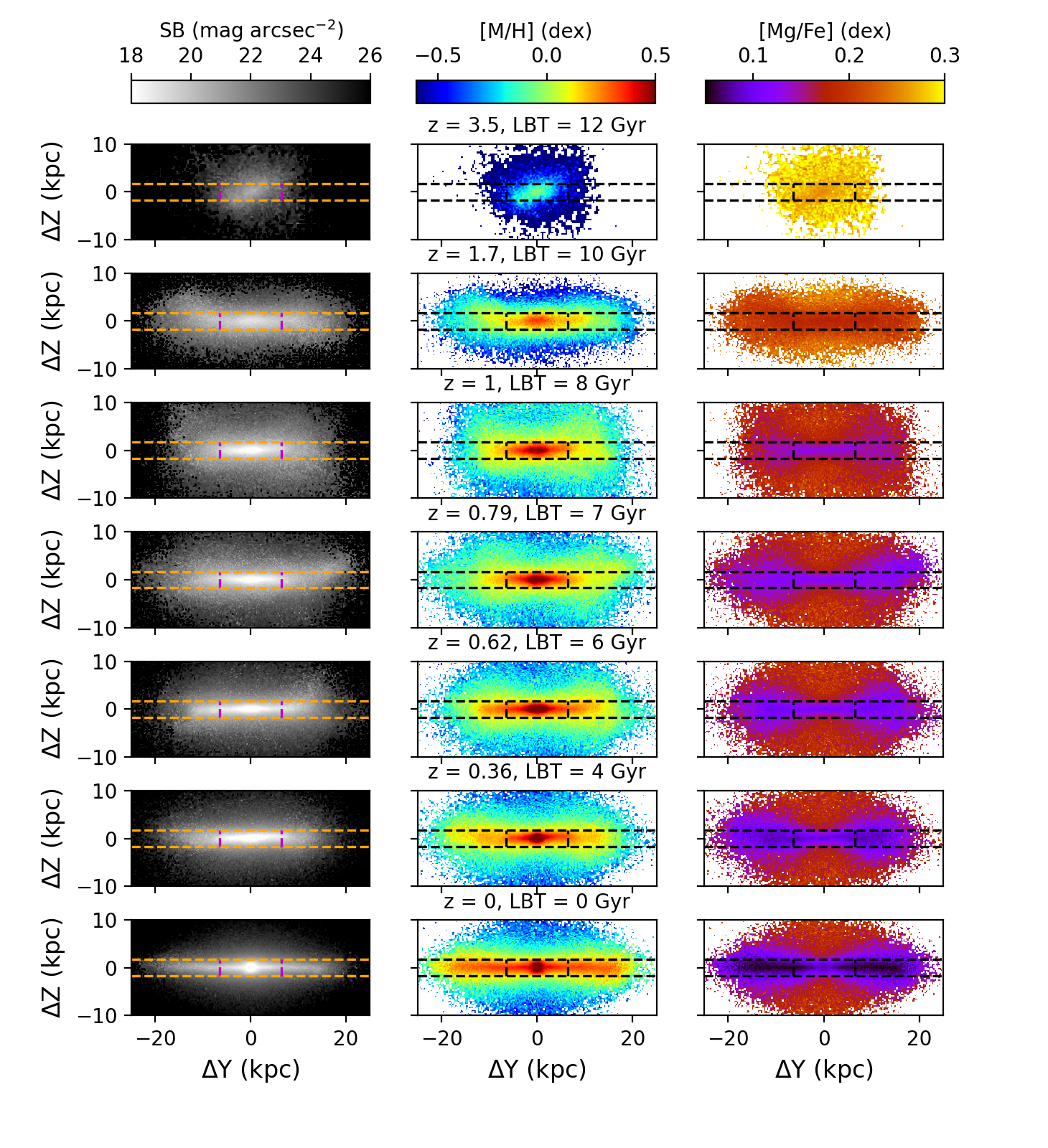}}
\caption{Same as Fig.~\ref{fig:snap_au7} but for the galaxy Au18.  
}
\label{fig:snap_au18}
\end{figure*}

We show here two specific examples, Au7 and Au18, respectively in Fig.~\ref{fig:snap_au7} and \ref{fig:snap_au18}. These two galaxies are representative of the sample as they show two typical  different types of evolution histories (although each galaxy has its own peculiarities): one with a  significant ex situ contribution and the other with a predominantly in situ growth. Snapshots for the rest of the galaxies are shown in Appendix~\ref{app:snap}. While we show here only surface brightness, [M/H] and [Mg/Fe], velocity maps (not shown) helped in the identification of (rotating-)disk formations. 
Au7, with a stellar mass of $4.9 \times 10^{10}$\msun\,at $z=0$, had a slower evolution than Au18, which is slightly more massive (with a stellar mass of $8.0\times 10^{10}$\msun). 

Au7 had formed only a few, metal-poor and [Mg/Fe]-enhanced stars at $z=3.5$, and had no disky shape. At $z=1.7$, it showed a small and relatively thick disk. 
{It had a similar vertical extension to the future thin disk at $z=0$ (indicated by the two black horizontal lines), and thus a significantly lower vertical extension than the thick disk at $z=0$. However, this primordial disk was radially much less extended than both the thin and thick disks at $z=0$ (see the vertical dashed black  lines as reference), leading to a similar thick shape to its shape at $z=0$. This primordial thick disk covered a region of approximate size $0.4 R_{opt} \times h_{scale}$, with the same shape (intended as the ratio between the vertical and horizontal size in projection) as our analyzed region at $z=0$ ($1.6 R_{opt} \times 4 h_{scale}$). 
} 
It had a stellar mass of about $5.4 \times 10^9$\msun\,at $z=1.7$, about 10\% of the final galaxy mass at $z=0$. 
No solar-metallicity stars were yet present in this second snapshot. Later on, the disky shape and rotation were distorted by mergers ($z=1$ and 0.79). Around $z=0.6$ the thin disk was developing from inside out embedded in the thick disk. A final important merger, tilting the disk as visible in the $z=0.36$ snapshot, provided a large amount of gas to allow for the full growth of the thin disk to the outskirts at $z=0$. 

Au18 formed its disks and grew much faster than Au7. At $z=3.5$, it was already more extended than Au7. Its fast formation timescale led to a global higher [Mg/Fe] abundance, as well as a higher [M/H] in the inner region (solar values were already reached). The galaxy already started to show some amount of rotation. 
At $z=1.7$, its stellar mass was already about $\sim 4.7\times 10^{10}$\msun\,and most of its total mass at $z=0$ had already formed. It was approximately one order of magnitude more massive than Au7 at the same time. 
A primordial thinner metal-rich and a thicker metal-poor disks were already in place. From that time on, the thicker component evolved only slowly during the rest of the life of the galaxy. 
{While in Au7 the low density at early stages and the disturbed shapes due to the merger later on make it difficult to fit surface brightness profiles for different redshifts, we have done so for Au18. These fits confirm that thick-disk scale height varied very little (within their uncertainties) across the galaxy life. 
Thick-disk [Mg/Fe] and [M/H] remain almost unchanged after $z \sim 1$. }
On the other hand, the buildup of the metal-rich thin disk had just started, it grew inside out to its full final extension at $z=0$, became very metal rich and its [Mg/Fe] abundance decreased toward solar values. 

{At $z=0$, Au7 is thicker, as indicated by its $h_{scale}$, than Au18 (Table~\ref{tab:sample}). The thick disk is younger in Au7 than in Au18}, as displayed in Fig.~\ref{fig:sample_age_4pan}. Half of the {mass in the thick-disk region of} Au7 was accreted (Table~\ref{tab:accr}, as also suggested by the merger in Fig.~\ref{fig:snap_au7} at $z=1$ and 0.36). On the other hand, only $\sim 7$\% of the thick-disk mass in Au18 was accreted, and it was already mostly (in situ) formed at $z=1.7$. 
This is connected to the fact that, in general, the oldest more [Mg/Fe]-enhanced thick disks are found in more massive galaxies with a less intense merger history (at least after $z\sim 1$, see Fig.~\ref{fig:snap_au1} to \ref{fig:snap_au28}). 

\section{Discussion} \label{sec:disc}

\subsection{Origin of thick disks and their evolutionary connection with thin disks} \label{sub:disc:origin}

We have shown in previous sections that geometrically defined thick disks are relatively old, metal poor and enhanced in [Mg/Fe] in AURIGA simulations. This indicates that most of their masses were formed a long time ago from gas that had not been significantly enriched. 
In fact, Sect.~\ref{sub:snap} showed that a primordial thick rotating disk (which later kept growing) formed in most galaxies more than 10~Gyr ago (around redshift $z=2$). 
{This happened in a comparable timescale to what was found for the Milky Way \citep[e.g.,][]{Gallart2019, Belokurov2022, Conroy2022}.} 
Only in Au15 and Au19 we observe distorted shapes (and no clear rotation) in the snapshots at redshift $z=1.7$ (Fig.~\ref{fig:snap_au15} and \ref{fig:snap_au19}). 
However, the primordial thicker components had very different sizes, compared to the final thick disks at $z=0$, in different galaxies (see the horizontal dashed lines in the figures). Some had already extended thick disks (vertically and radially, e.g., Au2, Au17, and Au18), while others were smaller and either thinner (Au7) or thicker (e.g., Au3, Au10, and Au12) than the thick disk at $z=0$. 

In Sect.~\ref{sub:accr}, we have seen that all thick disks in the sample are made up of in situ and ex situ components, whose mass ratios can vary significantly for different galaxies. 
Most of the {mass in the thick-disk regions comes from in situ stars, with a mass fraction of accreted stars lower than 50\% in all cases} except in Au12 and Au7. However, a significant accreted component indicates that {geometric} thick disks grew thanks to the contribution from merging satellites across time. 
While the balance between in situ and ex situ formation strongly depends on the individual galaxy, mergers played a key role in the mass assembly of thick disks. 
{We would like to add a word of caution about the implications on the discussion about thick-disk formation of a pure geometric definition of them (chosen because our aim is to provide clues for the interpretation of observational results of edge-on galaxies). 
Geometric thick disks are defined by the region where their light dominates. However, in that region, there might be a contribution from stars belonging to other structural or kinematic components overlapped in the LOS and that are not possible to disentangle morphologically. These components include stellar halos, often defined in the Milky Way or in simulations using kinematic properties \citep[e.g.,][]{Naidu2020,Gallart2019,Monachesi2019}, as well as flaring thin disks (Sect.~\ref{sub:SPmaps}). These are very challenging to be disentangled in observed edge-on galaxies. } 

{Here, we take advantage of the combination of spatial three-dimensional information with kinematic properties in simulations, to quantify the contamination from stars belonging to other (non disk-like) kinematic components in the geometric thick disks.
In general, stellar-halo stars are thought to have been mostly accreted from satellites (about 80\% of halo stars were accreted in the Milky Way, \citealt{Naidu2020}, and about 90\% in AURIGA galaxies, 
\citealt{Monachesi2019}). We take the opportunity to discuss a potential contamination of halo stars in the ex situ component of thick disks. }
{Within the analyzed thick-disk region, we disentangled the fractions of stars with thick-disk kinematic properties, with orbital circularities $\epsilon > 0.5$ (following \citealt{Yi2023}), or associated with spheroidal components such as the stellar halo ($\epsilon < 0.5$). 
Stars with $\epsilon < 0.5$ make up on average 38\% of the mass in our geometric thick disks. However, it can be as low as 13\%, it is less than 35\% in half of the thick disks,  while in about one third of the galaxies this fraction is more than 50\% (up to 63\% in one case). Thus, the degree of contamination can be significant depending on the specific case. 
Regarding the accreted stars in the thick-disk region, the mass fraction of the ones with $\epsilon < 0.5$ can be as low as 14\% and as high as 79\%, on average 57\% in our sample. Hence, the contamination from ex situ halo stars in the accreted component in thick-disk region can be insignificant in some cases (since it is much lower than typical differences in the accreted fraction from galaxy to galaxy), or dominant. 
This contamination, while it is challenging to be corrected, should be discussed in observations of edge-on galaxies, where thick disks are also defined geometrically. 
A potential way forward would come from combining a geometrical definition with de-projected kinematic properties extracted with orbit-superposition methods \citep{Zhu2018}. 
Despite this word of caution, our results show that the impact of mergers through direct accretion of stars to the thick-disk-dominated region is important.
}

{In addition to star accretion}, we expect mergers also to impact on the in situ component of thick disks, through the dynamical heating of stars closer to the midplane and in situ star formation after gas accretion. 
The vertical dynamical heating of AURIGA stellar disks by mergers was previously discussed by \citet{Grand2016,Grand2020} and \citet{Monachesi2019}. 
Gas accretion has also a strong impact in the baryonic cycle and chemical evolution of galaxies, as shown in AURIGA simulations by \citet{Grand2019}. About half of this gas comes from accreted satellite galaxies. 
In a companion paper, we will show the star-formation histories of thick disks and will discuss further the in situ star formation and ex situ contribution of thick disks.
Thus, the processes responsible for the formation of thick disks are both internal and external, with a combination of the three main scenarios presented in Sect.~\ref{sec:intro}. 
After the formation of a primordial thick disk at high redshift, 
this kept growing, through accretion, in situ star formation, and dynamical heating.

Regarding the thin disks, in some galaxies they initially formed at high redshift. Galaxies with a faster evolution at their early stages, implying that they also formed extended thick disks very early (e.g., Au2, Au17, and Au18), started to form a thin metal-rich disk in the inner region about 10~Gyr ago ($z=1.7$). 
Our results suggest that while thick disks systematically start forming at early times, the formation of thin disks can either follow the primordial thick-disk formation or can be delayed until later times.

\subsection{Comparisons with observations}

\subsubsection{Morphology of thin and thick disks: Stellar disks are thicker in simulations} \label{sub:morph_comp_obs}

\begin{figure}
\centering
\resizebox{0.54\textwidth}{!}
{\includegraphics[scale=1., trim={0.9cm 0cm 0cm 0.8}]
{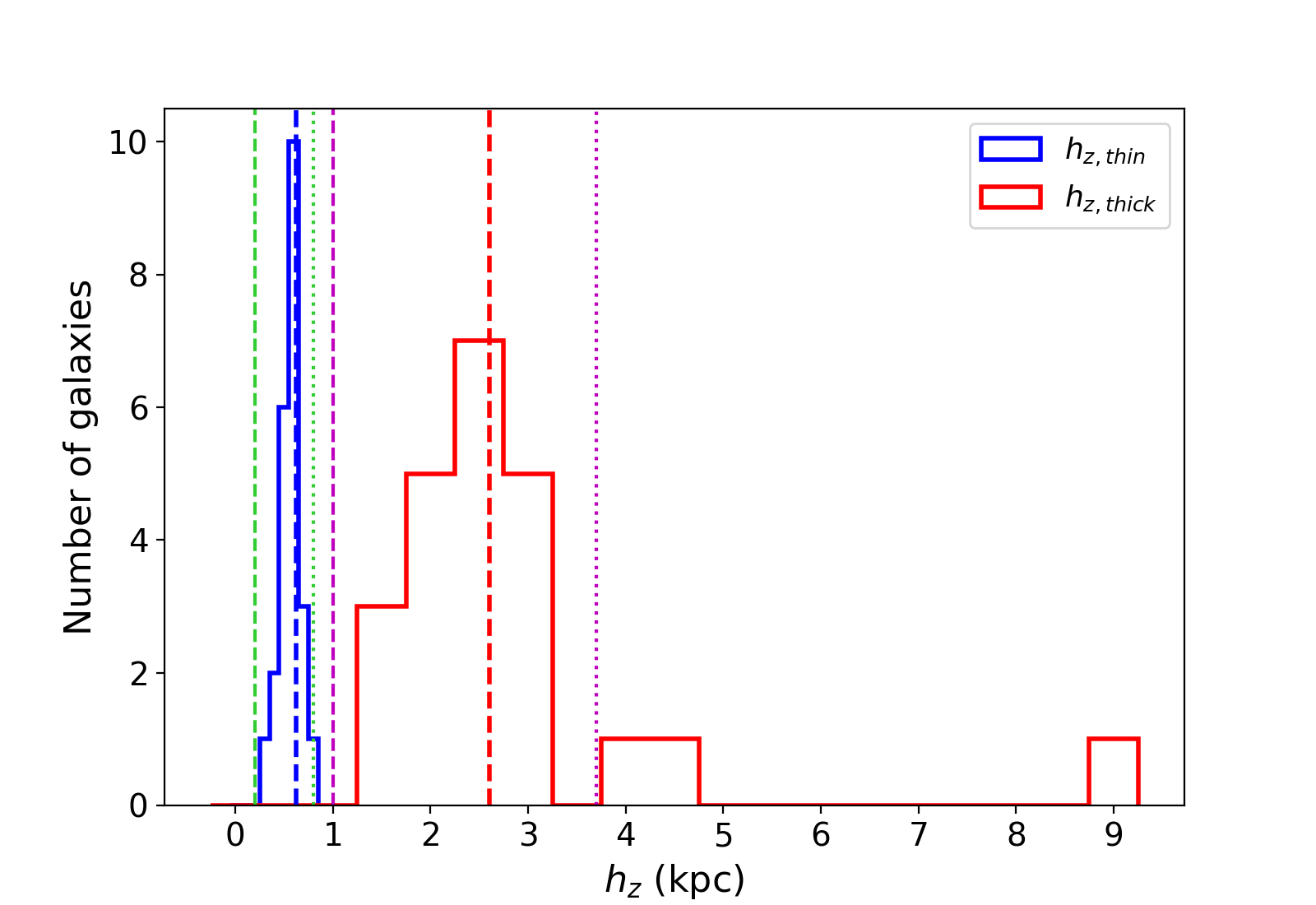}}
\caption{Frequency histograms of $h_{z,thin}$ and $h_{z,thick}$ in our AURIGA sample, with 0.1 and 0.5~kpc bins, respectively (solid blue and red lines). These distributions have a weighted average value of 0.6 and 2.6~kpc, respectively, indicated as vertical dashed blue and red lines. Vertical
green and magenta lines indicate the average (dashed) and maximum (dotted) thin- and thick-disk values from observations in the \citet{Comeron2018} sample. 
}
\label{fig:hz_hist}
\end{figure}

{We show in Fig.~\ref{fig:hz_hist} the distributions, in terms of frequency histograms, of the thick and thin-disk scale heights of the 24 galaxies in our AURIGA sample, obtained as described in Sect.~\ref{sub:morph_decomp} and gathered in Table~\ref{tab:thick}. Thick disks mostly have scale heights between 1.6 and 3.5~kpc (with three outliers, see also below), with an average value (weighted by the uncertainties) of 2.6~kpc. Thin disks have scale heights between 370 and 810~pc, on average 620~pc (see also Table~\ref{tab:thick}). 
}
We compare here the thickness of our thick and thin disks with the 124 galaxies morphologically fitted by \citet{Comeron2018} that have two clear distinct regions dominated by the thick or the thin disk. We find on average thicker thick and thin disks in AURIGA simulations than in this observed sample. 
{
Thick disks in the sample from \citet{Comeron2018} have a smaller average scale height ($\sim 1$~kpc, magenta dashed vertical line in Fig.~\ref{fig:hz_hist}), but a maximum scale height of $\sim 3.7$~kpc (magenta dotted vertical line). All except three thick disks from AURIGA have scale heights below this upper limit. We have not found any correlation between $h_{z,thick}$ and the fraction of stars with a halo-like kinematics in the thick-disk-dominated region (Sect.~\ref{sub:disc:origin}), reason why we exclude a potential contamination from halo stars as responsible for the higher scale heights of these three galaxies. 
Observed thin disks in \citet{Comeron2018} have very similar upper limit (green dotted vertical line in Fig.~\ref{fig:hz_hist}) as our thin disks, but an average scale height of $\sim 200$~pc, much lower than ours (respectively vertical dashed green and blue   lines).
}

While the fact that stellar disks in AURIGA are thicker than observed disks was already pointed out \citep{Grand2017,Gomez2017b}, this may be partially due to some limitations of current cosmological simulations \citep[e.g.,][]{Martig2014b}. 
One potential issue is numerical heating. Due to the resolution limitations of the simulations (in particular the mass of dark-matter particles, often more massive than baryonic particles), and to equipartition of energy, an additional two-body scattering increases random motions of the star particles \citep{Ludlow2021}. 
As said by \citet{Ludlow2021}, AURIGA simulations are unlikely to be strongly impacted by this effect, as this is a more common problem of cosmological volume simulations, with more massive dark-matter particles. \citet{Grand2017} showed in their Fig.~24 and 25 that changing resolution does not have a strong impact on the thickness of disks. 
However, the comparison between scale heights at solar radius in the AURIGA simulations of different resolution levels (for simulations used here in \citealt{Grand2017} and for higher resolution in \citealt{Grand2018b}) shows that resolution has some impact on disk scale heights. 

Another potential culprit is the interstellar medium model. Gas disks are already very often too thick in simulated galaxies \citep[e.g.,][]{Trayford2017}.
This is related to the fact that, to limit computational costs (and to avoid the spurious fragmentation of gas structures), the interstellar medium does not explore the full temperature - pressure parameter space in most current cosmological simulations. 
An artificial ``pressure floor'' leads to a longer Jeans length at the star-formation threshold, so that stellar disks cannot be much thinner than that. This results in thicker disks than in observed galaxies \citep{Bahe2016,Trayford2017}. 
This has been widely discussed in the literature \citep[e.g.,][]{Gabor2017,Crain2017, Sullivan2018, Weinberger2023} and it is illustrated for AURIGA simulations in \citeauthor{Kelly2022} (\citeyear{Kelly2022}, in their Fig.~11, compared to APOSTLE simulations). 
Nevertheless, it is worth noting that \citet{Marinacci2017} showed that the thickness of HI disks in AURIGA might be in agreement with HI observations. 
Finally, stellar feedback likely plays an important role in setting the scale heights of cool gas \citep[e.g.,][]{Roskar2014,Marinacci2016} and remains an active area of study.

\subsubsection{Spectroscopic observations of edge-on galaxies}

Throughout the paper we have used  an approach that is oriented to facilitate the comparison of these results with observations of edge-on galaxies. 
As mentioned in Sect.~\ref{sec:intro}, detailed observational studies of thick disks through IFS are limited to a few galaxies, in particular for spiral galaxies \citep{Yoachim2008b, Comeron2015, Martig2021}. 
We should be cautious with comparing our results, from AURIGA Milky Way-mass spiral galaxies,  with observations of galaxies of different morphology, mass, and environment. 
Furthermore, observational stellar-population parameters of thick disks in different studies were sometimes derived with different methods and we should also be cautious in comparing them to each other. Despite all this, most observations displayed older, more metal-poor and [Mg/Fe]-enhanced thick disks than their corresponding thin disks, independently of the morphological type and environment of their host galaxy \citep{Yoachim2008b, Comeron2015, Pinna2019b, Martig2021}. 
For these reasons, the following discussion is mostly based on a qualitative comparison. 

Thick disks in our study, with ages between 5 and 9~Gyr, are on average not as old as most observed thick disks. In particular, we compare qualitatively our sample to the only two edge-on spiral galaxies with clearly distinct thick and thin disks, with published two-dimensional stellar-population analyses from IFS data. 
One is NGC\,5746, an Sb galaxy of similar mass to the most massive AURIGA galaxies, with a $\sim 12$-Gyr-old thick disk \citep{Martig2021}. Another is the Sc galaxy ESO\,533-4, of similar mass to the least massive AURIGA galaxies, and  with a $\sim 10$-Gyr-old thick disk \citep{Comeron2015}. Both thick disks are older than all AURIGA thick disks, and the difference is larger than the estimated uncertainties in observations ($\sim 1$~Gyr in \citealt{Martig2021}). 
Some lenticular galaxies show old thick disks of similar ages as NGC\,5746 and ESO\,533-4 \citep{Comeron2016,Pinna2019a,Pinna2019b}, while others show thick disks  of similar ages of our youngest thick disks in AURIGA \citep{Kasparova2016,Kasparova2020}. 

The fact that young stars strongly flare in AURIGA disks (Fig.~\ref{fig:sample_age}), contaminating the thick-disk heights with younger stars, could be contributing to the fact that the average age of thick disks is never as old as the inner thick disk (see, in Fig.~\ref{fig:sample_age}, the age difference between the inner and the outer Voronoi bins in the thick disk of, e.g., Au9). 
In observations, such clear and strong flares have not been detected in age maps, because the outer region of the disks is not covered by the data \citep[e.g.,][]{Martig2021}, because of the lower spatial resolution in the faint outskirts, or just because they are not as strong in most real galaxies. In some cases they were instead observed in metallicity maps, suggesting that age differences may be present but sometimes difficult to be recovered (e.g., more than metallicity differences; see the discussion in \citealt{Pinna2019a,Pinna2019b}). 
Finally, another reason why thick disks are younger in AURIGA could be that star formation occurs in thicker gas disks even at low redshift, as discussed in Sect.~\ref{sub:morph_comp_obs}. 
This could also explain the fact that thick-disk metallicities are also on average higher than observed, reaching sometimes solar or slightly super-solar values. 
However, the two thick disks observed in NGC\,5746 and ESO\,533-4 show [M/H] values in the range covered by our simulation sample. While \citet{Comeron2015} did not provide [Mg/Fe] values for ESO\,533-4, NGC\,5746 shows higher values than our thick-disk sample. 
 
Regarding thin disks, they have shown a larger variety of different stellar-population properties 
than thick disks in observations. Furthermore, \citet{Pinna2019b,Pinna2019a} showed that their formation and evolution, happening in general later than thick disks, is more affected by the environment of their host galaxy than thick disks. 
Some observed thin disks have similar properties to the ones in our simulations. However, very old thin disks have been also observed in lenticular galaxies in galaxy clusters \citep{Comeron2016,Pinna2019a}. These are absent in our sample, made up of relatively isolated spiral galaxies. 
On the other hand, the few measured thin-disk [Mg/Fe] values in observations \citep{Pinna2019b,Pinna2019a,Martig2021} are in good agreement with the values in the right panels of Fig.~\ref{fig:tTdisks_mass} and \ref{fig:tTdisks_faccr}. 
We have already mentioned several potential causes of the identified differences between stellar-population properties in observed and simulated thin and thick disks. In addition to that, observational effects that can impact the recovery of stellar-population parameters due to, for example, projection, systematics related to the SSP models and their age, [M/H] or [Mg/Fe] resolution, and instrument spatial and spectral resolution, as well as the presence of dust (although not expected in large amounts in thick disks), were not corrected in observations and not taken into account in simulations. 
Furthermore, recent studies have suggested that full-spectrum fitting techniques tend to bias the results toward older ages \citep[e.g.,][]{Pinna2019a}. 
While producing mock spectra is beyond the scope of this paper, it might be the future way to go to produce results that are more quantitatively comparable to observations.

While accreted stars are easy to be mapped in simulations where we can track the particles and classify them into in situ and ex situ, this is challenging to be done in observations. \citet{Pinna2019a,Pinna2019b} and \citet{Martig2021} identified accreted stellar populations from their distribution in the age - metallicity plane, and mapped their spatial distribution. While this method is completely different from what is done here, they also found that accretion has a much more significant impact (measured in terms of mass fraction) in the thick-disk region, similarly to our results (Fig.~\ref{fig:sample_accrmf}). On the other hand, the mass density of accreted stars is higher in a central region of a few kiloparsecs both in all our simulated galaxies (Fig.~\ref{fig:sample_accrmassdens}) and in the observed ones. 
{Since no kinematic cut was applied in the observations or simulations, stars in the thick-disk-dominated region may in both cases contain some contamination from stars with halo kinematics (see Sect.~\ref{sub:disc:origin}).}

\section{Conclusions} \label{sec:concl}
The origin of thick disks and the connection with thin disks is still a matter of debate despite recent progress. Detailed observational studies (in particular from IFS) are so far limited to individual galaxies or very small samples, and simulation studies do not usually provide results that are directly comparable  to observations. 
We analyzed the properties of thick and thin disks, defined purely geometrically, in 24 galaxies from the AURIGA suite of zoom-in cosmological simulations. 
We projected the star particles of the 24 galaxies in an edge-on view, applying Voronoi binning. 
We fitted vertical surface brightness profiles in the $V$ band with two disk components, enabling a geometrical definition of the thick and thin disks. 
We analyzed the age, total metallicity [M/H], and [Mg/Fe] abundance of the galaxies seen edge-on. 
After classifying the star particles into in situ and ex situ components, we also mapped the distribution of accreted stars. 
From these results, we drew the following conclusions:
\begin{itemize}
    \item Thick disks are older, more metal poor, and more [Mg/Fe] enhanced than thin disks, indicating that most of their mass was formed at earlier times from gas that was not significantly enriched. They are on average $\sim 3$~Gyr older, $\sim 0.25$~dex more metal poor, and $\sim 0.06$ more [Mg/Fe]-enhanced than thin disks.
    \item In most galaxies, primordial thick disks were already in place 10~Gyr ago ($z=1.7$), with a shape and size sometimes very different from those at $z=0$, depending on the specific galaxy.
    \item Most thick disks had most of their mass formed in situ, with a significant fraction (on average $\sim 22$\%) of accreted stars contributing to their growth. Two galaxies had half of their thick disks made up of ex situ stars. In only four galaxies did satellite debris contribute less than 10\%. These results suggest that, in general, accretion plays a very important role in the growth of {geometric} thick disks, although internal formation is the dominant mechanism. 
    \item Accreted stars are mostly distributed in the inner region of galaxies (where they have higher mass densities). However, their mass fraction is much more significant in the thick-disk-dominated regions. A very similar trend was found in observations. 
    \item {Geometric thick disks can be contaminated, significantly in some cases,  by stars with halo kinematic properties that cannot be disentangled in observations of edge-on external galaxies. The halo contamination is in general greater in the accreted component of geometric thick disks. }
    \item Thick disks in galaxies with higher accreted fractions tend to be younger. Mergers contributed younger stars than the in situ component, especially relatively recent ones. 
    \item Mergers also cause additional thickening in preexisting disks, scattering cooler stars to greater distances from the midplane and thus contributing to the growth of the thick disk. 
    \item Thin disks, which are young, metal-rich, and poorly enhanced in [Mg/Fe] in all simulations, start to form at high redshifts only in some galaxies. 
    In others, with recent more active merger histories, they formed only at late times.
\end{itemize}
To sum up, thick disks result from a combination of external processes with the internal evolution of the galaxy. The balance between these processes, as well as the time of formation of thin metal-rich disks, depends on the specific galaxy and its merger history. 
In one companion paper, we will show the star-formation histories of thick disks in this sample and further discuss the in situ star formation and ex situ contribution in thick disks. In another companion paper, we will investigate bimodalities in the [Mg/Fe] - [M/H] plane, their correspondence with geometrically defined thick and thin disks, and how bimodality can be recovered in observations. 
While this work will be essential for the interpretation of future observations of Milky Way-mass galaxies, shedding more light on the properties and origin of thick disks and their connection with thin disks, additional simulations with a larger variety of galaxy masses, morphological types, and environments will be needed to achieve a complete view of how thick and thin disks form. Finally, comparing stellar properties from the direct star-particle analysis with results from mock spectra obtained from the simulations, and comparing the mock spectra with observations, will be the way forward toward understanding IFS observations. 

\section*{Acknowledgements}
RG acknowledges financial support from the Spanish Ministry of Science and Innovation (MICINN) through the Spanish State Research Agency, under the Severo Ochoa Program 2020-2023 (CEX2019-000920-S), and support from an STFC Ernest Rutherford Fellowship (ST/W003643/1).
FF acknowledges support from STFC grants ST/T000244/1 and ST/X001075/1. 
FAG acknowledges support from ANID FONDECYT Regular 1211370, the Max Planck Society through a “Partner Group” grant and ANID Basal Project FB210003.
We would like to thank Sylvia Plöckinger for the useful discussions.

\section*{}

\bibliographystyle{aa}
\bibliography{biblio_thickdisk}

\begin{appendix} 
\section{Results from the morphological decomposition into two disk components}

We provide in Table~\ref{tab:thick} the average heights ($z_{tT}$) beyond which the light of the thick disk dominates over the thin-disk light, as well as the scale heights of the thin and the thick disks (respectively $h_{z,thin}$ and $h_{z,thick}$), resulting from the fits described in Sect.~\ref{sub:morph_decomp}. Uncertainties ($\delta z_{tT}$, $\delta h_{z,thin}$ and $\delta h_{z,thick}$), also provided in Table~\ref{tab:thick}, were calculated propagating the $1\sigma$ uncertainties obtained from the Bayesian approach for single radial bins. 
For Au1, fitted with only one disk component (Sect.~\ref{sub:morph_decomp}) as also indicated in the second column of Table~\ref{tab:thick}, we provide the scale height and its uncertainty of this full-disk component. 
\onecolumn
\begin{table}
\caption{Parameters from the morphological decomposition into a thin and a thick disk.}
\centering
\begin{tabular}{cccccccc}
\hline\hline
Galaxy name & $N_{disks}$ & $z_{tT}$ (kpc) & $\delta z_{tT}$ (kpc)& $h_{z,thin}$ (kpc)& $\delta h_{z,thin}$ (kpc)& $h_{z,thick}$ (kpc)& $\delta h_{z,thick}$(kpc)\\
\hline
Au1$^{(1)}$   &  1 & -   & -   & -   & -   & 1.60 & 0.09 \\
Au2   &  2 & 2.0 & 0.3 & 0.66 & 0.08 & 2.66 & 0.21 \\
Au3   &  2 & 2.0 & 0.4 & 0.51 & 0.06 & 1.85 & 0.20 \\
Au5   &  2 & 2.4 & 0.2 & 0.53 & 0.04 & 2.81 & 0.20 \\
Au6   &  2 & 2.1 & 0.2 & 0.53 & 0.05 & 2.56 & 0.19 \\
Au7   &  2 & 2.6 & 0.2 & 0.78 & 0.07 & 4.26 & 0.25 \\
Au8   &  2 & 1.1 & 0.4 & 0.63 & 0.16 & 2.38 & 0.16 \\
Au9   &  2 & 1.9 & 0.2 & 0.47 & 0.04 & 2.20 & 0.17 \\
Au10  &  2 & 1.8 & 0.3 & 0.56 & 0.07 & 2.77 & 0.17 \\
Au12  &  2 & 3.9 & 0.2 & 0.77 & 0.04 & 9.48 & 0.80 \\
Au14  &  2 & 3.5 & 0.3 & 0.81 & 0.06 & 4.81 & 0.32 \\
Au15  &  2 & 2.8 & 0.2 & 0.68 & 0.05 & 3.43 & 0.24 \\
Au16  &  2 & 2.4 & 0.3 & 0.70 & 0.07 & 3.14 & 0.24 \\
Au17  &  2 & 1.8 & 0.6 & 0.61 & 0.10 & 1.82 & 0.16 \\
Au18  &  2 & 1.7 & 0.2 & 0.48 & 0.05 & 2.15 & 0.18 \\
Au19  &  2 & 2.4 & 0.2 & 0.63 & 0.06 & 2.89 & 0.23 \\
Au21  &  2 & 2.9 & 0.3 & 0.72 & 0.06 & 3.22 & 0.26 \\
Au22  &  2 & 1.2 & 0.2 & 0.37 & 0.06 & 2.30 & 0.13 \\
Au23  &  2 & 2.0 & 0.2 & 0.52 & 0.05 & 2.25 & 0.18 \\
Au24  &  2 & 2.3 & 0.2 & 0.65 & 0.06 & 3.29 & 0.23 \\
Au25  &  2 & 1.3 & 0.7 & 0.70 & 0.14 & 1.65 & 0.13 \\
Au26  &  2 & 1.0 & 0.3 & 0.65 & 0.14 & 2.65 & 0.16 \\
Au27  &  2 & 2.8 & 0.2 & 0.68 & 0.05 & 2.75 & 0.23 \\
Au28  &  2 & 1.3 & 0.2 & 0.57 & 0.08 & 3.24 & 0.17 \\
\hline
\end{tabular}
\label{tab:thick}
\begin{tablenotes}
\item {\footnotesize Notes. Columns from left to right: galaxy name, number of disk components ($N_{disks}$) identified in the analyzed $1.6 R_{opt} \times 4h_{scale}$ region, distance from the midplane beyond which the thick-disk light dominates over the thin-disk light ($z_{tT}$), its uncertainty ($\delta z_{tT}$), thin-disk scale height ($h_{z,thin}$), its uncertainty ($\delta h_{z,thin}$), thick-disk scale height ($h_{z,thick}$) and its uncertainty ($\delta h_{z,thick}$). 
{(1) Au1, with no clear double-disk structure, was fitted with only one disk component and $h_{z,thick}$ and $\delta h_{z,thick}$ refer to its scale height and uncertainty.
}
\normalsize
}
\end{tablenotes}
\end{table}
\newpage
\twocolumn
\section{Additional maps}

\subsection{Stellar kinematic maps in an edge-on projection}\label{app:kin_edgeon}
\begin{landscape}
\begin{figure}
\centering
\resizebox{1.3\textwidth}{!}
{\includegraphics[scale=1.5]{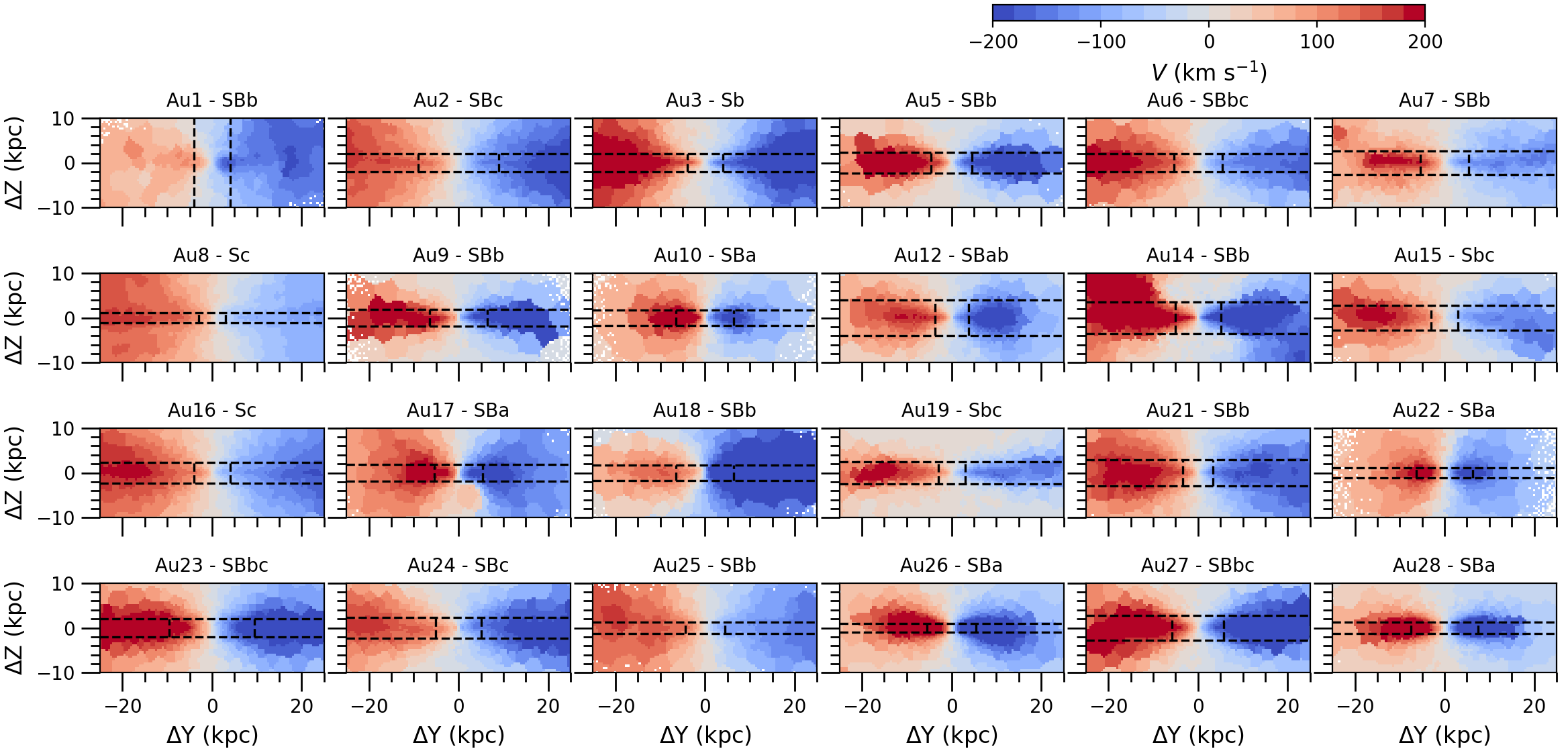}}
\caption{Mean-velocity ($V$) maps of the 24 galaxies in our sample, seen edge on. For display purposes, we show a central area of the same 50~kpc$\times$20~kpc size for all galaxies. Horizontal dashed black lines indicate the regions where the thin disk (within the two lines) and the thick disk dominate (above and below the region between the two lines). Vertical dashed black lines enclose the central region dominated by a bar or a classical bulge. Hubble types from Table~\ref{tab:sample} are indicated on top of each panel. 
Pixels of 0.5~kpc$\times$0.5~kpc size, Voronoi binned to a target number of particles of 900 star particles per bin, were used to plot these maps. 
}
\label{fig:sample_vel}
\end{figure}
\end{landscape}

\begin{landscape}
\begin{figure}
\centering
\resizebox{1.3\textwidth}{!}
{\includegraphics[scale=1.5]{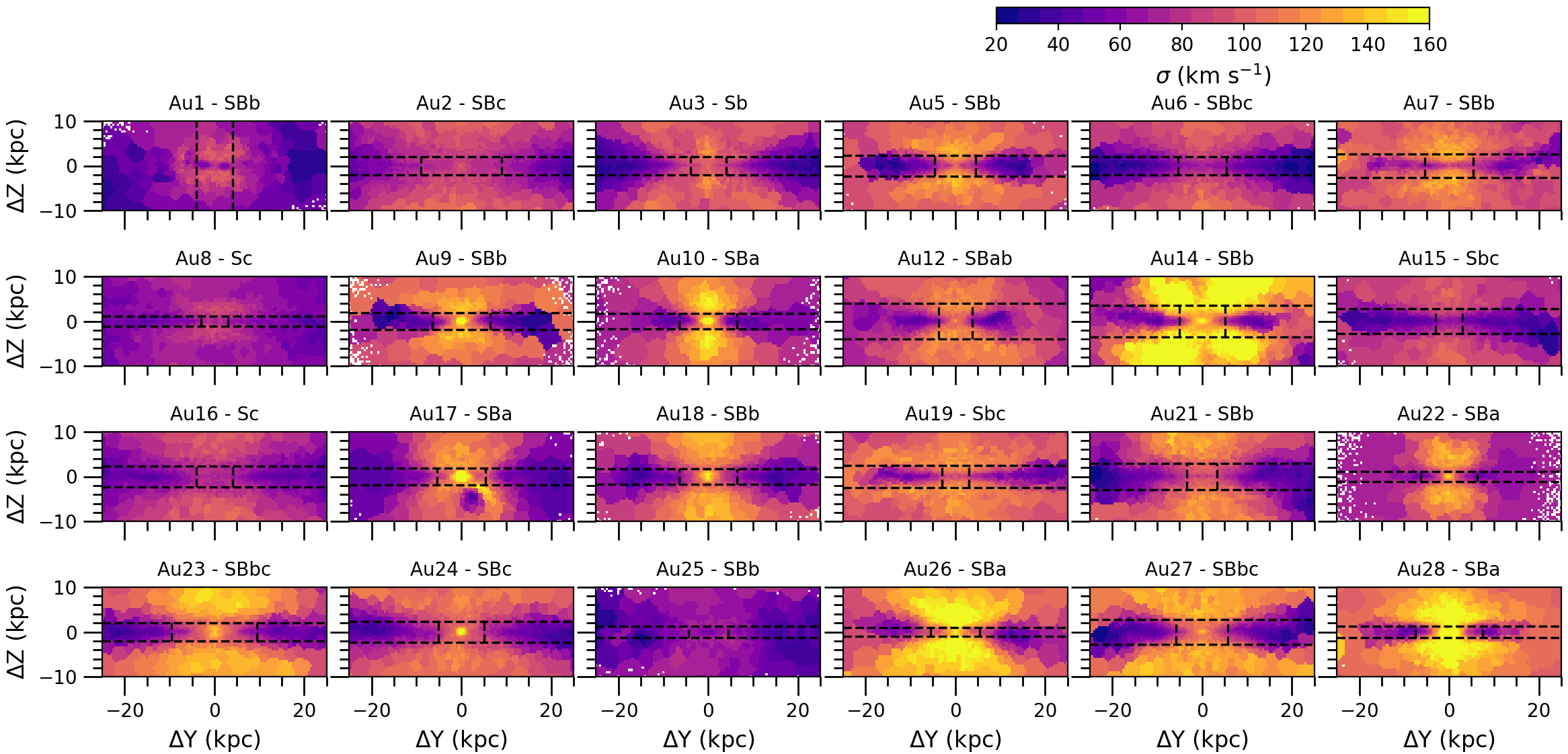}}
\caption{Velocity-dispersion ($\sigma$) maps of the 24 galaxies in our sample, seen edge on. For display purposes, we show a central area of the same 50~kpc$\times$20~kpc size for all galaxies. Horizontal dashed black lines indicate the regions where the thin disk (within the two lines) and the thick disk dominate (above and below the region between the two lines). Vertical dashed black lines enclose the central region dominated by a bar or a classical bulge. Hubble types from Table~\ref{tab:sample} are indicated on top of each panel. 
Pixels of 0.5~kpc$\times$0.5~kpc size, Voronoi binned to a target number of particles of 900 star particles per bin, were used to plot these maps. 
}
\label{fig:sample_sigma}
\end{figure}
\end{landscape}

Kinematic parameters shown here were calculated by fitting the line-of-sight velocity distribution (LOSVD) of star particles with a Gauss-Hermite expansion, which is widely used for observations  \citep{vanderMarel1993,Gerhard1993,Cappellari2004}. We show here the maps of the first four moments of the LOSVD, often  shown in observations of edge-on galaxies \citep{Guerou2016,Pinna2019a,Pinna2019b,Martig2021, Sattler2023}. 
As in the rest of the maps shown in this paper, we map a region extended 25~kpc radially and 10~kpc in height, with a total size of 50~kpc$\times$20~kpc. 
Velocity ($V$) maps, in Fig.~\ref{fig:sample_vel}, show strong rotation in the thin-disk region and slower rotation in the thick disks, with larger (e.g., in Au9) or smaller (Au3) thick-disk rotation lags, depending on the galaxy. 
Velocity dispersion ($\sigma$) maps show in Fig.~\ref{fig:sample_sigma} that stars in the thin-disk midplane region are in general kinematically cooler than stars in the region dominated by thick disks. However, while differences between the two components are clear in most galaxies (see, e.g., Au5 and Au28), they are not clear in some others with similar thick- and thin-disk properties (such as Au8 and Au25; see also Sect.~\ref{sub:disc:origin}). In some galaxies, a much higher velocity dispersion is seen in the central region (see, e.g., Au10 and Au28), mostly related to the presence of a (boxy or classical) bulge and showing a clear ``X'' shape in Au14. 

\begin{landscape}

\begin{figure}
\centering
\resizebox{1.3\textwidth}{!}
{\includegraphics[scale=1.5]{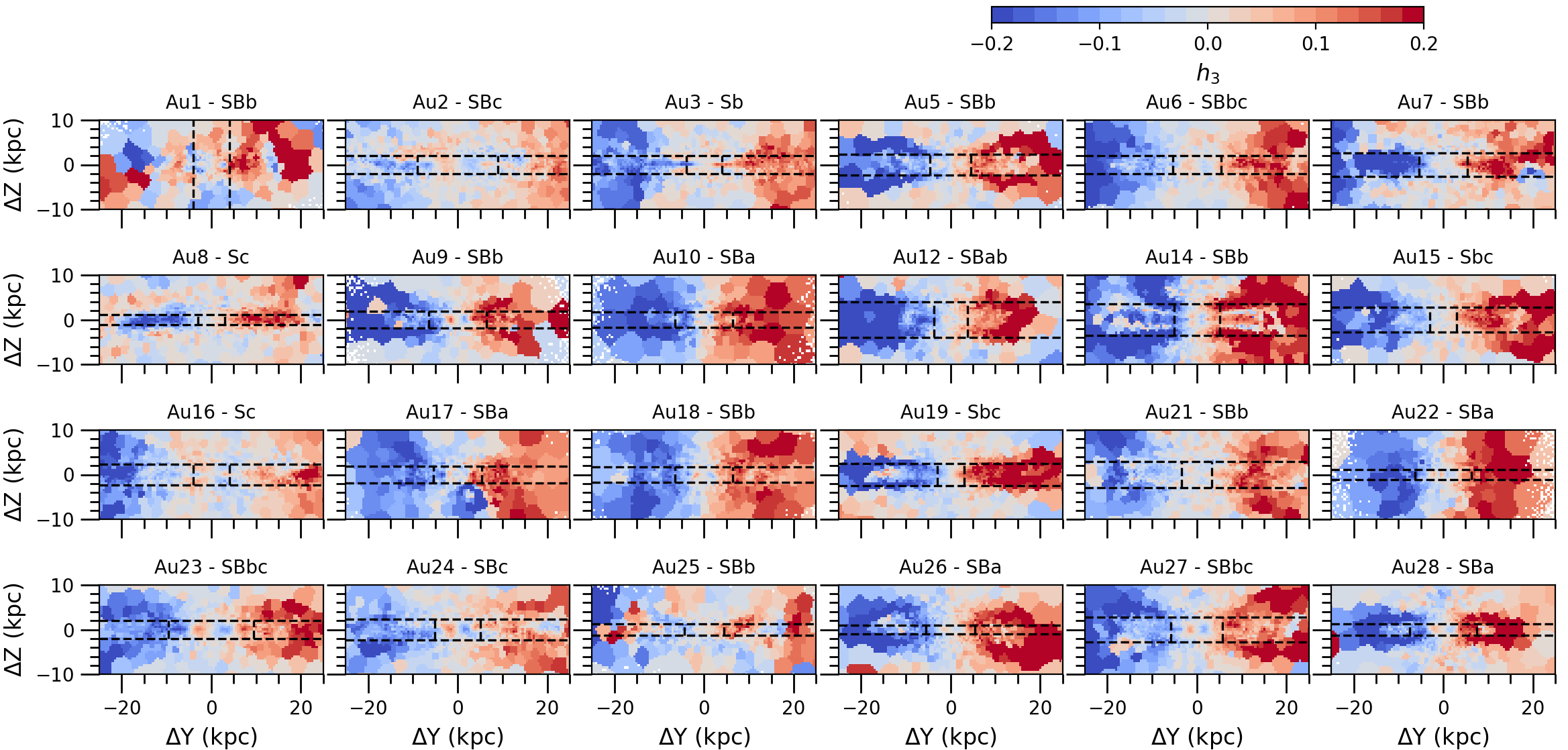}}
\caption{Skewness ($h_3$) maps of the 24 galaxies in our sample, seen edge on. For display purposes, we show a central area of the same 50~kpc$\times$20~kpc size for all galaxies. Horizontal dashed black lines indicate the regions where the thin disk (within the two lines) and the thick disk dominate (above and below the region between the two lines). Vertical dashed black lines enclose the central region dominated by a bar or a classical bulge. Hubble types from Table~\ref{tab:sample} are indicated on top of each panel. 
Pixels of 0.5~kpc$\times$0.5~kpc size, Voronoi binned to a target number of particles of 900 star particles per bin, were used to plot these maps. 
}
\label{fig:sample_h3}
\end{figure}
\end{landscape}

\begin{landscape}
\begin{figure}
\centering
\resizebox{1.3\textwidth}{!}
{\includegraphics[scale=1.5]{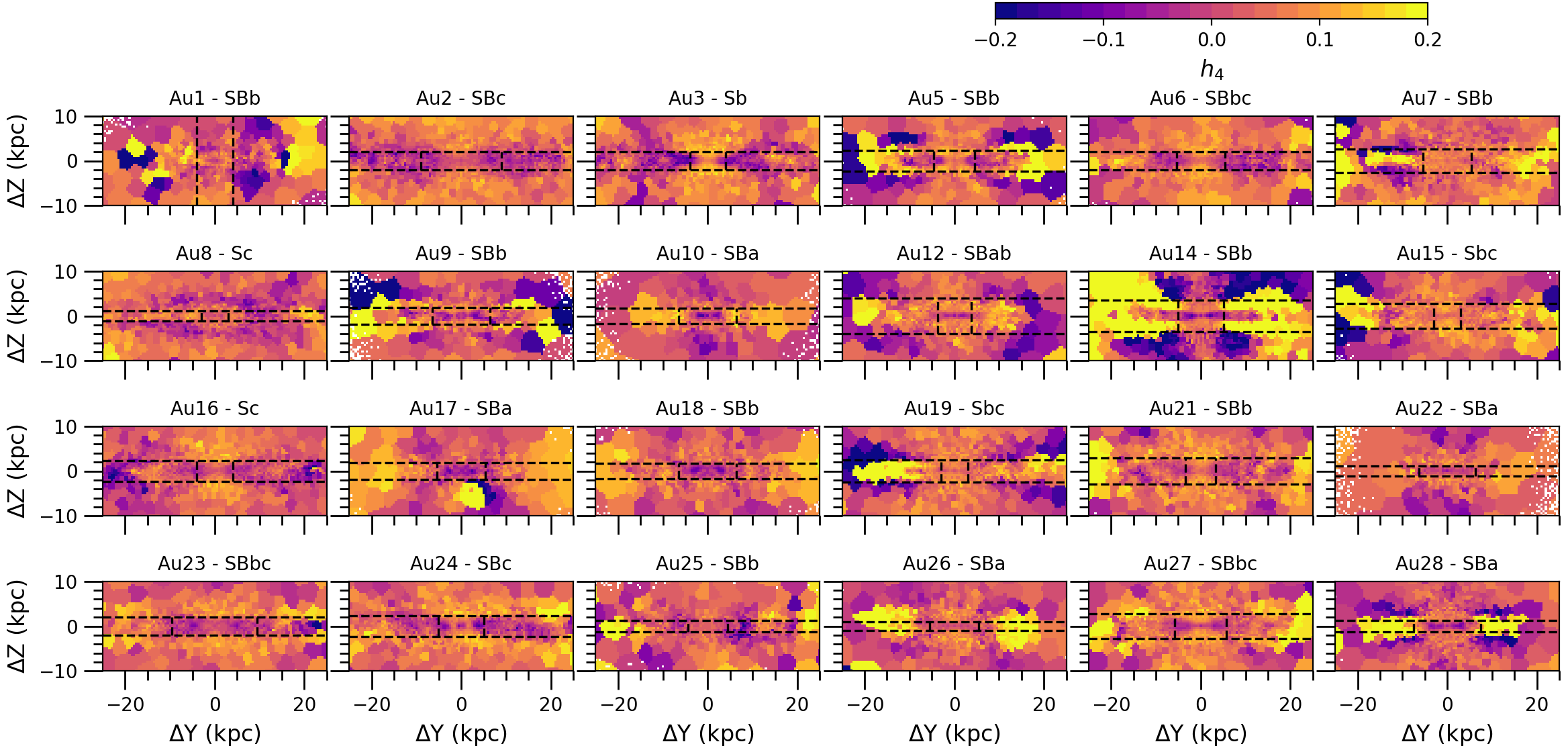}}
\caption{Kurtosis ($h_4$) maps of the 24 galaxies in our sample, seen edge on. For display purposes, we show a central area of the same 50~kpc$\times$20~kpc size for all galaxies. Horizontal dashed black lines indicate the regions where the thin disk (within the two lines) and the thick disk dominate (above and below the region between the two lines). Vertical dashed black lines enclose the central region dominated by a bar or a classical bulge. Hubble types from Table~\ref{tab:sample} are indicated on top of each panel. 
Pixels of 0.5~kpc$\times$0.5~kpc size, Voronoi binned to a target number of particles of 900 star particles per bin, were used to plot these maps. 
}
\label{fig:sample_h4}
\end{figure}
\end{landscape}

Skewness ($h_3$) is shown in Fig.~\ref{fig:sample_h3} and is characterized by a clear anticorrelation with $V$ in regions dominated by disk-like rotation (as observed by, e.g., \citealt{Krajnovic2008,Pinna2019a,Pinna2019b}). The highest absolute values of $h_3$ can be found either in the thin- or the thick-disk regions, depending on the galaxy. 
In the central region of barred galaxies (see in particular Au2, Au6, Au9, Au14, Au16, Au18, Au23, Au24, Au27, Au28), $h_3$ is correlated with $V$, as it is expected in bar-dominated regions characterized by cylindrical rotation (see also, e.g., \citealt{Pinna2019a}). 
Kurtosis ($h_4$), in Fig.~\ref{fig:sample_h4}, shows different patterns in different galaxies. Transitions from negative values (indicating a LOSVD with a broader symmetric profile than a pure Gaussian) to positive values (indicating a LOSVD with a narrower symmetric profile than a pure Gaussian), or vice versa, are observed from the inner to the outer regions of thin and thick disks (see, e.g., Au3, Au5, Au9, and Au14) or across the vertical direction (as in Au2, Au5, Au19, Au24, and Au28). 

While it is not clear in mock images that thin disks flare (Fig.~\ref{fig:sample_images}), kinematic maps (Figs.~\ref{fig:sample_vel} to \ref{fig:sample_h4}) show that a disk component defined by fast-rotating, dynamically cool stars flares. 
While most of these patterns in Figs.~\ref{fig:sample_vel} to \ref{fig:sample_h4} were also found in observations, some features may be more visible in results from simulations, where the signal-to-noise is not affecting the penalization of higher moments (here $h_3$ and $h_4$) as it happens in observations, biasing toward lower values \citep{Pinna2019a, Pinna2019b, Pinna2021}. In particular, the wealth of structures displayed here in the $h_4$ maps has not been observed in observations of thick and thin disks in edge-on galaxies.



\newpage

\subsection{Stellar-population maps for the full sample of 24 galaxies} \label{sub:SPmaps_all}

\begin{landscape}
\begin{figure}
\centering
\resizebox{1.3\textwidth}{!}
{\includegraphics[scale=1.5]{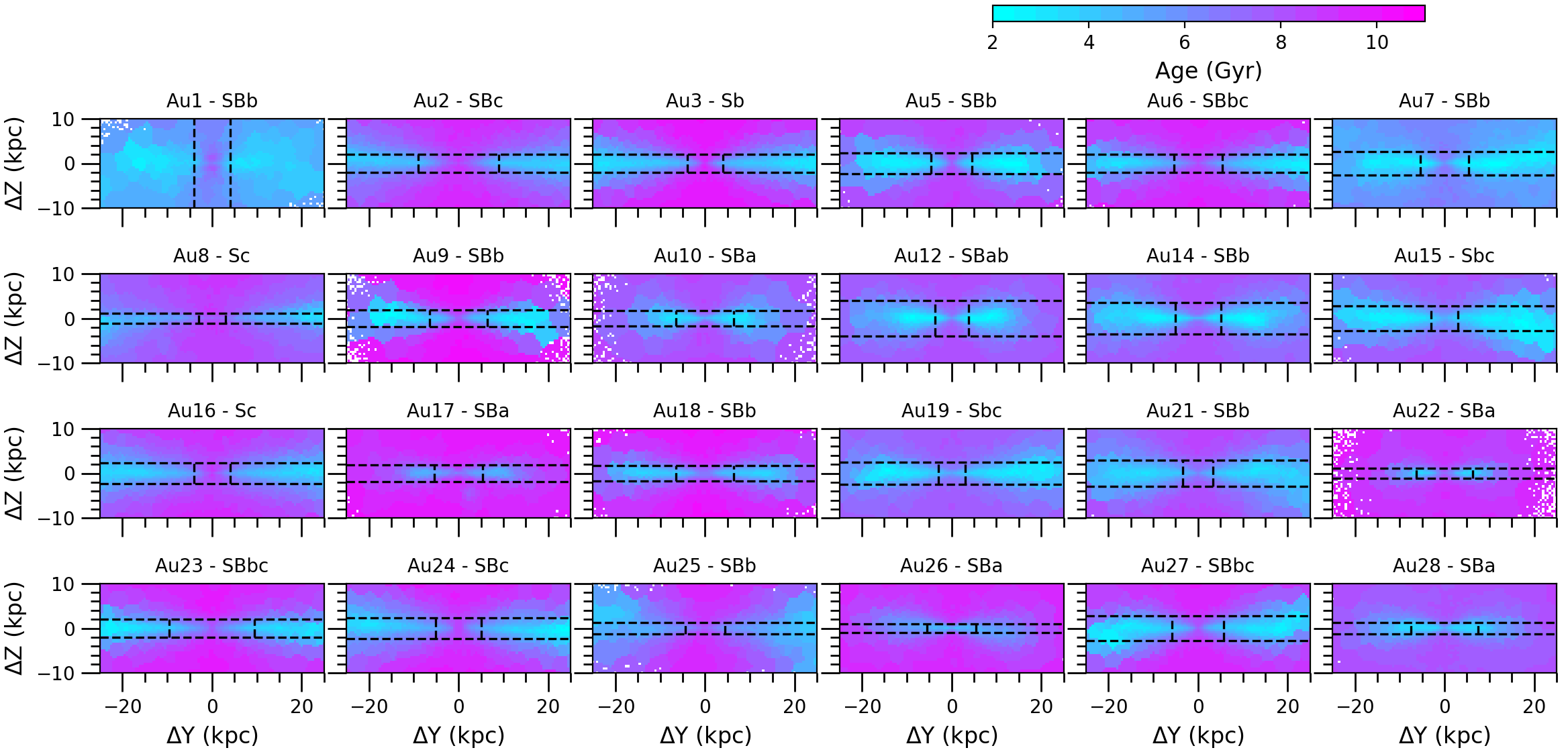}}
\caption{Mass-weighted stellar age maps of the 24 galaxies in our sample, seen edge on. For display purposes, we show for all galaxies a central area of the same 50~kpc$\times$20~kpc size. Horizontal dashed black lines indicate the regions where the thin disk (within the two lines) and the thick disk dominate (above and below the region between the two lines). Vertical dashed black lines enclose the central region dominated by a bar or a classical bulge. Names of the AURIGA halos and Hubble types from Table~\ref{tab:sample} are indicated on top of each panel.
Ages were integrated and projected into pixels of 0.5~kpc$\times$0.5~kpc size and Voronoi binned to obtain at least 900 star particles per bin. 
}
\label{fig:sample_age}
\end{figure}
\end{landscape}

\begin{landscape}
\begin{figure}
\centering
\resizebox{1.3\textwidth}{!}
{\includegraphics[scale=1.5]{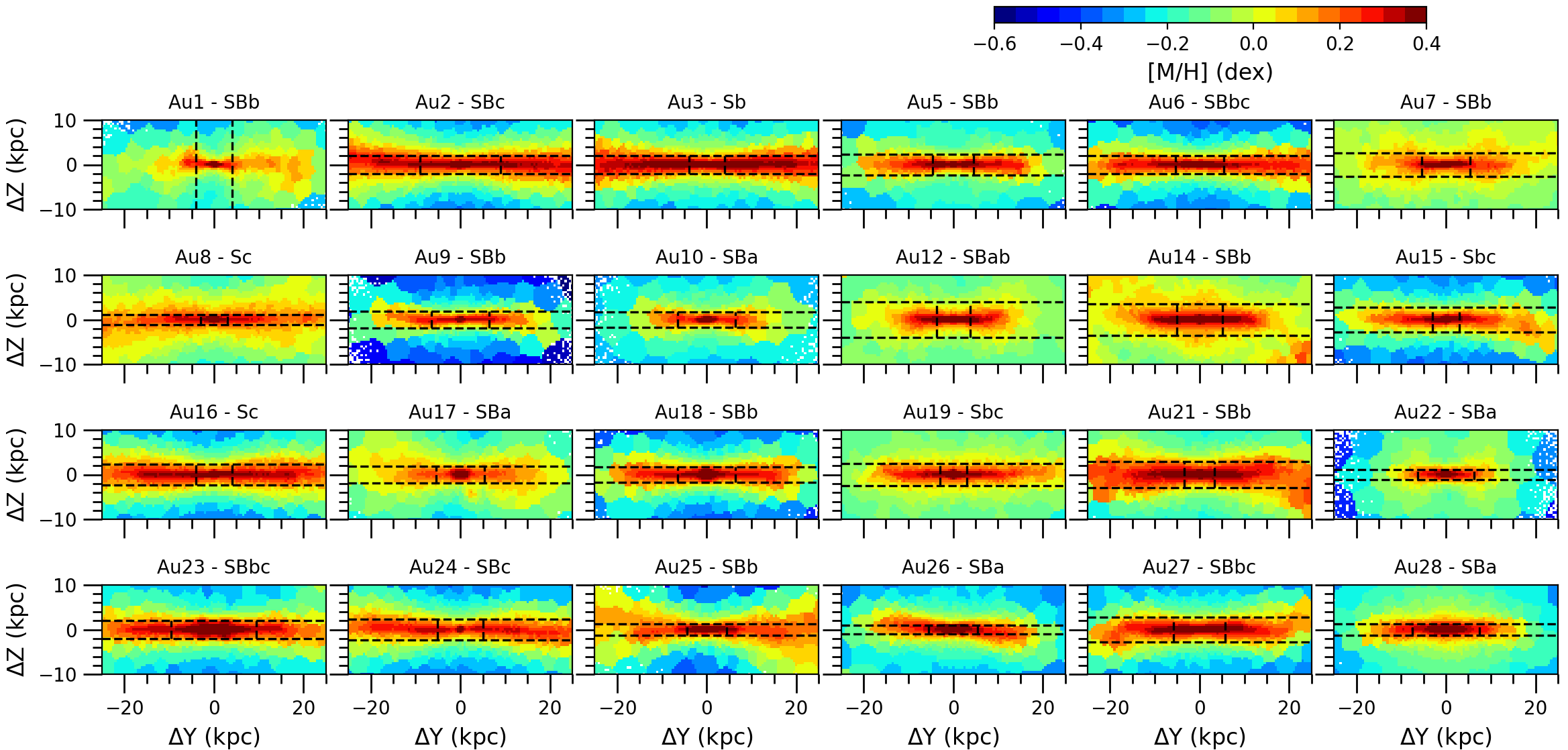}}
\caption{Mass-weighted total stellar metallicity [M/H] maps of the 24 galaxies in our sample, seen edge on. For display purposes, we show for all galaxies a central area of the same 50~kpc$\times$20~kpc size. Horizontal black dashed lines indicate the regions where the thin disk (within the two lines) and the thick disk dominate (above and below the region between the two lines). Vertical dashed black lines enclose the central region dominated by a bar or a classical bulge. Names of the AURIGA halos and Hubble types from Table~\ref{tab:sample} are indicated on top of each panel. 
Metallicities were integrated and projected into pixels of 0.5~kpc$\times$0.5~kpc size and Voronoi binned to obtain at least 900 star particles per bin. 
}
\label{fig:sample_met}
\end{figure}
\end{landscape}

\begin{landscape}
\begin{figure}
\centering
\resizebox{1.3\textwidth}{!}
{\includegraphics[scale=1.5]{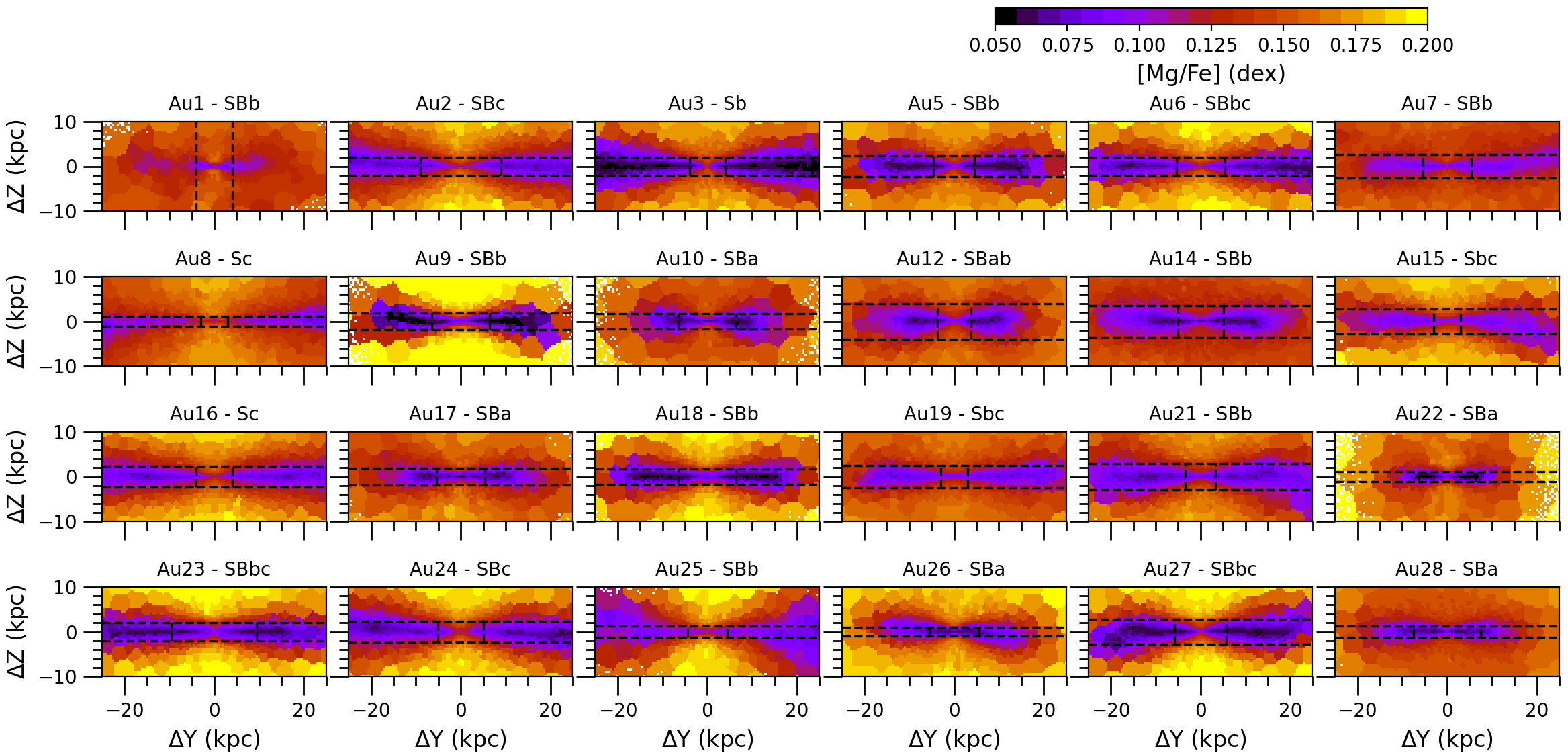}}
\caption{Mass-weighted stellar [Mg/Fe]-abundance maps of the 24 galaxies in our sample, seen edge on. For display purposes, we show for all galaxies a central area of the same 50~kpc$\times$20~kpc size. Horizontal dashed black lines indicate the regions where the thin disk (within the two lines) and the thick disk dominate (above and below the region between the two lines). Vertical dashed black lines enclose the central region dominated by a bar or a classical bulge. Names of the AURIGA halos and Hubble types from Table~\ref{tab:sample} are indicated on top of each panel. 
[Mg/Fe] abundances were integrated and projected into pixels of 0.5~kpc$\times$0.5~kpc size and Voronoi binned to obtain at least 900 star particles per bin. 
}
\label{fig:sample_mg}
\end{figure}
\end{landscape}

We provide here maps of age (Fig.~\ref{fig:sample_age}), metallicity (Fig.~\ref{fig:sample_met}) and [Mg/Fe] abundance (Fig.~\ref{fig:sample_mg}) for the full sample of 24 galaxies. 

\subsection{Maps of accreted stars for the full sample of 24 galaxies} \label{sub:accr_maps_all}

\begin{landscape}
\begin{figure}
\centering
\resizebox{1.3\textwidth}{!}
{\includegraphics[scale=1.5]{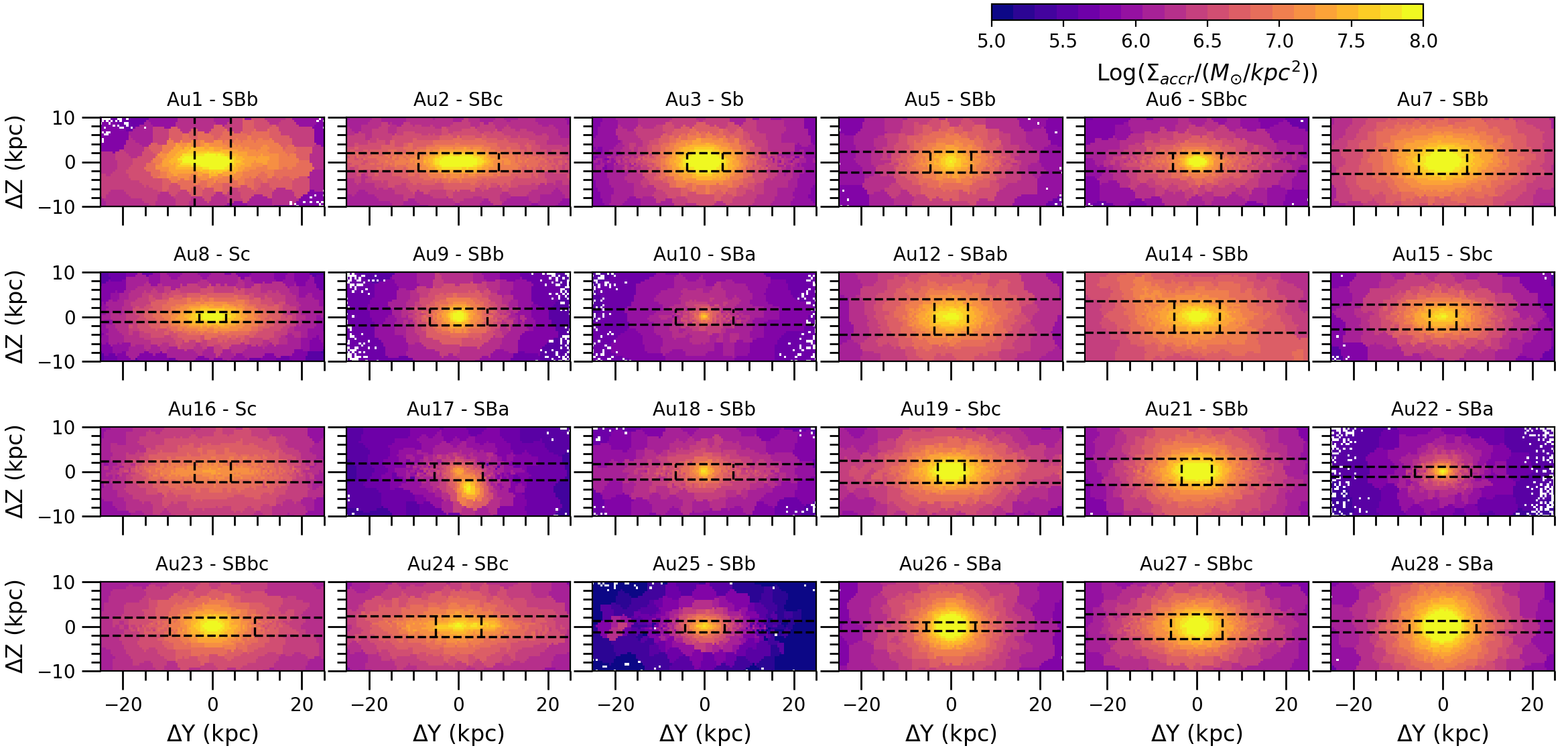}}
\caption{Maps of the mass surface density of accreted stars, in logarithmic scale, of the 24 galaxies in our sample, seen edge on. For display purposes, we show for all galaxies a central area of the same 50~kpc$\times$20~kpc size.  Horizontal dashed black lines indicate the regions where the thin disk (within the two lines) and the thick disk dominate (above and below the region between the two lines). Vertical dashed black lines enclose the central region dominated by a bar or a classical bulge. Names of the AURIGA halos and Hubble types from Table~\ref{tab:sample} are indicated on top of each panel. 
Pixels of 0.5~kpc$\times$0.5~kpc size, Voronoi binned to a target number of particles of 900 star particles per bin, were used to plot these maps. 
}
\label{fig:sample_accrmassdens}
\end{figure}
\end{landscape}

\begin{landscape}
\begin{figure}
\centering
\resizebox{1.3\textwidth}{!}
{\includegraphics[scale=1.5]{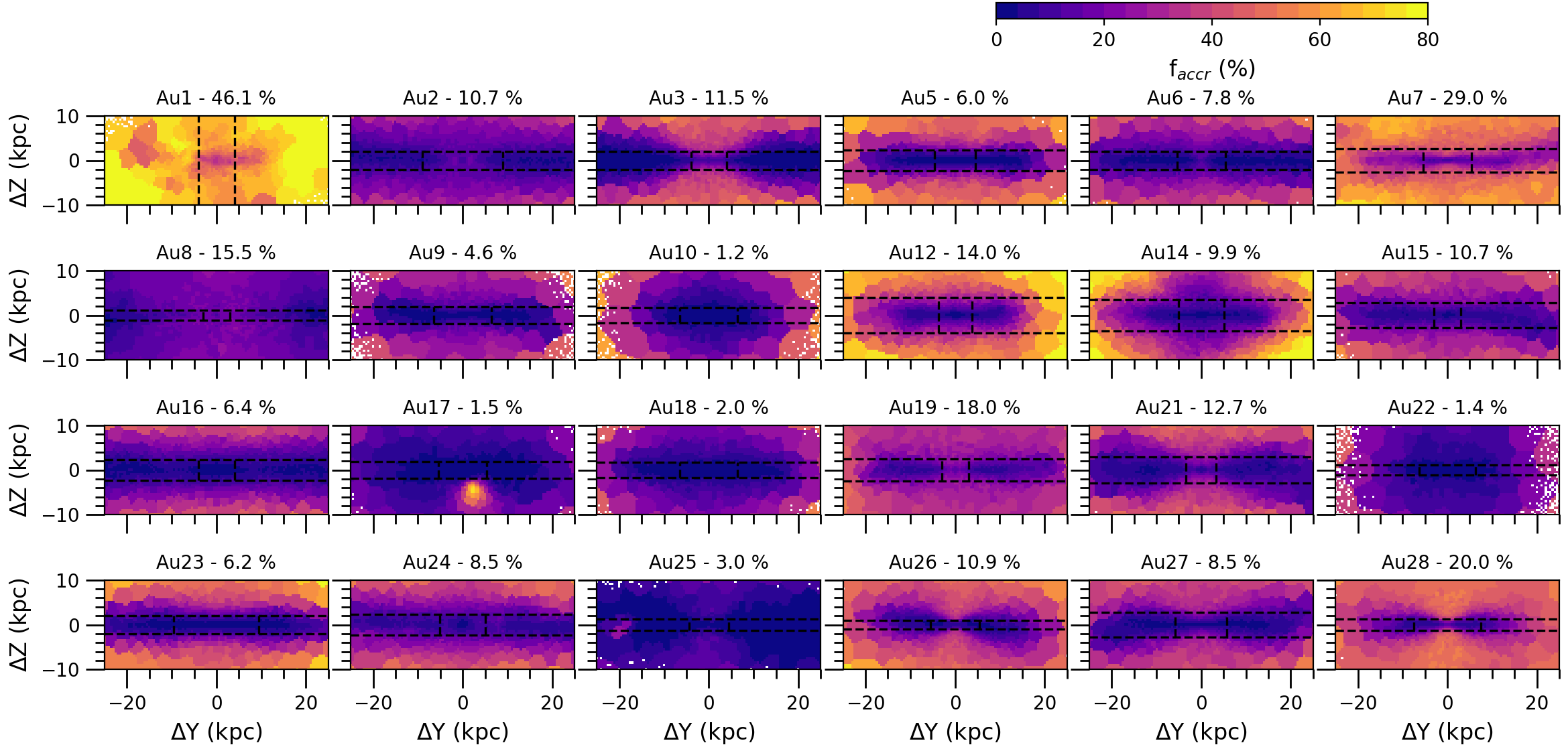}}
\caption{Maps of the mass fraction of accreted stars of the 24 galaxies in our sample, seen edge on. For display purposes, we show for all galaxies a central area of the same 50~kpc$\times$20~kpc size. Horizontal dashed black lines indicate the regions where the thin disk (within the two lines) and the thick disk dominate (above and below the region between the two lines). Vertical dashed black lines enclose the central region dominated by a bar or a classical bulge. Names of the AURIGA halos and the total mass fraction of accreted stars (calculated in the analyzed region, of size $1.6 R_{opt} \times 4h_{scale}$, Sect.~\ref{sub:region}) are indicated on top of each panel. 
Pixels of 0.5~kpc$\times$0.5~kpc size, Voronoi binned to a target number of particles of 900 star particles per bin, were used to plot these maps. 
}
\label{fig:sample_accrmf}
\end{figure}
\end{landscape}

We show here maps of the mass surface density (Fig.~\ref{fig:sample_accrmassdens}) and mass fraction (Fig.~\ref{fig:sample_accrmf}) of accreted stars for the full sample of 24 galaxies. 


\subsection{Time evolution} \label{app:snap}
\begin{figure*}
\centering
\resizebox{1.\textwidth}{!}
{\includegraphics[scale=1.5]{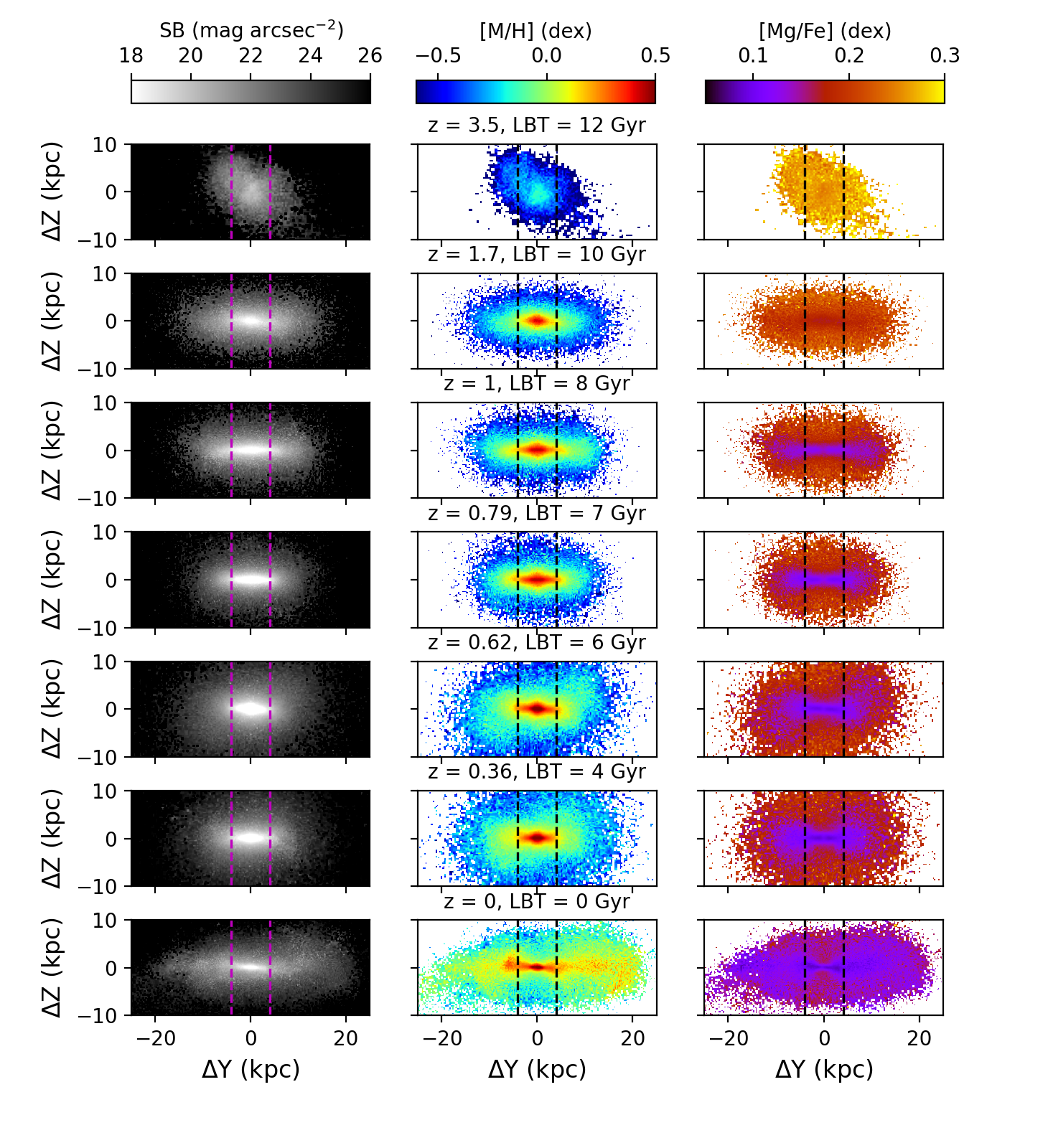}}
\caption{Same as Fig.~\ref{fig:snap_au7} but for the galaxy Au1.
}
\label{fig:snap_au1}
\end{figure*}

\begin{figure*}
\centering
\resizebox{1.\textwidth}{!}
{\includegraphics[scale=1.5]{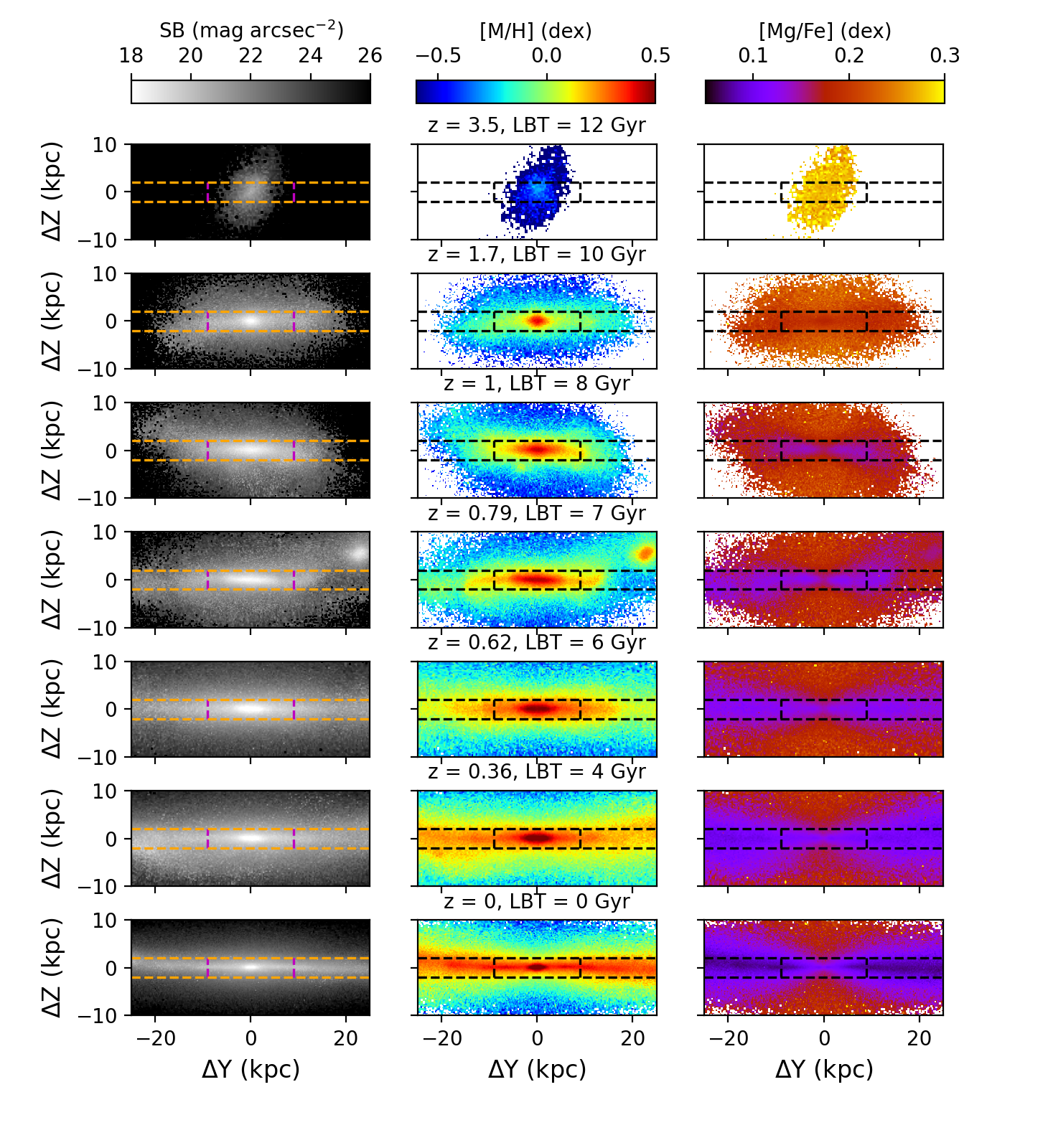}}
\caption{Same as Fig.~\ref{fig:snap_au7} but for the galaxy Au2.
}
\label{fig:snap_au2}
\end{figure*}

\begin{figure*}
\centering
\resizebox{1.\textwidth}{!}
{\includegraphics[scale=1.5]{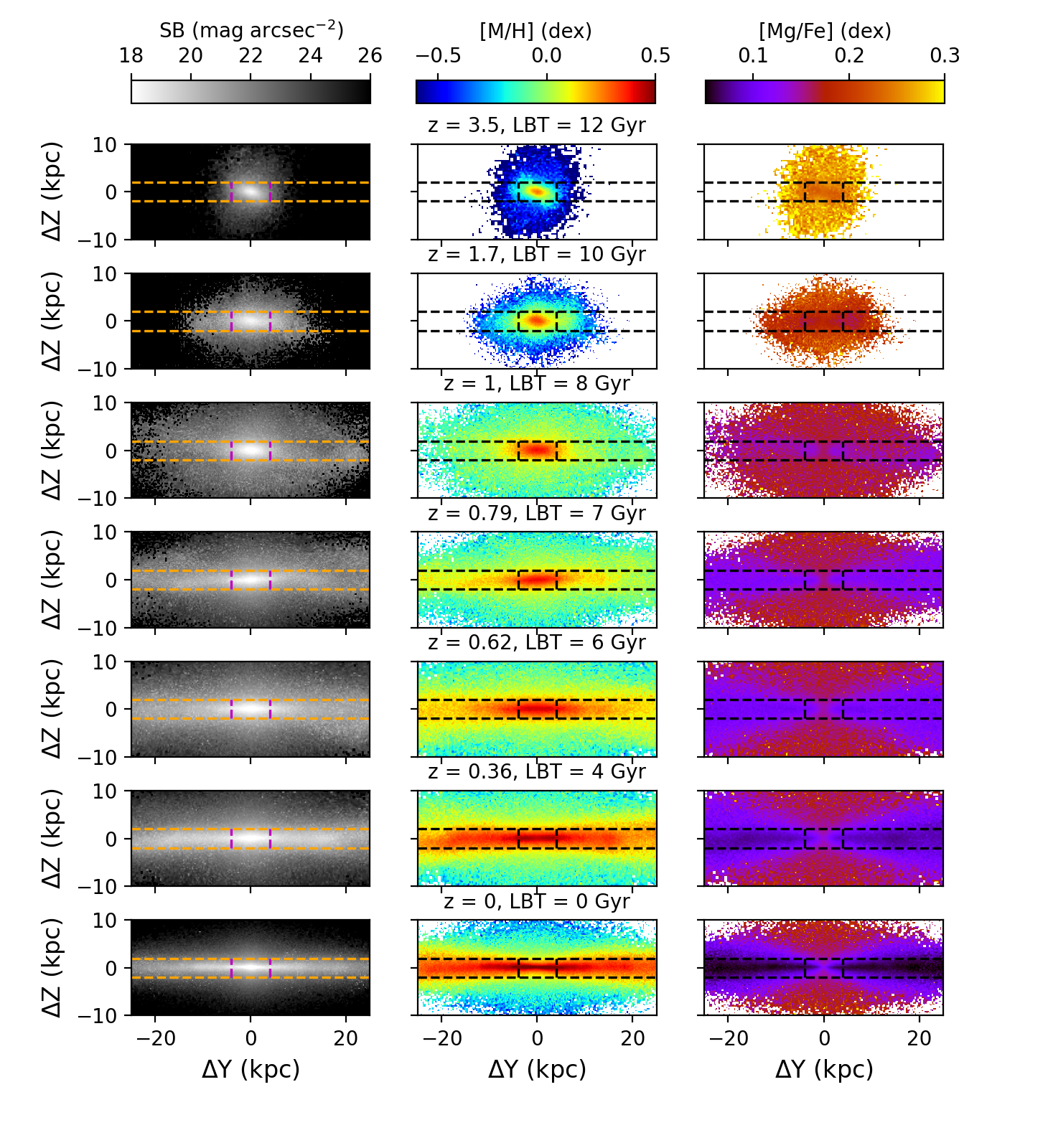}}
\caption{Same as Fig.~\ref{fig:snap_au7} but for the galaxy Au3. 
}
\label{fig:snap_au3}
\end{figure*}

\begin{figure*}
\centering
\resizebox{1.\textwidth}{!}
{\includegraphics[scale=1.5]{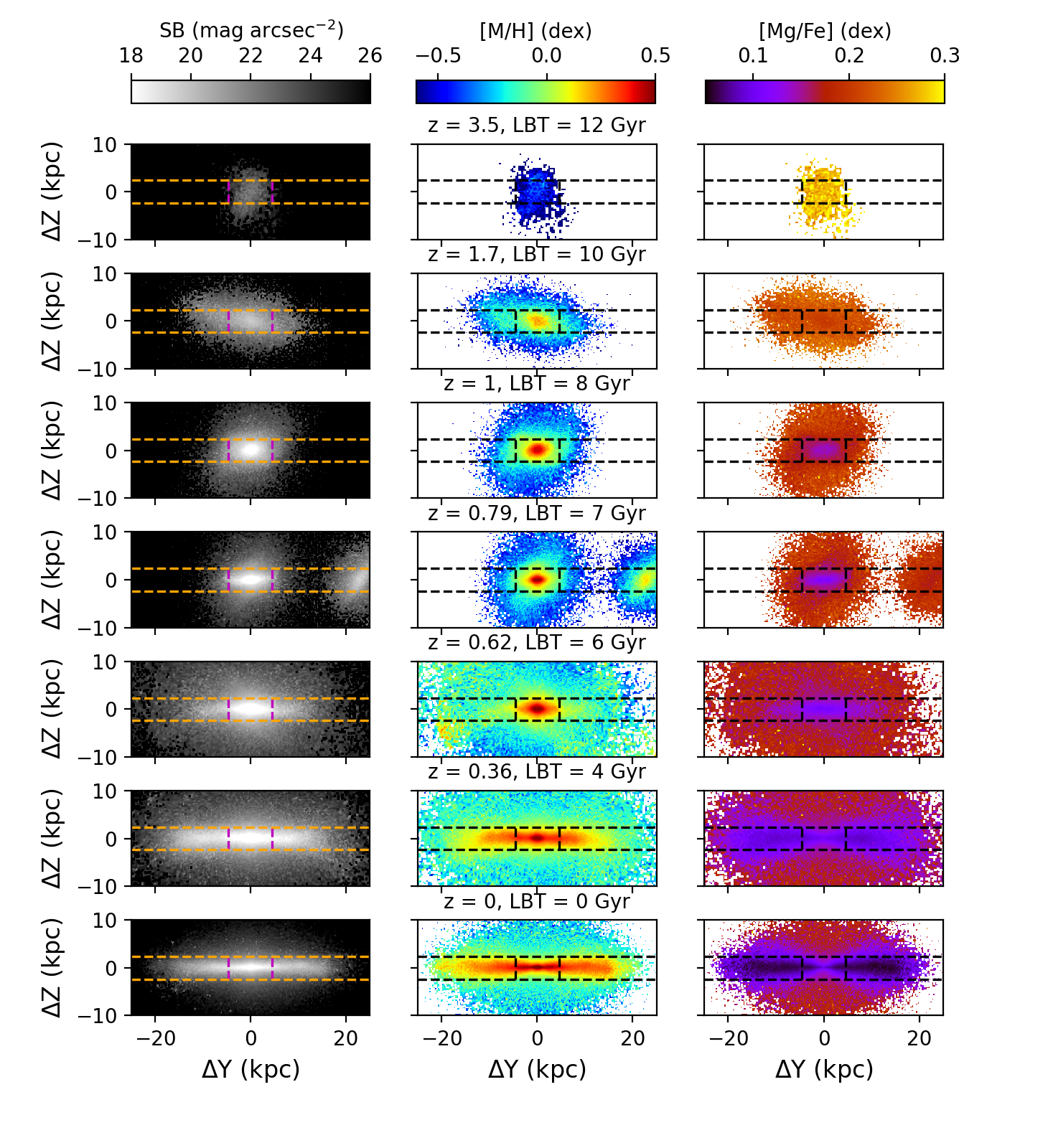}}
\caption{Same as Fig.~\ref{fig:snap_au7} but for the galaxy Au5. 
}
\label{fig:snap_au5}
\end{figure*}

\begin{figure*}
\centering
\resizebox{1.\textwidth}{!}
{\includegraphics[scale=1.5]{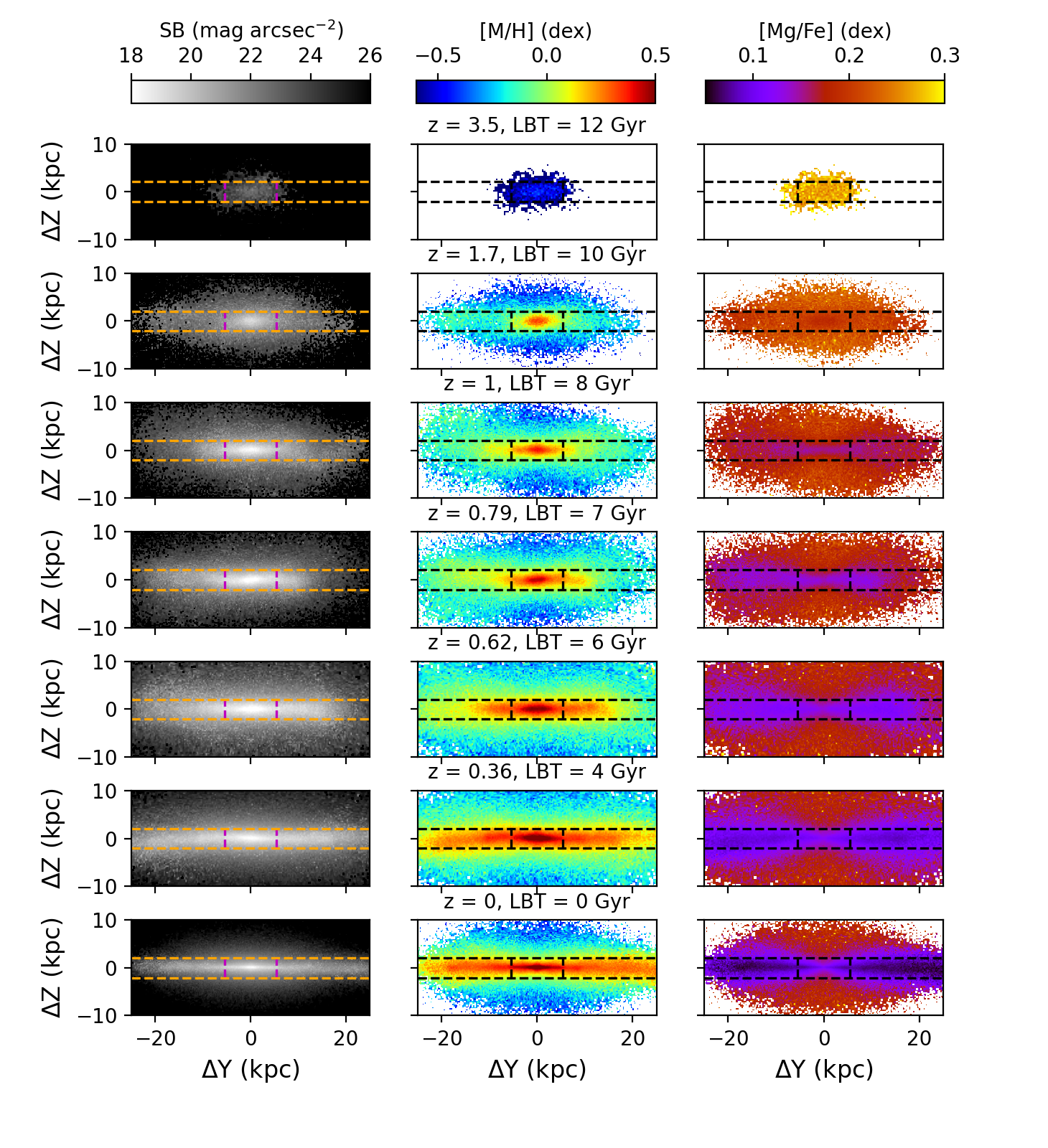}}
\caption{Same as Fig.~\ref{fig:snap_au7} but for the galaxy Au6. 
}
\label{fig:snap_au6}
\end{figure*}

\begin{figure*}
\centering
\resizebox{1.\textwidth}{!}
{\includegraphics[scale=1.5]{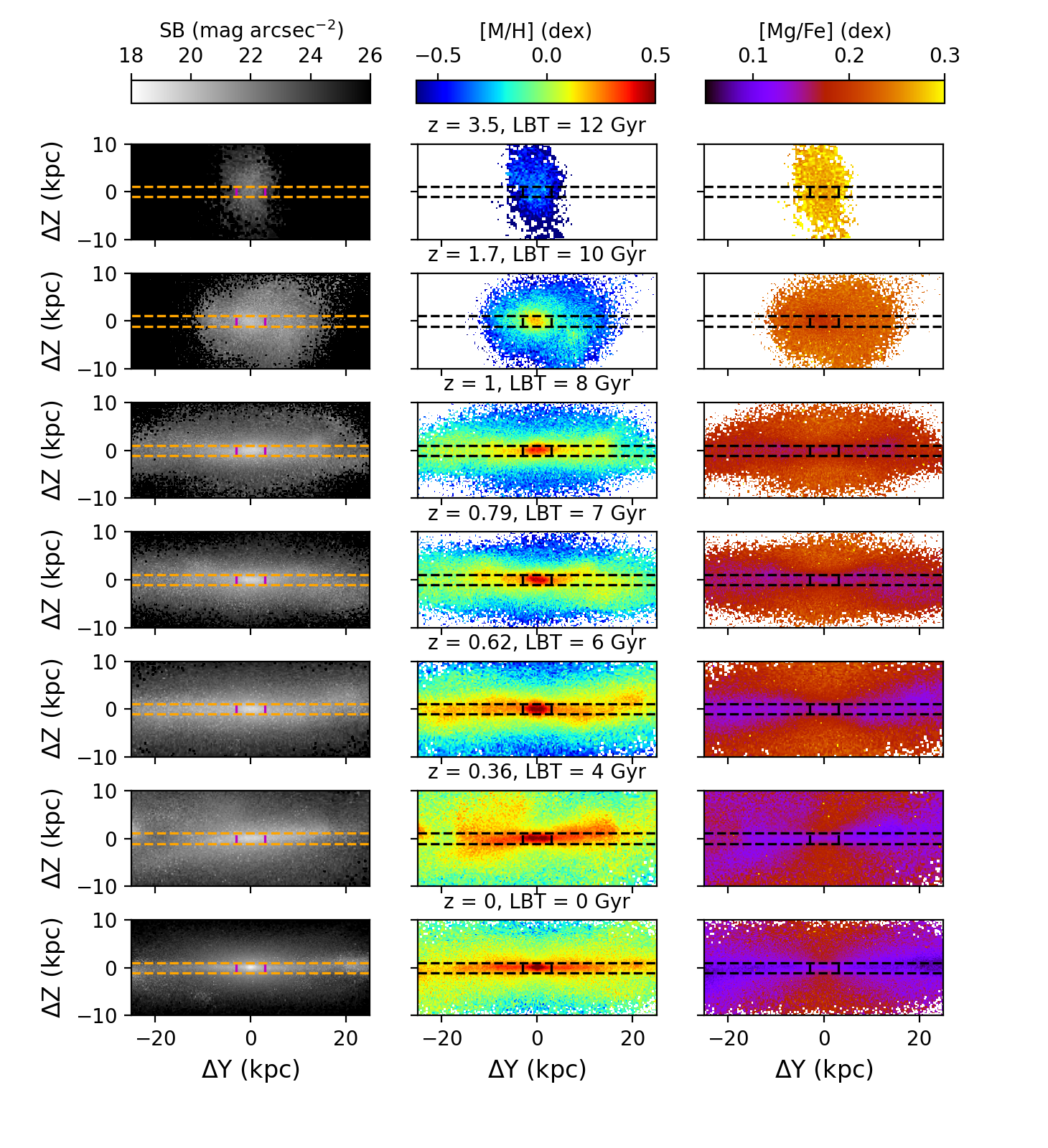}}
\caption{Same as Fig.~\ref{fig:snap_au7} but for the galaxy Au8. 
}
\label{fig:snap_au8}
\end{figure*}

\begin{figure*}
\centering
\resizebox{1.\textwidth}{!}
{\includegraphics[scale=1.5]{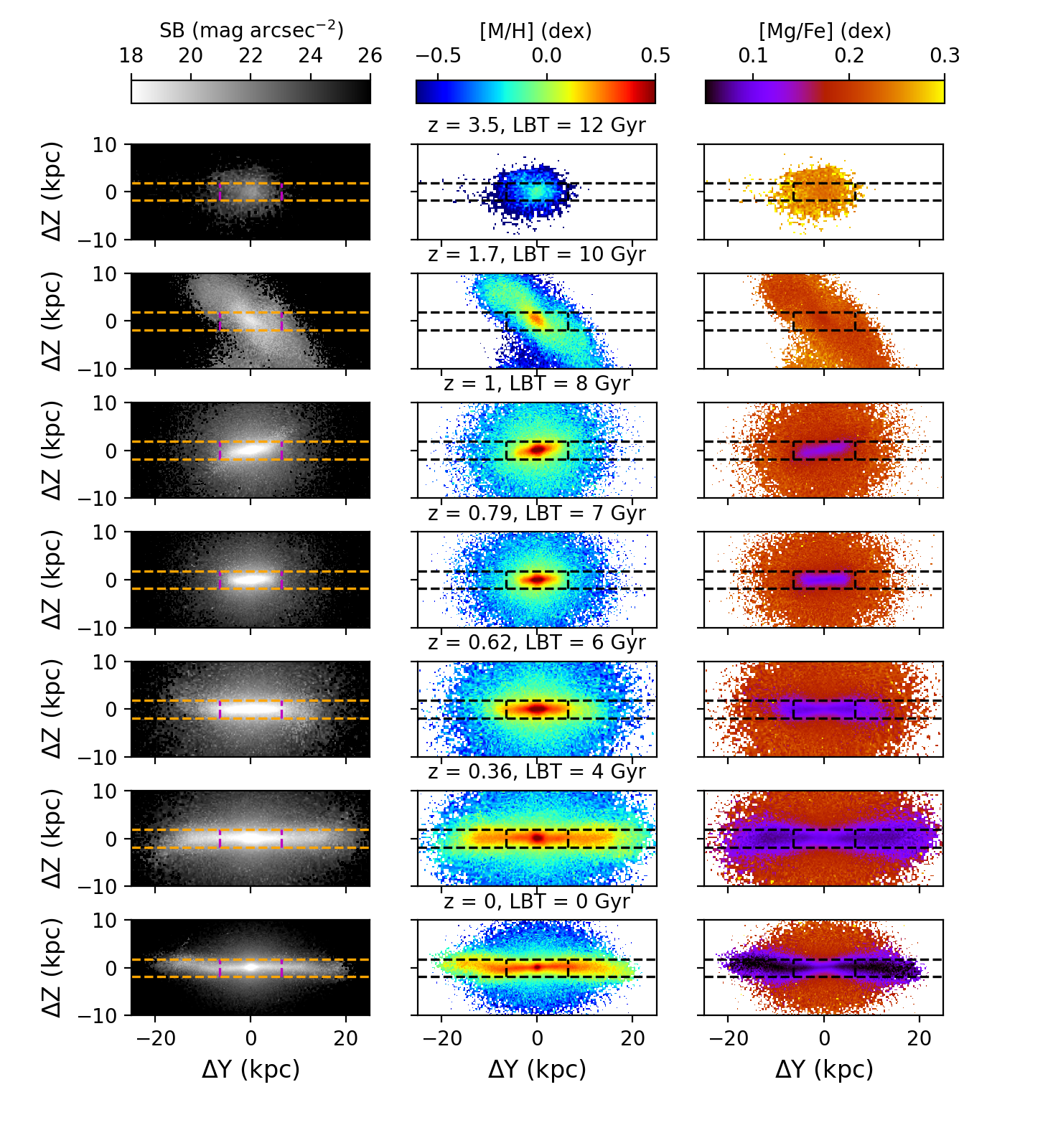}}
\caption{Same as Fig.~\ref{fig:snap_au7} but for the galaxy Au9. 
}
\label{fig:snap_au9}
\end{figure*}

\begin{figure*}
\centering
\resizebox{1.\textwidth}{!}
{\includegraphics[scale=1.5]{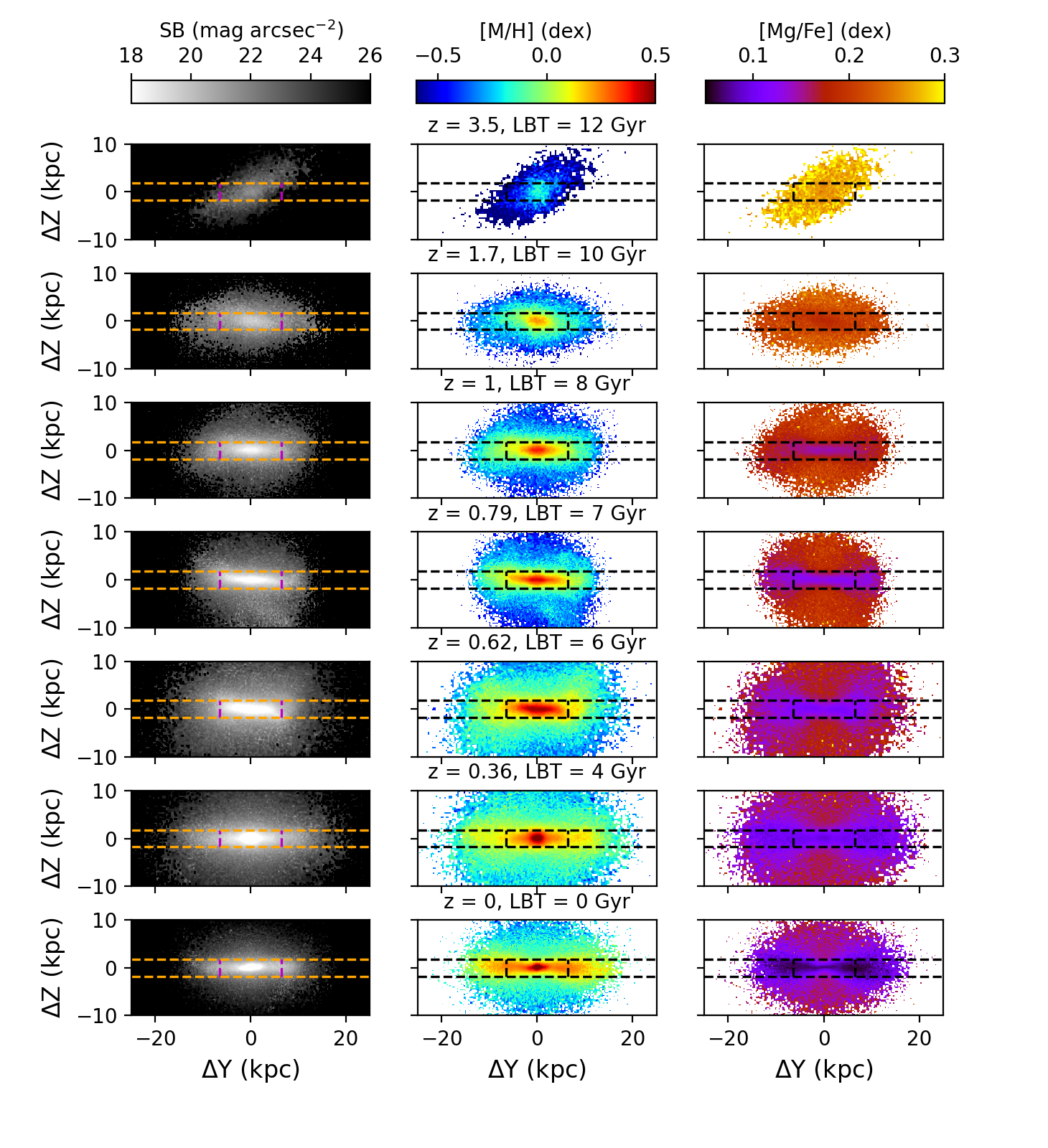}}
\caption{Same as Fig.~\ref{fig:snap_au7} but for the galaxy Au10. 
}
\label{fig:snap_au10}
\end{figure*}

\begin{figure*}
\centering
\resizebox{1.\textwidth}{!}
{\includegraphics[scale=1.5]{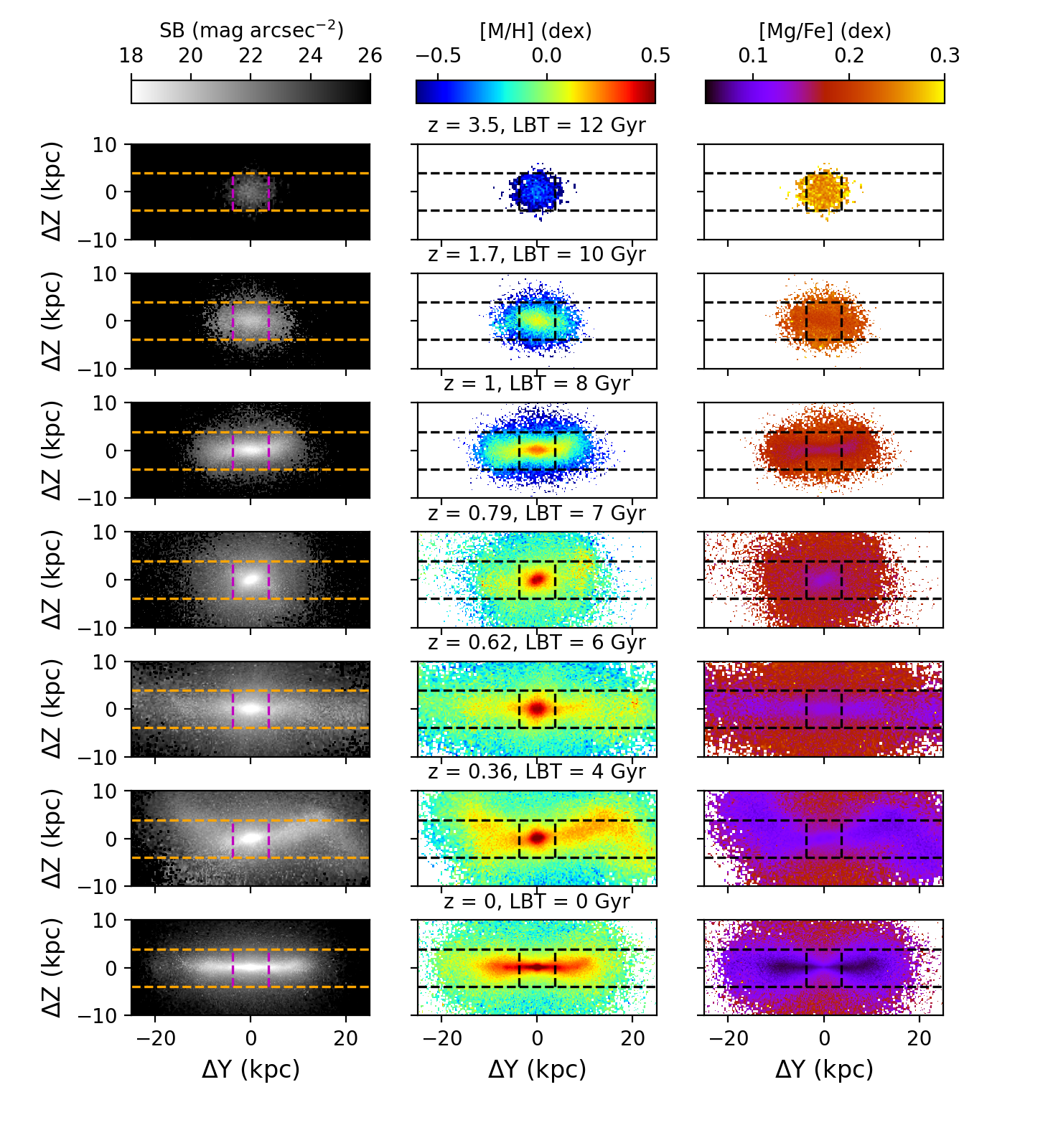}}
\caption{Same as Fig.~\ref{fig:snap_au7} but for the galaxy Au12. 
}
\label{fig:snap_au12}
\end{figure*}

\begin{figure*}
\centering
\resizebox{1.\textwidth}{!}
{\includegraphics[scale=1.5]{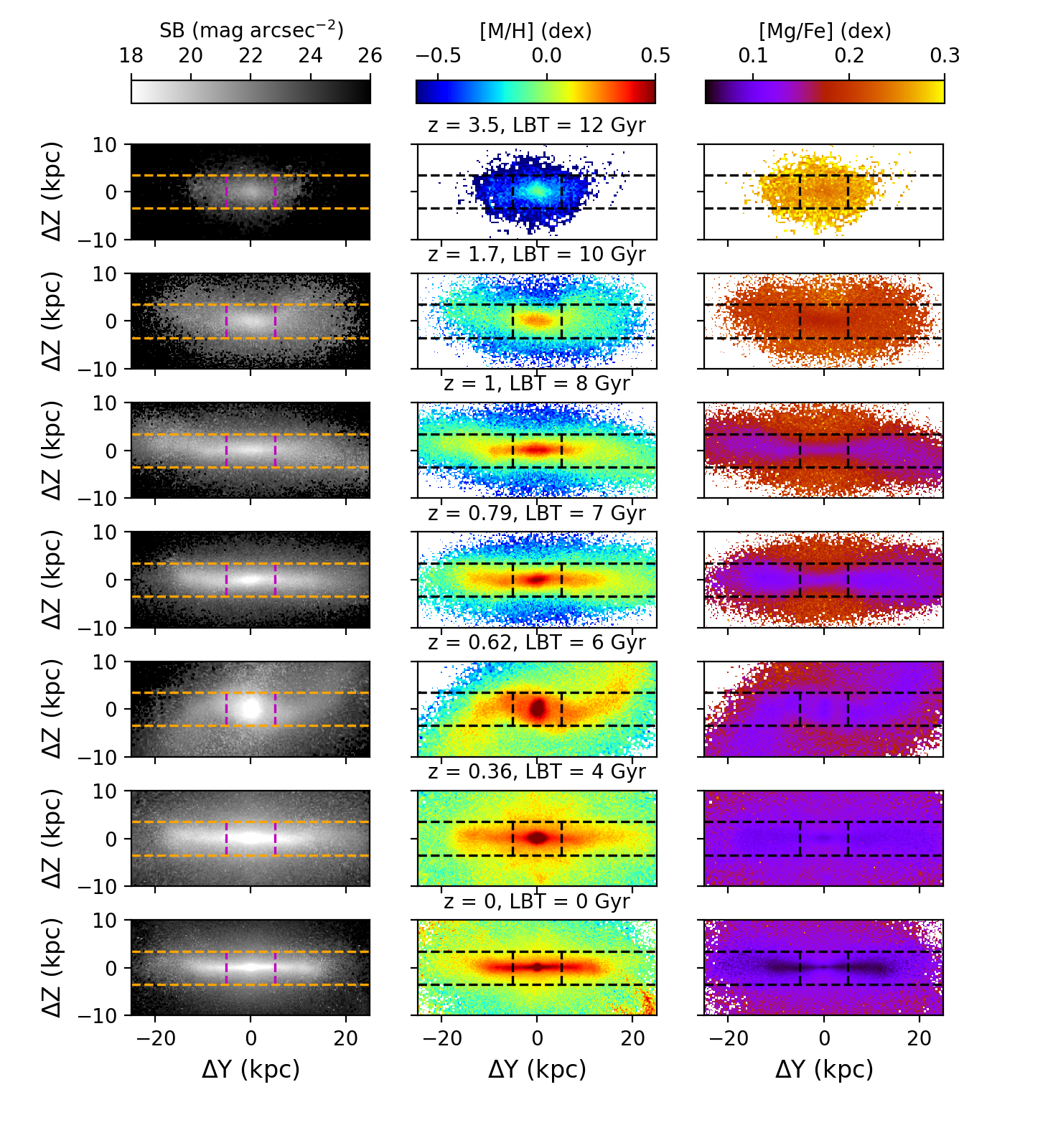}}
\caption{Same as Fig.~\ref{fig:snap_au7} but for the galaxy Au14. 
}
\label{fig:snap_au14}
\end{figure*}

For completeness, we show here the time evolution of all galaxies in our sample that were not shown in Sect.~\ref{sub:snap}. 
Au1 (Fig.~\ref{fig:snap_au1}) never grew an extended thin metal-rich disk and does not have a clear double-disk structure at $z=0$ (Sect.~\ref{sub:morph_decomp}).
Very recently (see the transition between redshift $z=0.36$ and $z=0$), its global growth was rapidly enhanced by an important merger, which also led to the scattering and distortions of the more metal-rich stars. This galaxy has a mass fraction of accreted stars of $\sim 46$\% at $z=0$ (Table~\ref{tab:accr}). 
At early times, Au2 and Au9 formed  a relatively metal-poor (thick) disk (which was already in place at $z=1.7$), with a metal-rich core that later expanded and formed the thin disk in an inside-out fashion (Fig.~\ref{fig:snap_au2} and \ref{fig:snap_au9}). The snapshot at $z=1.7$ in Fig.~\ref{fig:snap_au9} is an example of a failed alignment of the galaxy (see Sect.~\ref{sub:snap}). 
Au3, Au5, Au12 and Au14 (respectively Figs.~\ref{fig:snap_au3}, \ref{fig:snap_au5}, \ref{fig:snap_au12}, and \ref{fig:snap_au14}) show originally ($z=1.7$) a thick structure that grew inside out, while the metal-rich disk was formed only later on. 
In Au5 and Au12 this clearly happened after a merger between $z=0.62$ and 0.79, while at $z=0.36$ this thin disk was still disturbed in Au12. Au14 formed it at $z\sim 1$, it was disturbed by a merger around $z=0.62$, but it had recovered its thin and smooth shape at $z=0.36$. Au3 showed a metal-rich core since $z=3.5$ (12~Gyr ago), when it still had a spheroidal morphology. 
Au6 formed a small, thick, metal-poor disk (which had at $z=3.5$ the thickness of the thin disk at $z=0$ but a much smaller radial extension), which grew very fast in radius and height. It started to form a metal-rich thin disk, inside out, around $z=1$ (Fig.~\ref{fig:snap_au6}). 
Au7 had a similar but slightly slower initial evolution, with at least two mergers (at $z\lesssim1$ and $z\lesssim0.36$), which disturbed the initial disky morphology and prevented the formation of a massive thin disk until very recent times ($z<0.36$, Fig.~\ref{fig:snap_au7}). 
In Au8, the first metal-poor structure was spheroidal and a disky shape appeared only around $z \sim 1$ (Fig.~\ref{fig:snap_au8}). 
Au10, with a similar appearance to Au3 and Au5 at early times, did not grow much in radius, while it formed a thin metal-rich disk embedded in the thick disk (Fig.~\ref{fig:snap_au10}). 

In summary, AURIGA galaxies generally first formed a metal-poor, small spheroidal or thick-disk structure that then grew inside out and bottom up. They formed a metal-rich structure a bit later on (still very early in some galaxies, and much later in others), which grew inside out as a thin disk. 
All galaxies experienced the expected chemical enrichment \citep[e.g.,][]{Worthey1992} with time. They started forming the first stars with low metallicity and high [Mg/Fe], and the latter decreased only when and where metallicity increases. In different regions at different snapshots, all galaxies show anticorrelation between [M/H] and [Mg/Fe].

\begin{figure*}
\centering
\resizebox{1.\textwidth}{!}
{\includegraphics[scale=1.5]{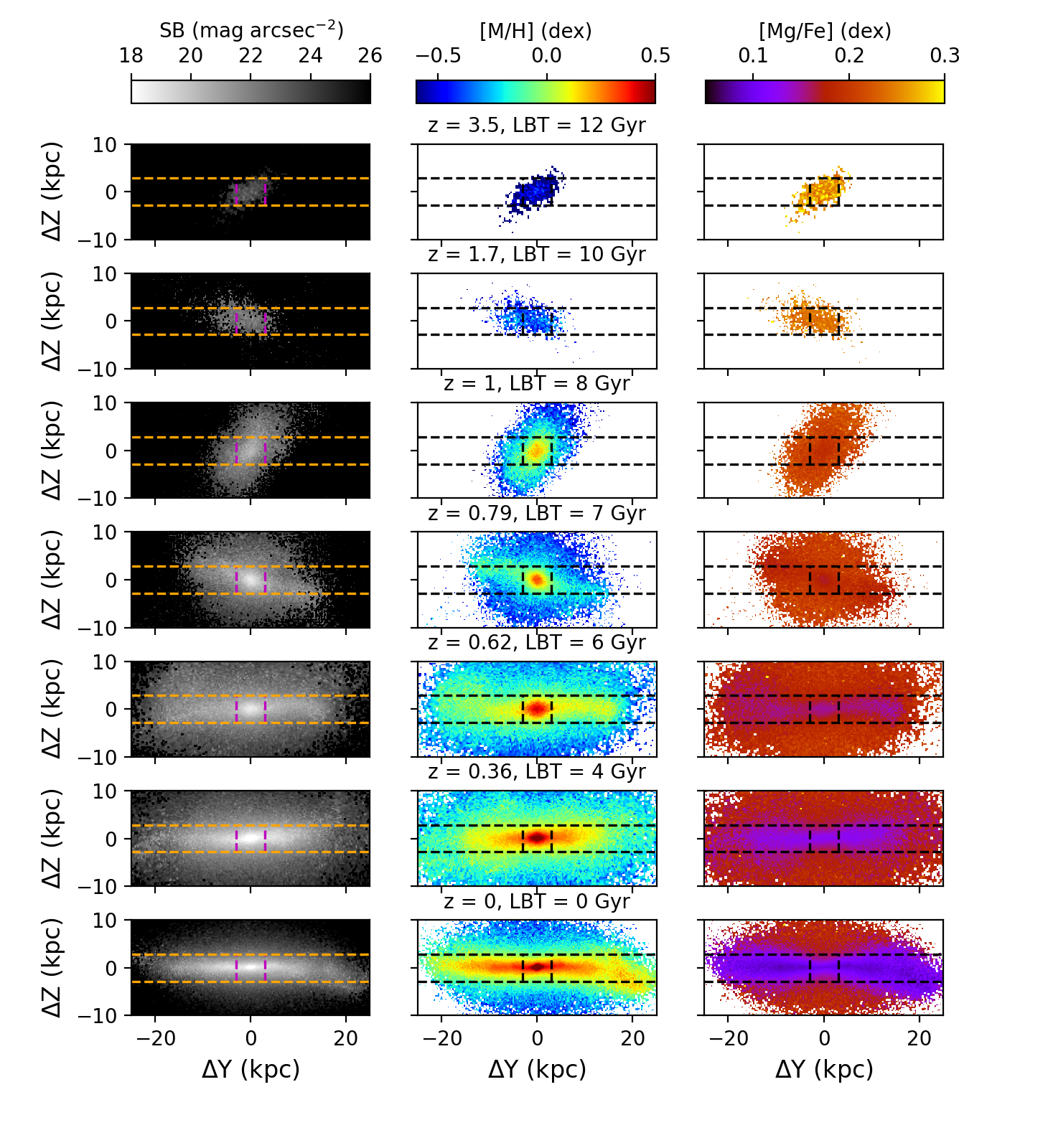}}
\caption{Same as Fig.~\ref{fig:snap_au7} but for the galaxy Au15. 
}
\label{fig:snap_au15}
\end{figure*}

\begin{figure*}
\centering
\resizebox{1.\textwidth}{!}
{\includegraphics[scale=1.5]{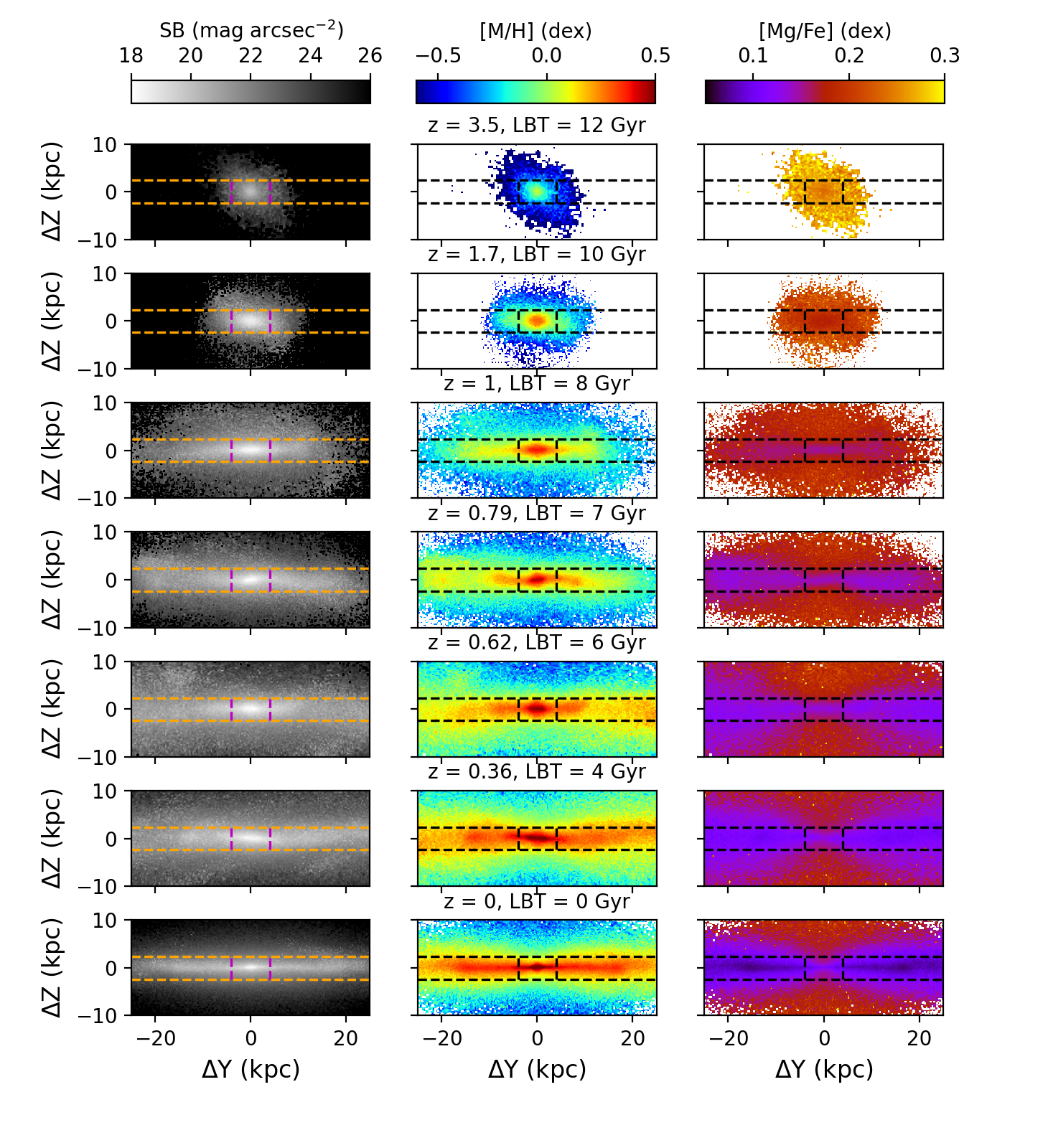}}
\caption{Same as Fig.~\ref{fig:snap_au7} but for the galaxy Au16. 
}
\label{fig:snap_au16}
\end{figure*}


\begin{figure*}
\centering
\resizebox{1.\textwidth}{!}
{\includegraphics[scale=1.5]{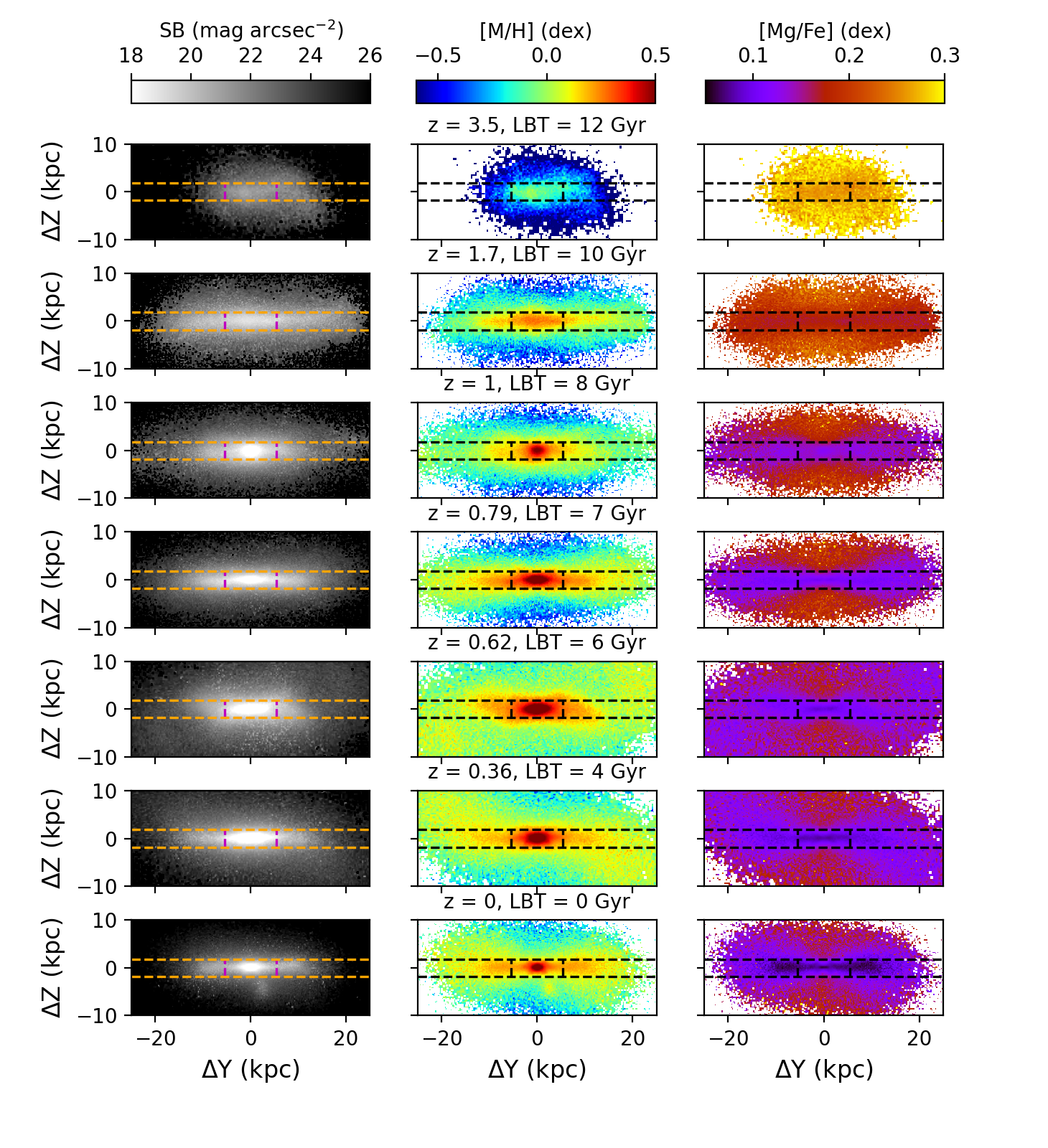}}
\caption{Same as Fig.~\ref{fig:snap_au7} but for the galaxy Au17. 
}
\label{fig:snap_au17}
\end{figure*}



\begin{figure*}
\centering
\resizebox{1.\textwidth}{!}
{\includegraphics[scale=1.5]{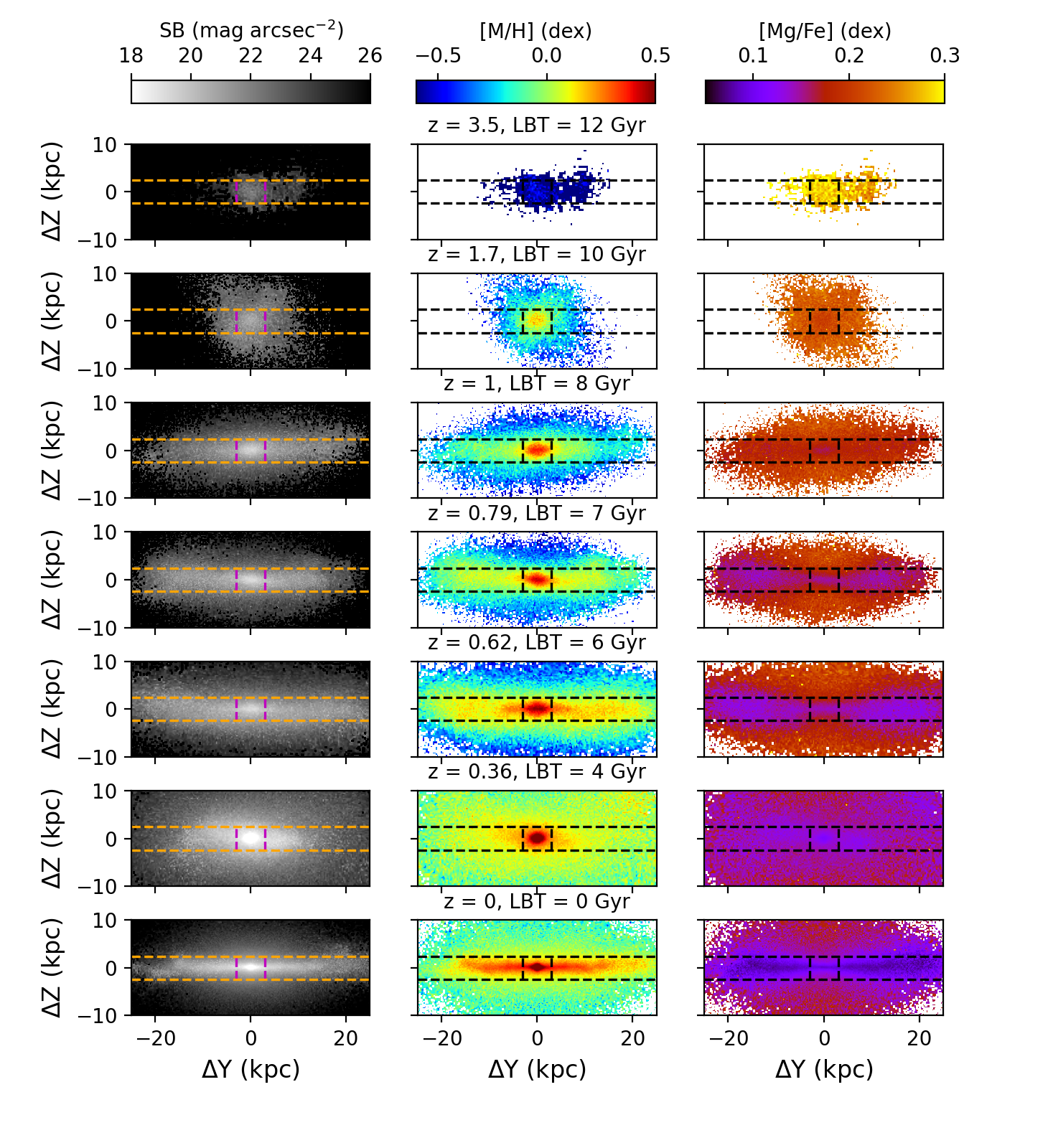}}
\caption{Same as Fig.~\ref{fig:snap_au7} but for the galaxy Au19. 
}
\label{fig:snap_au19}
\end{figure*}


\begin{figure*}
\centering
\resizebox{1.\textwidth}{!}
{\includegraphics[scale=1.5]{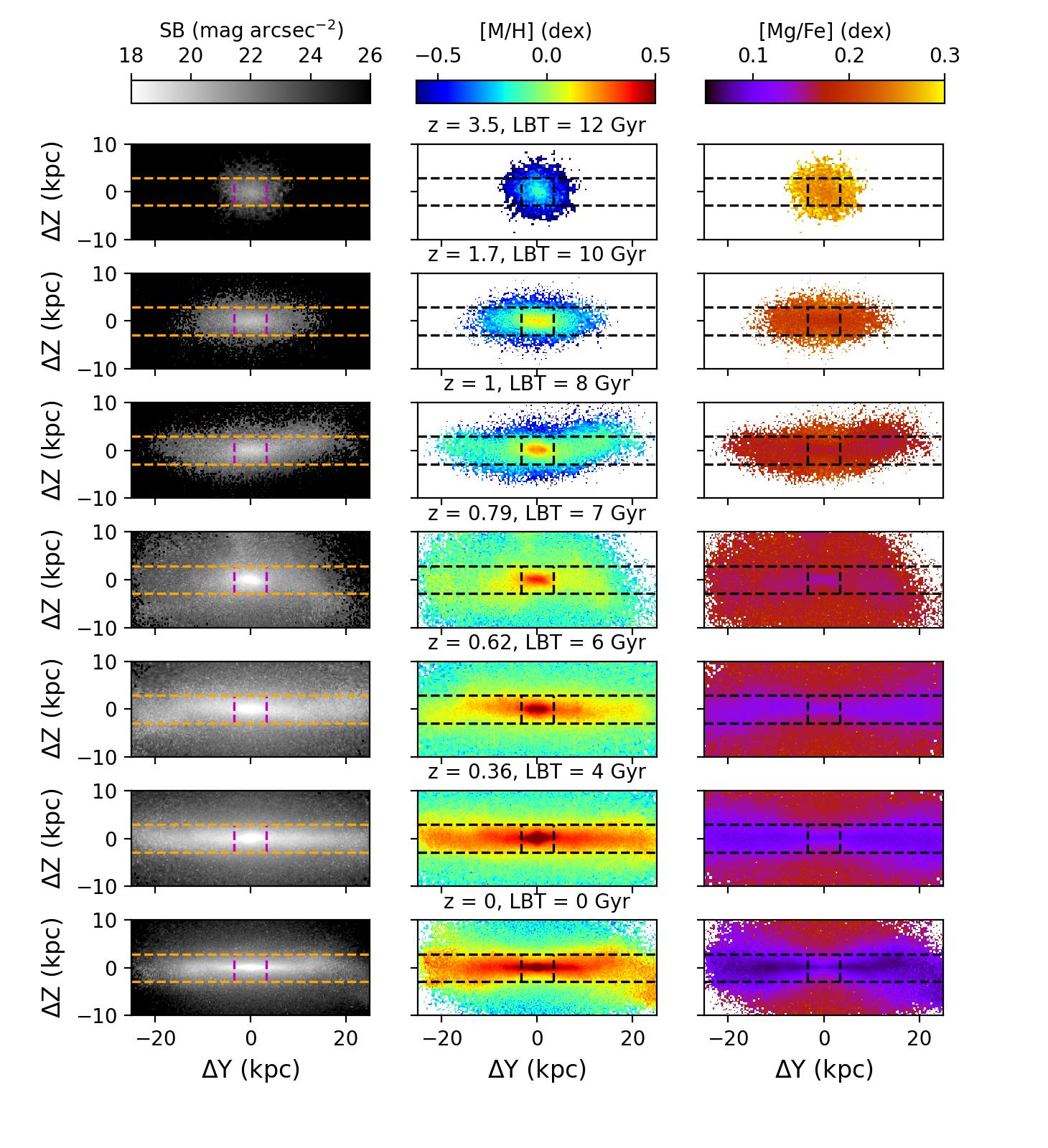}}
\caption{Same as Fig.~\ref{fig:snap_au7} but for the galaxy Au21. 
}
\label{fig:snap_au21}
\end{figure*}


\begin{figure*}
\centering
\resizebox{1.\textwidth}{!}
{\includegraphics[scale=1.5]{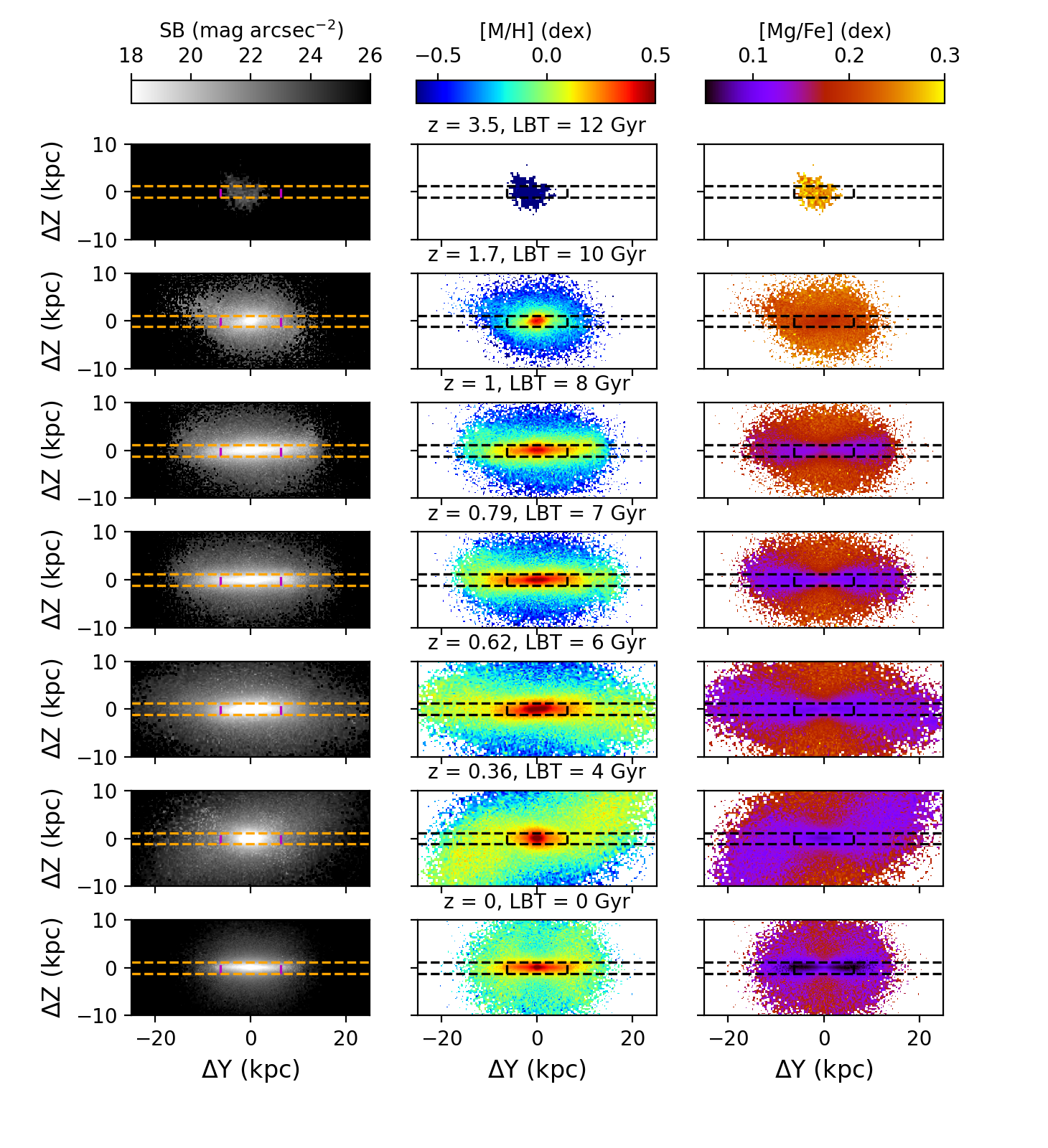}}
\caption{Same as Fig.~\ref{fig:snap_au7} but for the galaxy Au22. 
}
\label{fig:snap_au22}
\end{figure*}


\begin{figure*}
\centering
\resizebox{1.\textwidth}{!}
{\includegraphics[scale=1.5]{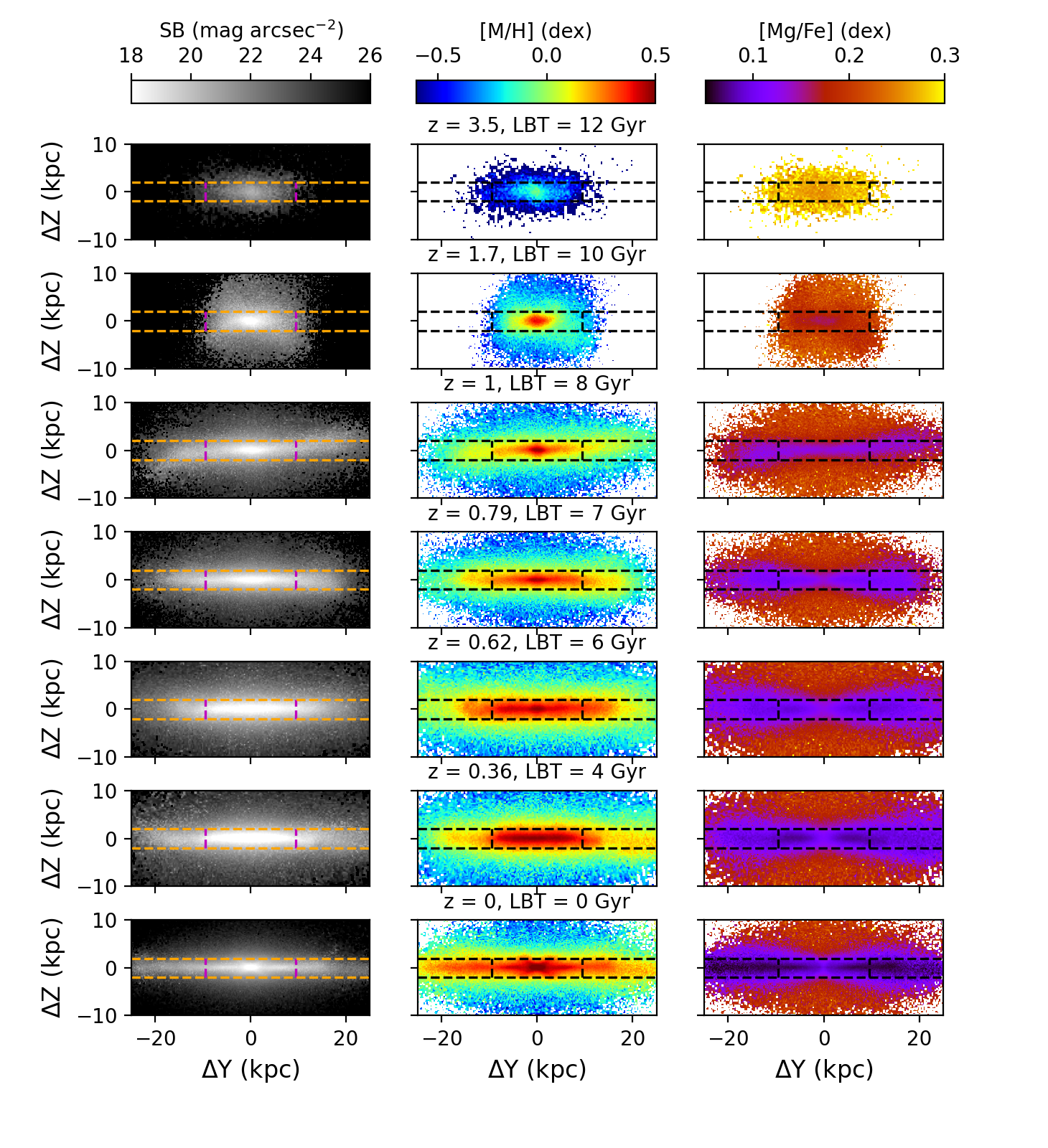}}
\caption{Same as Fig.~\ref{fig:snap_au7} but for the galaxy Au23. 
}
\label{fig:snap_au23}
\end{figure*}


\begin{figure*}
\centering
\resizebox{1.\textwidth}{!}
{\includegraphics[scale=1.5]{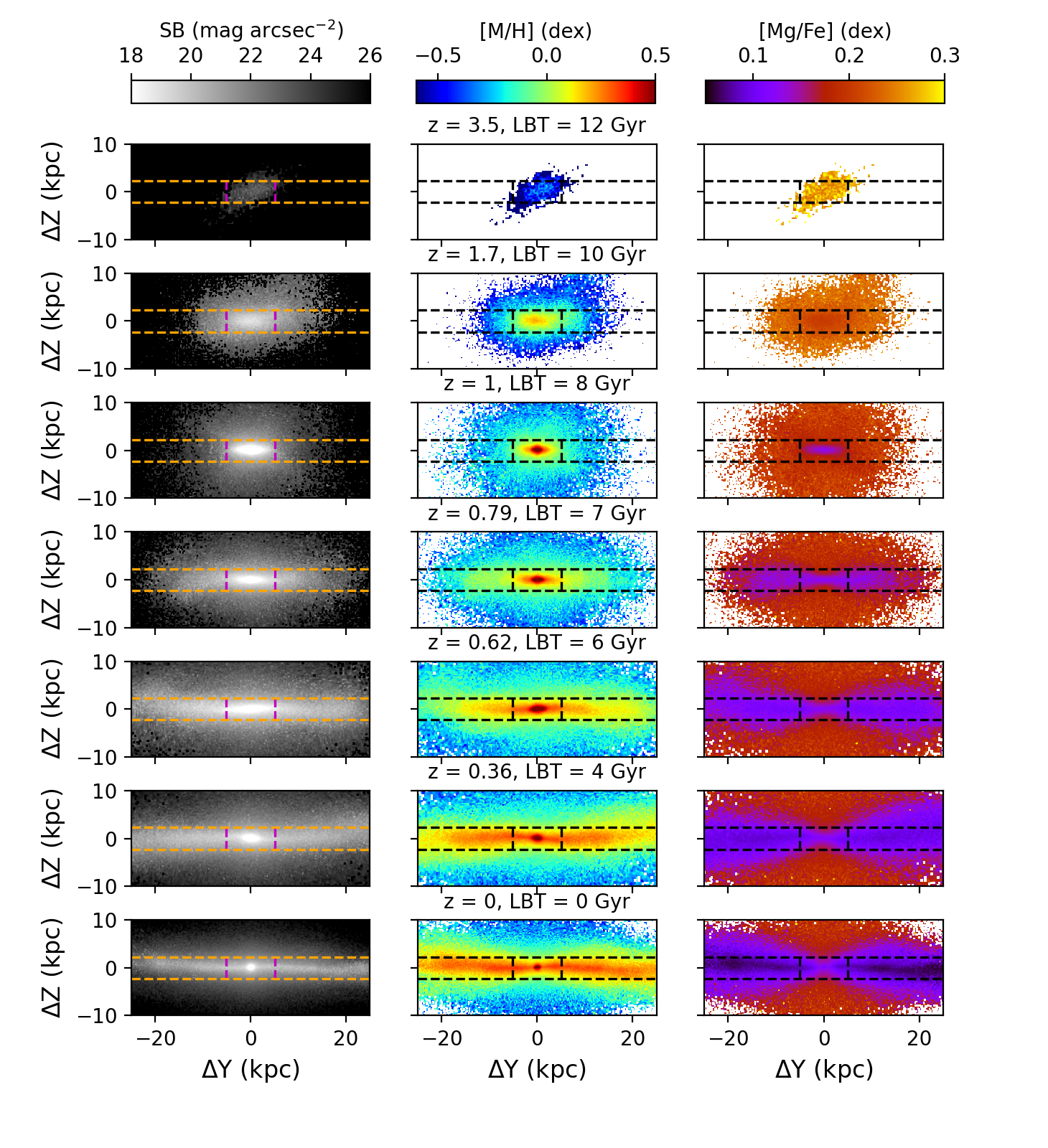}}
\caption{Same as Fig.~\ref{fig:snap_au7} but for the galaxy Au24. 
}
\label{fig:snap_au24}
\end{figure*}


\begin{figure*}
\centering
\resizebox{1.\textwidth}{!}
{\includegraphics[scale=1.5]{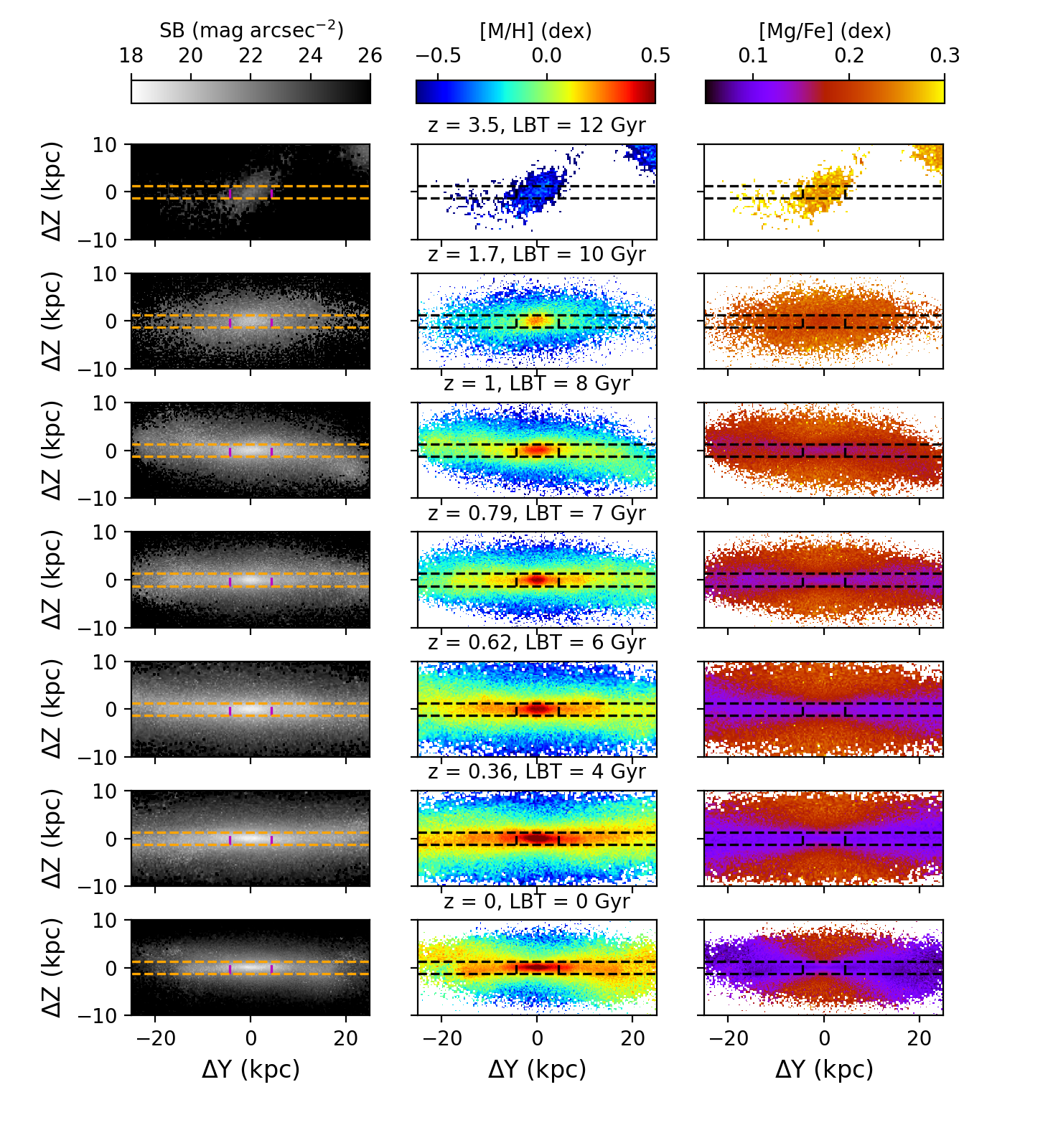}}
\caption{Same as Fig.~\ref{fig:snap_au7} but for the galaxy Au25. 
}
\label{fig:snap_au25}
\end{figure*}

\begin{figure*}
\centering
\resizebox{1.\textwidth}{!}
{\includegraphics[scale=1.5]{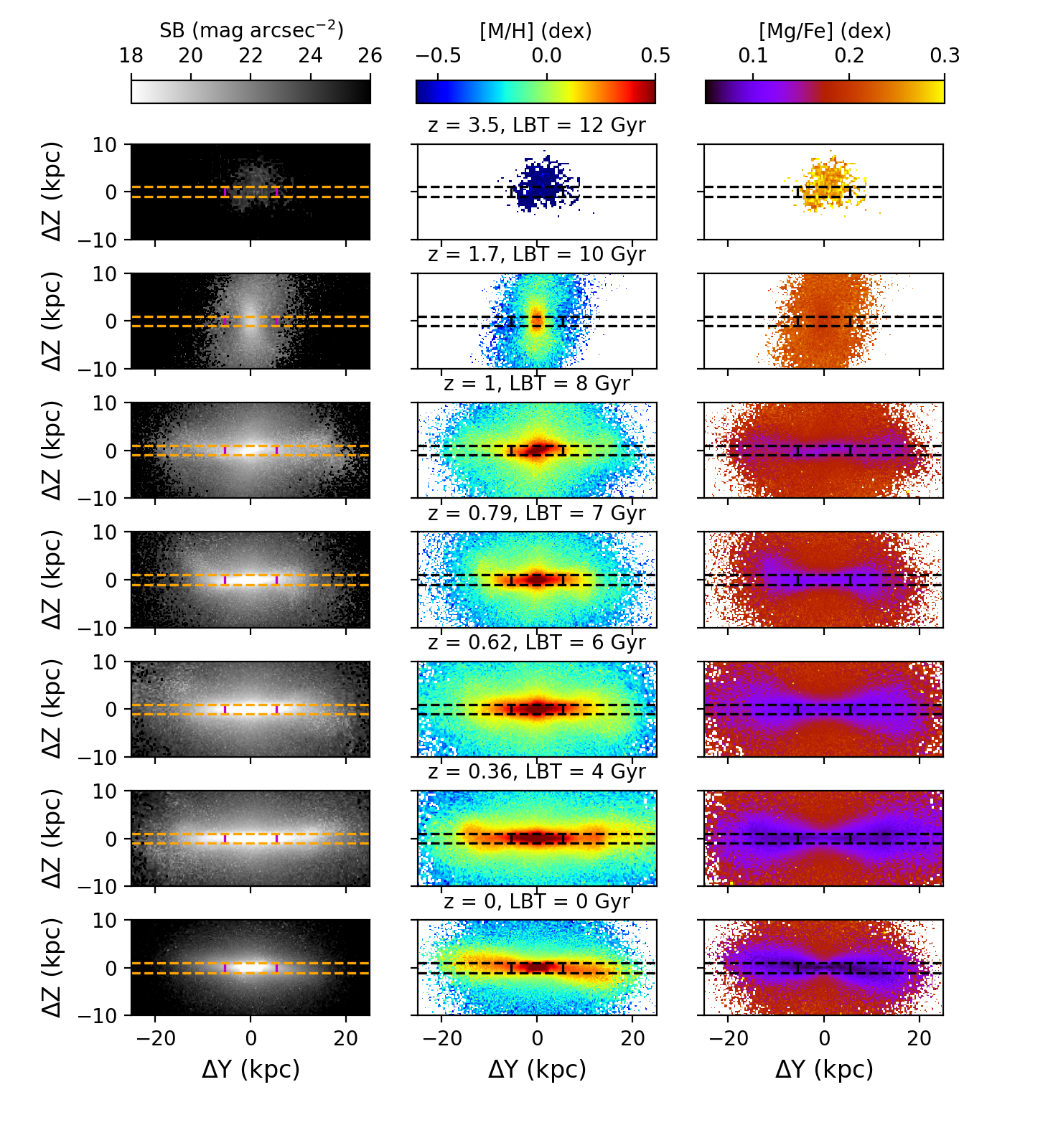}}
\caption{Same as Fig.~\ref{fig:snap_au7} but for the galaxy Au26. 
}
\label{fig:snap_au26}
\end{figure*}


\begin{figure*}
\centering
\resizebox{1.\textwidth}{!}
{\includegraphics[scale=1.5]{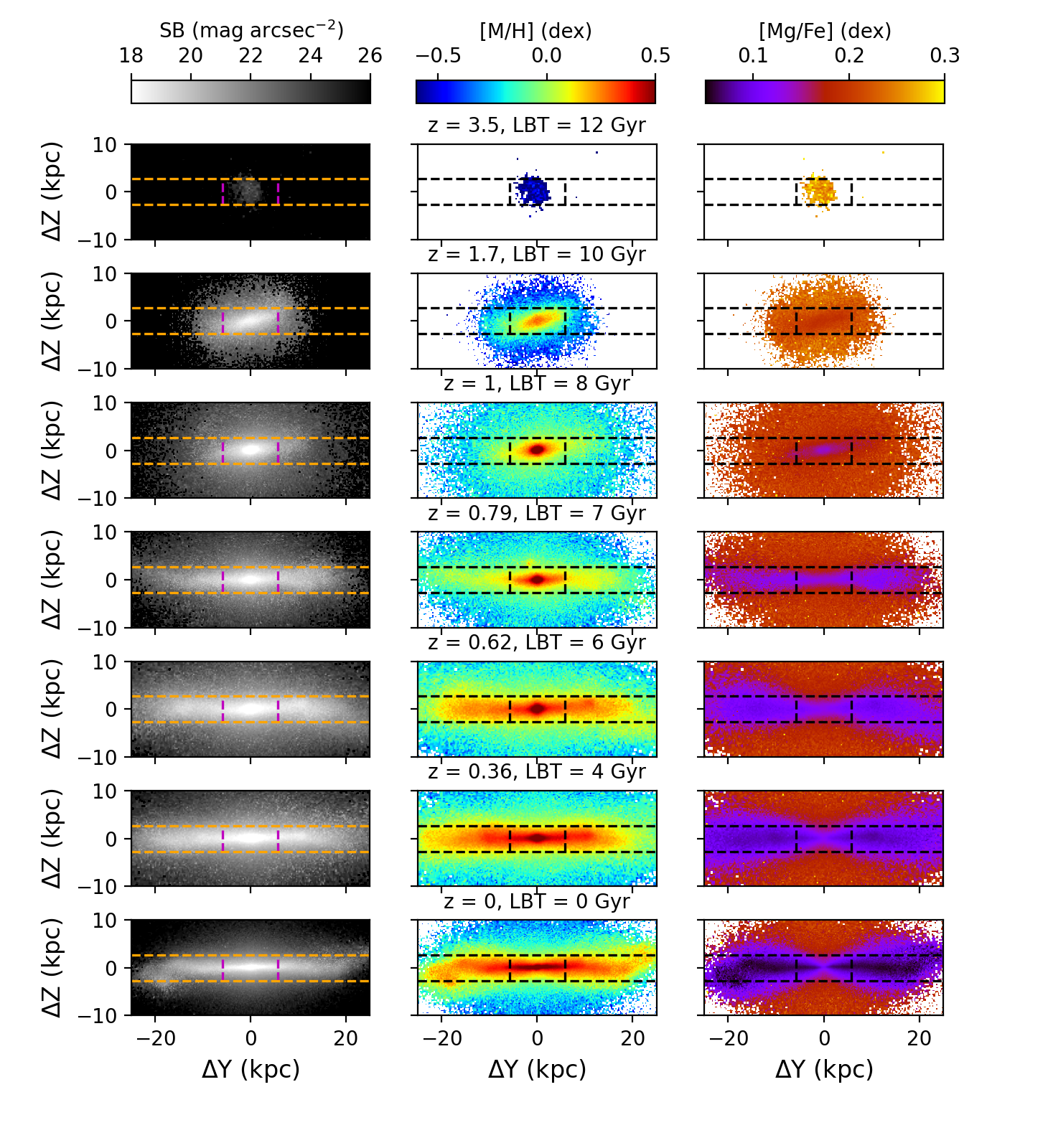}}
\caption{Same as Fig.~\ref{fig:snap_au7} but for the galaxy Au27. 
}
\label{fig:snap_au27}
\end{figure*}


\begin{figure*}
\centering
\resizebox{1.\textwidth}{!}
{\includegraphics[scale=1.5]{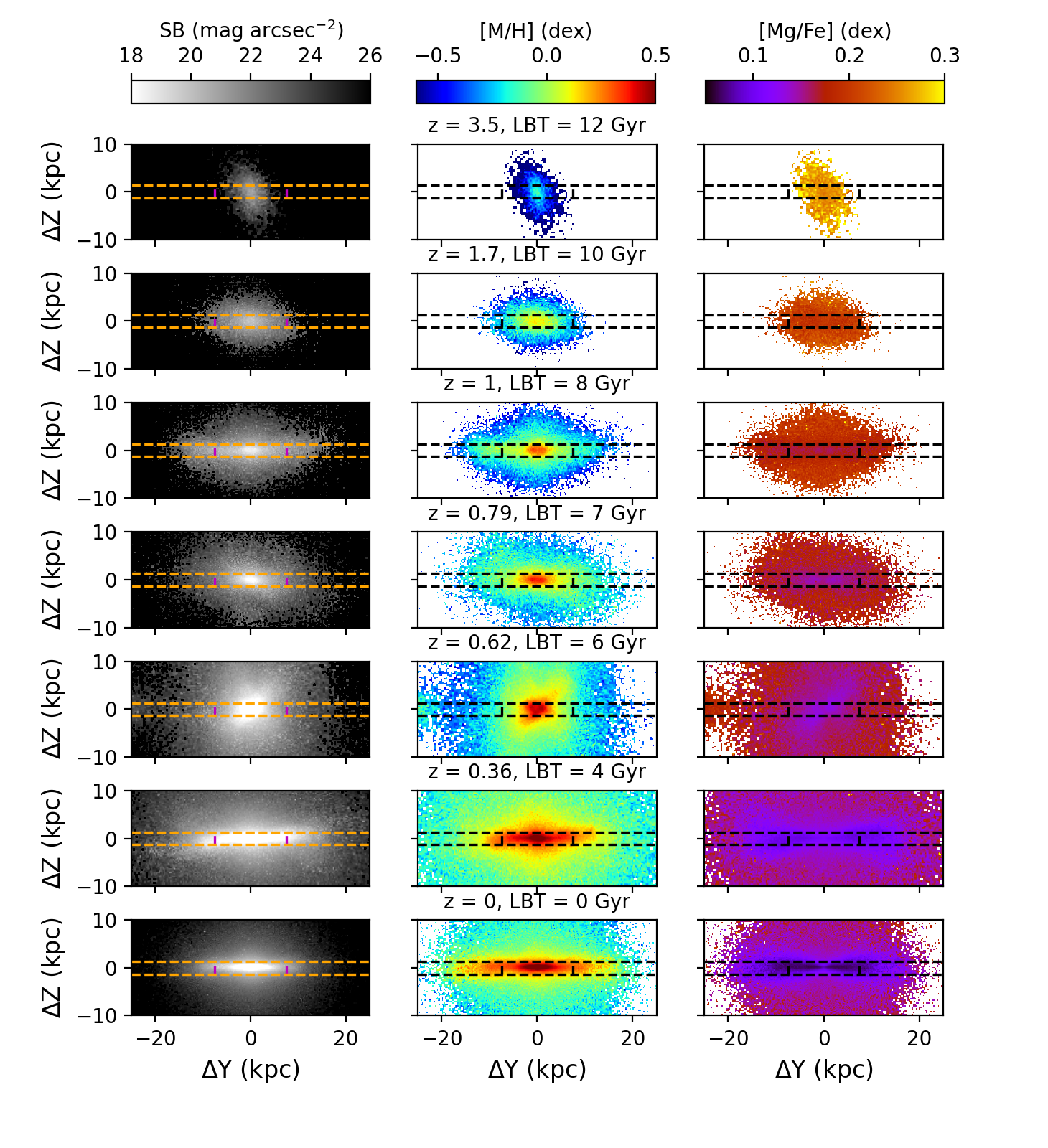}}
\caption{Same as Fig.~\ref{fig:snap_au7} but for the galaxy Au28.  
}
\label{fig:snap_au28}
\end{figure*}


\end{appendix}
\end{document}